\documentclass[a4paper,12pt]{article}
\usepackage{a4wide}
\usepackage[bookmarks=false,colorlinks]{hyperref}
\usepackage{graphicx,graphics,epsfig,color}
\usepackage{amsmath,amsfonts,amssymb,mathtools,bm}
\usepackage{rotating}
\usepackage{tabularx}
\usepackage{multicol,multirow}
\usepackage{latexsym}
\usepackage{relsize}
\usepackage[center]{subfigure}
\usepackage{float}
\usepackage{caption}
\usepackage{slashed}
\usepackage{lineno}
\usepackage{cite,mcite}
\usepackage{appendix}
\usepackage{url}
\hypersetup{urlcolor=blue, citecolor=blue}

\date{\today}
 
\normalsize

\def\Bbar{\overline{B}}
\def\Kbar{\overline{K}}
\def\Mbar{\overline{M}}
\def\bbar{\overline{b}}
\def\sbar{\overline{s}}

\def\lbar{\overline{\ell}}
\def\sq{{\tilde{q}}}
\def\sg{{\tilde{g}}}
\def\M{\mathcal{M}}
\def\A{\mathcal{A}}
\def\BR{\mathcal{B}}
\def\Re{\mathcal{R}e}
\def\Im{\mathcal{I}m}
\def\O{\mathcal{O}}
\def\Op{\mathcal{O}^{\,\prime}}
\def\OOp{\mathcal{O}^{\,(\prime)}}
\def\Cp{C^{\,\prime}}

\def\Cpc{C^{\,\prime\,*}}
\def\CSM{C^{\,\rm(SM)}}
\def\CNP{C^{\,\rm(NP)}}

\def\CpNP{C^{\,\prime\,\rm(NP)}}
\def\CCp{C^{\,(\prime)}}

\def\CCpNP{C^{\,(\prime)\,\rm(NP)}}

\def\CCpeff{C^{\,(\prime)\,\rm eff}}

\begin{document}
\thispagestyle{empty} 
\rightline{LPT-ORSAY 12-38}
\rightline{LAL 12-146}
\vspace{2.5cm} 
{\Large
\begin{center}
	{\bf Future prospects for the determination} \\
	{\bf of the Wilson coefficient $\Cp_{7\gamma}$}
\end{center}
}
\vspace{0.3cm}

\begin{center}
	{\sc D. Be\v{c}irevi\'c$^{\,a}$, E. Kou$^{\,b}$, A. Le Yaouanc$^{\,a}$ and A. Tayduganov$^{\,a,b}$} \\
	\vspace{0.3cm}
	\small
	\emph{$^a$ Laboratoire de Physique Th\'eorique, CNRS/Univ. Paris-Sud 11 (UMR 8627)}\\
	\emph{91405 Orsay, France}

	\emph{$^b$ Laboratoire de l'Acc\'el\'erateur Lin\'eaire, Univ. Paris-Sud 11, CNRS/IN2P3 (UMR 8607)} \\
	\emph{91405 Orsay, France}
\end{center}

\vskip4cm
\begin{center}
	\small{\bf Abstract}\\[3mm]
\end{center}
We discuss the possibilities of assessing a non-zero $\Cp_{7\gamma}$ from the direct and the indirect measurements of the photon polarization in the exclusive $b\to s\gamma^{(*)}$ decays. We focus on three methods and explore the following three decay modes: $B\to K^*(\to K_S\pi^0)\gamma$, $B\to K_1(\to K\pi\pi)\gamma$, and $B\to K^*(\to K\pi)\ell^+\ell^-$. By studying different New Physics scenarios we show that the future measurement of conveniently defined observables in these decays could provide us with the full determination of $C_{7\gamma}$ and $\Cp_{7\gamma}$. 

\vskip3cm
{\noindent\small PACS: 12.90.+b, 13.20.He}

\newpage


\section{Introduction}

The radiative decay $b\to s\gamma$ has been extensively studied as a probe of the flavour structure of the Standard Model (SM) as well as New Physics (NP), beyond the SM. While the majority of studies has been focused on the prediction of the decay rates of exclusive and inclusive $b\to s\gamma$ decays, relatively few studies of the right-handed currents in these decays have been made. In the SM, the emitted photon is predominantly left-handed in $b$, and right-handed in $\bbar$ decays. This is due to the fact that the dominant contribution comes from the chiral-odd dipole operator $\sbar_{L(R)}\sigma_{\mu\nu}b_{R(L)}$. As only left-handed quarks participate in weak interaction, this effective operator induces a helicity flip on one of the external quark lines, which results in a factor $m_b$ for $b_R\to s_L\gamma_L$, and a factor $m_s$ for $b_L\to s_R\gamma_R$. Hence, the emission of right-handed photons is suppressed by a factor $m_s/m_b$. This suppression can be lifted in some NP models where the helicity flip occurs on an internal line, which brings in a factor $m_{\rm NP}/m_b$ instead of $m_s/m_b$~\footnote{This so-called chiral enhancement occurs only in the $b\to s\gamma$ and $b\to sg$ processes with the photon and gluon being on-shell. In the supersymmetric models this issue was studied using the time-dependent $CP$-asymmetry in the $B\to\phi K_S$ decay as one can have a large NP effect in the penguin loop diagram that is suppressed in the box diagrams~\cite{Khalil:2002fm}. Another hint of NP is discussed in the direct $CP$-violation in the $D\to K^+K^-,~\pi^+\pi^-$ which allow a large NP contribution within the constraint from the $D-\overline{D}$ mixing~\cite{Brod:2011re}.}. If the amplitude for $b\to s\gamma_R$ is of the same order as the SM prediction, or the enhancement of $b\to s\gamma_R$ goes along with the suppression of $b\to s\gamma_L$, the impact on the branching ratio is small since the two helicity amplitudes add incoherently. This implies that there can be a substantial contribution of NP to $b\to s\gamma$ escaping detection when only branching ratios are measured. Therefore, the photon polarization measurement could provide a good test of the SM or at least a useful indication of NP. However, since in our work we are dealing with exclusive decays, the non perturbative QCD effects, which are always hard to calculate, can have a non negligible contribution to the right-handed amplitude and therefore must be taken into account.

In some NP models the right-handed contribution can be significantly enhanced. In the Minimal Supersymmetric Standard Model (MSSM), it is known that the squark mass matrices and the trilinear couplings of squarks to the Higgs bosons, coming from the soft supersymmetry breaking terms in the Lagrangian, are not diagonal in the quark basis, which makes possible for squarks to change their flavour and chirality. In other words, the chirality can be flipped on the squark line propagating inside the loop of $b\to s\gamma$ leading to a right-handed photon emission (see e.g. Refs.~\cite{Everett:2001yy,*Foster:2006ze,*Lunghi:2006hc,*Goto:2007ee} and references therein). In the class of Left-Right Symmetric Models (LRSM), large contributions to the $b\to s\gamma$ decay amplitude can arise from the mixing of the $W_L$ and $W_R$ gauge bosons as well as from the charged Higgs boson~\cite{Babu:1993hx}. These amplitudes are enhanced by the factor $m_t/m_b$ compared to the contributions in the SM. In the Grand Unification models a right-handed quark coupling can appear by introducing the right-handed neutrino enhancing the ``wrong" helicity amplitude.

Three methods have been proposed for the measurement of the photon polarization:\footnote{Additional methods can be devised by considering $\Lambda_b\to\Lambda^{(*)}\gamma$~\cite{Mannel:1997pc,*Hiller:2001zj,*Legger:2006cq,*Hiller:2007ur,*Mannel:2011xg}, $\Xi_b\to\Xi^*\gamma$~\cite{Oliver:2010im}.} 

\begin{itemize}
	\item An {\it indirect} determination of the photon polarization, proposed by Atwood {\it et al.}~\cite{Atwood:1997zr,Atwood:2007qh}, is the measurement of the time-dependent mixing-induced $CP$-asymmetry in the radiative neutral $B$-mesons decays $B\to f^{CP}\gamma$ (where $f^{CP}$ is the $CP$-eigenstate). Prominent examples are $B\to K^*(\to K_S\pi^0)\gamma$, and $B_s\to\phi\gamma$. Such measurements are expected to be made at future super $B$~factories, reduce the experimental error on the asymmetry parameter $S_{K_S\pi^0\gamma}$ down to~2\%~\cite{Bona:2007qt}.

	\item A {\it direct} determination, proposed by Gronau {\it et al.}~\cite{Gronau:2001ng,*Gronau:2002rz}, is based on the study of the angular distribution of the three-body final state, $K\pi\pi$, coming from the axial vector $K_1(1^+)$-meson decay, in $B\to K_1(\to K\pi\pi)\gamma$. In Ref.~\cite{Kou:2010kn} this method was improved by using a new variable $\omega$, which includes not only the angular dependence but also the dependence on the three-body Dalitz variables which can significantly improve the sensitivity of the measurement of the polarization parameter. 
	Recently measured by the Belle collaboration $\BR(B\to K_1(1270)\gamma)$~\cite{Yang:2004as} appeared to be comparable to $\BR(B\to K^*\gamma)$, which opened the possibility of measuring the photon polarization in $B\to K_1 \gamma$.

	\item Another {\it indirect} way to study the right-handed currents is based on the angular analysis in the semileptonic $B\to K^*(\to K\pi)\ell^+\ell^-$ decay, proposed in Refs.~\cite{Melikhov:1998cd,Kruger:2005ep} and in many subsequent works. In particular, two transverse asymmetries, $\A_T^{(2)}(q^2)$ and $\A_T^{(\rm im)}(q^2)$, which can be expressed in terms of parallel and perpendicular spin amplitudes of $K^*$, are highly sensitive to the $b\to s\gamma$ process at very low dilepton invariant mass squared $q^2=(p_{\ell^+}+p_{\ell^-})^2$. 
\end{itemize}

As we will show in Section~\ref{sec:NP_constraints}, these three methods, having their own advantages and disadvantages, can be complementary to each other. Combination of all three of them can in principle put a strong constraint on the short-distance $\CCp_{7\gamma}$ coefficients in a model-independent way which then can be used as a constraint in building the NP models. In this sense the content of this paper is a contribution to the broad effort in the particle physics community to search for NP through $b\to s$ exclusive decays~\cite{Alok:2010zd,*DescotesGenon:2011yn,*Drobnak:2011aa,*Buras:2012ts,*Mahmoudi:2012un,*Beaujean:2012uj,*Matias:2012xw,*deBruyn:2012wk,*Behring:2012mv,*Ghosh:2012dh,*Altmannshofer:2012ir,*Becirevic:2012fy}.

This paper is organized as follows. In Section~\ref{sec:b-sgamma}, we remind the reader the basic formalism of the $b\to s\gamma$ process and explain the importance of the photon polarization measurement. In Section~\ref{sec:methods}, we briefly introduce three methods proposed for the determination of the Wilson coefficients $\CCp_{7\gamma}$ and discuss the sensitivity of the future experiments, namely the super $B$~factories and LHCb, to the ratio $\Cp_{7\gamma}/C_{7\gamma}$ using different $B$-meson decay modes. In particular, we consider the processes: $\Bbar^0\to\Kbar^{*0}(\to K_S\pi^0)\gamma$, $\Bbar\to\Kbar_1(1270)\gamma\to(\Kbar\pi\pi)\gamma$ and $\Bbar^0\to\Kbar^{*0}(\to K^-\pi^+)\ell^+\ell^-$. In Section~\ref{sec:NP_constraints}, we combine different methods and illustrate the possible future constraints on $\Cp_{7\gamma}/C_{7\gamma}$.

\section{Photon polarization in the $b\to s\gamma$ process \label{sec:b-sgamma}}

In the SM, the quark level $b\to s\gamma$ vertex {\it without any QCD corrections}, reads
\begin{equation}
	\overline{s}\,\Gamma_\mu^{b\to s\gamma}\,b=\frac{e}{(4\pi)^2}\frac{g^2}{2M_W^2}V_{ts}^*V_{tb}F_2\overline{s}i\sigma_{\mu\nu}q^\nu\left(m_b\frac{1+\gamma_5}{2}+m_s\frac{1-\gamma_5}{2}\right)b \,,
	\label{eq:bsgamma_Amp}
\end{equation}
where $q=p_b-p_s$ with $p_b$ and $p_s$ four-momentum of the $b$ and $s$ quark, respectively, $F_2$ is the loop function~\cite{Inami:1980fz}. If we choose the three momentum direction in the $b$-quark rest frame, $q^\mu=(|\vec{q}|,0,0,|\vec{q}|)$, and define the right- and left-handed polarization vectors as
\begin{equation}
	\varepsilon_{R,L}^\mu=\mp\frac{1}{\sqrt2}\left(0,1,\pm i,0\right) \,,
	\label{eq:epsilonRL}
\end{equation}
one can then compute the helicity amplitude and explicitly show that
\begin{equation}
	\overline{s}_L\sigma_{\mu\nu}q^\nu b_R\varepsilon_R^{\mu*}=\overline{s}_R\sigma_{\mu\nu}q^\nu b_L\varepsilon_L^{\mu*}=0 \,.
	\label{eq:RL_LR_0}
\end{equation}

We therefore readily find that the first(second) term in Eq.~\eqref{eq:bsgamma_Amp} is non-zero only when we multiply by the left(right)-handed circular-polarization vector. In other words the first term in Eq.~\eqref{eq:bsgamma_Amp}, proportional to $m_b$, describes the $b_R\to s_L\gamma_L$, while the second one, proportional to $m_s$, describes the $b_L\to s_R\gamma_R$~\footnote{More intuitively, the outgoing photon polarization can be determined in the following way: due to the chiral structure of the coupling of the $W$ boson to quarks, the first term in Eq.~\eqref{eq:bsgamma_Amp} describes $b_R\to s_L$ transition. Since $b\to s\gamma$ is a two-body back-to-back decay in the $b$-quark rest frame, the helicity conservation implies that the photon must be left-handed. Correspondingly, the second term in Eq.~\eqref{eq:bsgamma_Amp} describes the right-handed photon emission.}. Since $m_s/m_b\simeq0.02\ll1$, {\it the photon in $b\to s\gamma$ in the SM is almost purely left-handed if strong interactions are switched off}. 

After integrating out the heavy degrees of freedom the effective Hamiltonian reads
\begin{equation}
	\mathcal{H}_{\rm eff}=-\frac{4G_F}{\sqrt2}V_{tb}V_{ts}^*\left[\sum_{i=1}^6 C_i(\mu)\O_i(\mu)+\sum_{i=7\gamma,\,8g,\,9,\,10}\biggl(C_i(\mu)\O_i(\mu)+\Cp_i(\mu)\Op_i(\mu)\biggr)\right] \,,
	\label{eq:Heff}
\end{equation}
where the short-distance physics is encoded in the $C_i^{\,(\prime)}$ Wilson coefficients that are calculated in perturbation theory, $\O_{1,\dots6}$ are the local four-quark operators; $\OOp_{7\gamma}$ and $\OOp_{8g}$ are the electro-magnetic and chromo-magnetic penguin operators respectively, $\OOp_{9,10}$ are the semileptonic operators. Using the operator basis from Ref.~\cite{Chetyrkin:1996vx}, the operators relevant to our discussions are
\begin{subequations}
	\begin{align}
		\O_{7\gamma} =& \frac{e}{16\pi^2}m_b\sbar_{\alpha L}\sigma^{\mu\nu}b_{\alpha R}F_{\mu\nu} \,, \quad\quad\quad~
		\Op_{7\gamma} = \frac{e}{16\pi^2}m_b\sbar_{\alpha R}\sigma^{\mu\nu}b_{\alpha L}F_{\mu\nu} \,, \\
		\O_{8g} =& \frac{g_s}{16\pi^2}m_b\sbar_{\alpha L}\sigma^{\mu\nu} t_{\alpha\beta}^a b_{\beta R} G_{\mu\nu}^a \,, \quad\quad
		\Op_{8g} = \frac{g_s}{16\pi^2}m_b\sbar_{\alpha R}\sigma^{\mu\nu} t_{\alpha\beta}^a b_{\beta L} G_{\mu\nu}^a \,, \\
		\O_9 =& \frac{e^2}{16\pi^2}(\sbar_{\alpha L}\gamma^\mu b_{\alpha L})(\lbar\gamma_\mu\ell) \,, \quad\quad\quad~ ~
		\Op_9 = \frac{e^2}{16\pi^2}(\sbar_{\alpha R}\gamma^\mu b_{\alpha R})(\lbar\gamma_\mu\ell) \,, \\
		\O_{10} =& \frac{e^2}{16\pi^2}(\sbar_{\alpha L}\gamma^\mu b_{\alpha L})(\lbar\gamma_\mu\gamma_5\ell) \,, \quad\quad ~
		\Op_{10} = \frac{e^2}{16\pi^2}(\sbar_{\alpha R}\gamma^\mu b_{\alpha R})(\lbar\gamma_\mu\gamma_5\ell) \,,
	\end{align}
	\label{eq:operators}
\end{subequations}
where $\alpha,\beta$ are the colour indices, $q_{R,L}=\frac{1}{2}(1\pm\gamma_5)q$, $\sigma^{\mu\nu}=\frac{i}{2}[\gamma^\mu,\gamma^\nu]$, $t^a$ ($a=1,\dots8$) are the $SU(3)$ colour generators, $F_{\mu\nu}$ and $G_{\mu\nu}^a$ denote the electromagnetic and chromomagnetic field strength tensors respectively. The renormalization scale $\mu$ is conventionally chosen at $m_b$. In the SM, $\Cp_{9,10}=0$ and
\begin{equation}
	\frac{\Cp_{7\gamma}}{C_{7\gamma}}\simeq\frac{\Cp_{8g}}{C_{8g}}\simeq\frac{m_s}{m_b}\simeq0.02\,.
	\label{mssurmb}
\end{equation}

The short-distance QCD effects induce the mixing of $\O_{7\gamma}$ with $\O_{1,\dots6}$, the effect of which can be absorbed by defining the so-called "effective" coefficients $\CCpeff_{7\gamma}$. For notational simplicity, {\it whenever $\CCp_{7\gamma}$ appears in what follows, $\CCpeff_{7\gamma}$, evaluated at the scale $\mu_b\simeq m_{b,\,{\rm pole}}=4.8$~{\rm GeV}, will be understood}. In our numerics we use $\CSM_{7\gamma}(\mu_b)=-0.304$~\cite{Bobeth:1999mk}.

In this way, the amplitude for the {\it exclusive} $\Bbar\to\Kbar^*\gamma$, $\Bbar\to\Kbar_1\gamma$, $B_s\to\phi\gamma$ decays, that we generically refer to as $\Bbar\to\Mbar\gamma$, can be written as
\begin{equation}
	\M(\Bbar\to\Mbar\gamma) = -\frac{4G_F}{\sqrt2}V_{tb}V_{ts}^*\left[C_{7\gamma}(\mu_b)\langle\Mbar\gamma|\O_{7\gamma}(\mu_b)|\Bbar\rangle + \Cp_{7\gamma}(\mu_b)\langle\Mbar\gamma|\Op_{7\gamma}(\mu_b)|\Bbar\rangle + \dots \right] \,,
	\label{eq:amp_b-sgamma_LLA}
\end{equation}
where the dots stand for the long-distance contributions of the other operators. The ones explicitly written could be thought to give the main contribution. However, for the ``wrong" helicity, precisely because of the strong suppression of $\Cp_{7\gamma}$ by the factor $m_s/m_b$, one has to be careful when considering the other operators, as well as the perturbative corrections to the Wilson coefficients themselves.


\section{Various methods for determination of $\Cp_{7\gamma}$ \label{sec:methods}}

In this Section, we introduce the methods to measure the photon polarization in $b\to s\gamma^{(*)}$, and discuss the sensitivity of the future experiments, namely the super $B$~factories and LHCb, to the photon polarization in the $b\to s\gamma$ process by comparing and combining several methods in various $B$-meson decay modes. We also discuss the advantages and disadvantages of the direct method, based on the radiative $B\to K_1(1270)\gamma\to(K\pi\pi)\gamma$ decay, with respect to the indirect ones.

\subsection{Methods invoking $CP$-asymmetries \label{sec:ACP_method}}

An indirect method to measure the photon polarization is to study the time-dependent $CP$-asymmetry in the radiative decays of the neutral $B$-mesons. This asymmetry arises from the interference between $B(\Bbar)\to f^{CP}\gamma$ and $B(\Bbar)\to\Bbar(B)\to f^{CP}\gamma$ amplitudes where $f^{CP}$ is the final hadronic self-conjugate state. Since the $B(\Bbar)$-meson decays predominantly into a photon with right(left)-handed helicity, the dominant amplitudes of $B(\Bbar)\to f^{CP}\gamma_{R(L)}$ and $B(\Bbar)\to\Bbar(B)\to f^{CP}\gamma_{L(R)}$ can not interfere quantum-mechanically as the photon helicity is, in principle, a measurable quantity. Thus, in the SM, the time-dependent asymmetry, generated by the $B-\Bbar$ mixing, is expected to be zero up to $O(m_s/m_b)$ corrections. However, if NP induces a non-negligible contribution to the helicity-suppressed amplitudes with ``wrong" helicity, $B(\Bbar)\to f^{CP}\gamma_{L(R)}$, one can have a significant deviation of asymmetry from zero. That would constitute a clean signal for NP.

For the generic radiative decay of the neutral $B$-meson into any hadronic self-conjugate state $f^{CP}$, $B(t)\to f^{CP}\gamma$, {\it neglecting direct $CP$-violation and the small width difference between two $B$-mesons}, the $CP$-asymmetry is given by~\cite{Atwood:1997zr}\footnote{In fact, the non-negligible width difference $\Delta\Gamma_s$ in $B_s$-mesons leads to one more measurable quantity, also sensitive to the right-handed currents (e.g. see Ref.~\cite{Muheim:2008vu}). Eq.~\eqref{eq:ACP} is more complicated in the case of $B_s\to\phi\gamma$. For simplicity, we neglect this term proportional to $\sinh\left(\frac{\Delta\Gamma_s}{2}t\right)$, what is an excellent approximation in $B$-decays but not in $B_s$-decays, but keep in mind its significance in $B_s\to\phi\gamma$.}
\begin{equation}
	\A_{CP}(t) \equiv \frac{\Gamma(\Bbar(t)\to f^{CP}\gamma)-\Gamma(B(t)\to f^{CP}\gamma)}{\Gamma(\Bbar(t)\to f^{CP}\gamma)+\Gamma(B(t)\to f^{CP}\gamma)} \approx S_{f^{CP}\gamma}\sin(\Delta mt) \,,
	\label{eq:ACP}
\end{equation}
with
\begin{equation}
	S_{f^{CP}\gamma} \equiv \xi\frac{2\Im[e^{-i\phi_M}\,\M(\Bbar\to f^{CP}\gamma_L)\M(\Bbar\to f^{CP}\gamma_R)]}{|\M(\Bbar\to f^{CP}\gamma_L)|^2+|\M(\Bbar\to f^{CP}\gamma_R)|^2} \approx \xi\frac{2\Im[e^{-i\phi_M}\,C_{7\gamma}\Cp_{7\gamma}|}{|C_{7\gamma}|^2+|\Cp_{7\gamma}|^2} \,,
	\label{eq:Sfgamma}
\end{equation}
where $\xi(=\pm1)$ is the $CP$-eigenvalue of $f^{CP}$, $\phi_M$ is the phase in the $B-\overline{B}$ mixing, which in the SM is $\phi_d=2\beta\simeq43^\circ$, and $\phi_s\simeq0$, for the $B_d$ and $B_s$ mixing, respectively.

We should emphasise that the measurement of $\A_{CP}(t)$ allows us to determine the ratio of two amplitudes $\M(B\to f^{CP}\gamma_{L,R})$ {\it together with the $CP$ violating phase $\phi_M$ but not separately}. Thus, the fraction of the right-handed polarization can be obtained from this measurement only by having the value of the $B-\Bbar$ mixing phase.

Due to smallness of the right-handed amplitude in $b\to s\gamma$, the SM predicts
\begin{equation}
	S_{f^{CP}\gamma}^{\rm SM}\approx-2\frac{m_s}{m_b}\sin\phi_M \,.
	\label{}
\end{equation}
More specifically, for the $B\to K^*(\to K_S\pi^0)\gamma$ decay, the SM prediction reads~\cite{Ball:2006eu}
\begin{equation}
	S_{K_S\pi^0\gamma}^{\rm SM}=-(2.3\pm1.6)\% \,,
	\label{}
\end{equation}
which is to be compared with the current world average for the asymmetry in the $B\to K^*(\to K_S\pi^0)\gamma$~\cite{Asner:2010qj},
\begin{equation}
	S_{K_S\pi^0\gamma}^{\rm exp}=-0.16\pm0.22\,.
	\label{eq:SKSpi0gamma_exp}
\end{equation}
That last error is expected to be improved at the super $B$~factories to~$2\%$~\cite{Bona:2007qt}. The LHCb experiment will measure the $CP$-asymmetry in $B_s\to\phi\gamma$. Based on the MC simulation for 2~fb$^{-1}$, it is claimed in Ref.~\cite{LHCB-ROADMAP4-001} that LHCb will be able to determine the $S_{f^{CP}\gamma}$ parameter with an accuracy of the order of~0.2. In other words, only large NP contribution could be observed via this quantity at LHCb.

\subsection{GGPR and DDLR-inspired methods using $B\to K_1(\to K\pi\pi)\gamma$ \label{sec:GGPR_DDLR}}

Unless the mixing-induced $CP$-asymmetry is measured, the $B\to K^*(\to K_S\pi^0)\gamma$ decay provides no helicity information for the following reason: since the photon helicity is parity-odd and we measure only the momenta of the photon and the final hadrons, we can not form a hadronic quantity that would be also parity-odd. On the other hand, in the case of three-body decay of $K_1$ (i.e. $B\to K_1\gamma\to K\pi\pi\gamma$) one can form a triple product of three momenta. For example, $\vec{p}_\gamma\cdot(\vec{p}_\pi\times\vec{p}_K)$ is a pseudo-scalar; then applying the parity transformation it will have the different sign for left- and right-handed photons.

Gronau {\it et al.}~\cite{Gronau:2001ng,*Gronau:2002rz} proposed to study the angular distribution in the $B\to K_1(\to K\pi\pi)\gamma$ decay and extract the polarization parameter $\lambda_\gamma$,
\begin{equation}
	\lambda_\gamma\equiv\frac{|\M(\Bbar\to\Kbar_{1R} \gamma_R)|^2-|\M(\Bbar\to \Kbar_{1L}\gamma_L)|^2}{|\M(\Bbar\to \Kbar_{1}\gamma_R)|^2+|\M(\Bbar\to \Kbar_{1}\gamma_L)|^2} \approx \frac{|\Cp_{7\gamma}|^2-|C_{7\gamma}|^2}{|\Cp_{7\gamma}|^2+|C_{7\gamma}|^2} \,.
	\label{eq:lambdagamma}
\end{equation}
They proposed to measure the observable proportional to $\lambda_\gamma$, called up-down asymmetry (in the following we call it the GGPR method), defined as
\begin{equation}
	\A_{\rm up-down} = \frac{\int_0^1\,d\cos\theta\frac{d\Gamma}{d\cos\theta}-\int^0_{-1}\,d\cos\theta\frac{d\Gamma}{d\cos\theta}}{\int_{-1}^1\,d\cos\theta\frac{d\Gamma}{d\cos\theta}} \,,
	\label{eq:Aup-down}
\end{equation}
where $\theta$ is the angle between the $z$-axis and the vector orthogonal to the plane spanned by $K\pi\pi$ in the $K_1$ rest frame. Note that the $z$-axis is chosen opposite to the direction of the photon. This asymmetry allows us to determine $\lambda_\gamma$ directly from the measurement of the observed asymmetry of the total number of the events with the photons emitted above and below the $K\pi\pi$-plane in the $K_1$ rest frame.

In our previous work \cite{Kou:2010kn} we investigated the feasibility of determining the photon polarization $\lambda_\gamma$ using the $B\to K_1(1270)\gamma\to K\pi\pi\gamma$ decay and improved the GGPR method by introducing a {\it new} variable $\omega$ that contains information on the Dalitz distribution (in the following we call it the DDLR method). {\it In this way the experimental sensitivity to $\lambda_\gamma$ is significantly increased}. Please see Ref.~\cite{Kou:2010kn} for more details.

For an accurate determination of $\lambda_\gamma$, one needs modelling of the hadronic $K_1\to K\pi\pi$ decays. Contrary to Refs.~\cite{Gronau:2001ng,*Gronau:2002rz} we pointed out the complex hadronic structure of $K_1(1270)\to K\pi\pi$, that we studied in great detail in Ref.~\cite{Tayduganov:2011ui}.

It turns out that the probability density function, or equivalently the properly normalized differential decay width distribution, is linearly dependent on the polarization parameter $\lambda_\gamma$,
\begin{equation}
	\begin{split}
		W(m_{K\pi}^2,m_{\pi\pi}^2,\cos\theta) =& \frac{1}{\Gamma}\frac{d^3\Gamma(B\to K_1\gamma\to K\pi\pi\gamma)}{dm_{K\pi}^2dm_{\pi\pi}^2d\cos\theta} \\
		=& f(m_{K\pi}^2,m_{\pi\pi}^2,\cos\theta)+\lambda_\gamma\,g(m_{K\pi}^2,m_{\pi\pi}^2,\cos\theta) \,,
	\end{split}
	\label{eq:PDF}
\end{equation}
where $f$ and $g$ functions parametrize the $K_1$ helicity amplitudes and can be found in Refs.~\cite{Kou:2010kn,Tayduganov:2011phd}. Introducing a new single variable $\omega$,
\begin{equation}
	\omega(m_{K\pi}^2,m_{\pi\pi}^2,\cos\theta)\equiv\frac{g(m_{K\pi}^2,m_{\pi\pi}^2,\cos\theta)}{f(m_{K\pi}^2,m_{\pi\pi}^2,\cos\theta)} \,,
	\label{eq:omega}
\end{equation}
and writing the normalization condition for $W(m_{K\pi}^2,m_{\pi\pi}^2,\cos\theta)$ as
\begin{equation}
	\begin{split}
		&\int W(m_{K\pi}^2,m_{\pi\pi}^2,\cos\theta)dm_{K\pi}^2dm_{\pi\pi}^2d\cos\theta =1 \\
		&= \int_{-1}^1 d\omega \underbrace{\int W(m_{K\pi}^2,m_{\pi\pi}^2,\cos\theta)\,\delta\left[\omega-\frac{g(m_{K\pi}^2,m_{\pi\pi}^2,\cos\theta)}{f(m_{K\pi}^2,m_{\pi\pi}^2,\cos\theta)}\right]dm_{K\pi}^2dm_{\pi\pi}^2d\cos\theta}_{W^\prime(\omega)} \,, \\
	\end{split}
	\label{eq:PDF_norm}
\end{equation}
one can identify a new probability distribution of $\omega$,
\begin{equation}
	W^{\,\prime}(\omega)=\varphi(\omega)(1+\lambda_\gamma\omega) \,,
	\label{eq:PDF_omega}
\end{equation}
where $\varphi(\omega)$ reads
\begin{equation}
	\varphi(\omega) = \int f(m_{K\pi}^2,m_{\pi\pi}^2,\cos\theta)\,\delta\left[\omega-\frac{g(m_{K\pi}^2,m_{\pi\pi}^2,\cos\theta)}{f(m_{K\pi}^2,m_{\pi\pi}^2,\cos\theta)}\right]dm_{K\pi}^2dm_{\pi\pi}^2d\cos\theta \,.
\end{equation}
It is a complicated function that depends on hadronic model parameters. As proved in Ref.~\cite{Tayduganov:2011phd}, $\varphi(\omega)$ turns out to be an even function of $\omega$. Therefore, using the maximum likelihood method, $\lambda_\gamma$ can be expressed as the ratio of odd over even moments,
\begin{equation}
	\lambda_\gamma=\frac{\langle\omega^{2n-1}\rangle}{\langle\omega^{2n}\rangle} \,, \quad \langle\omega^n\rangle=\int_{-1}^1 \omega^n W^{\,\prime}(\omega)d\omega \quad (n\ge1) \,.
\end{equation}

\begin{figure}[t!]\centering
	\includegraphics[width=0.45\textwidth]{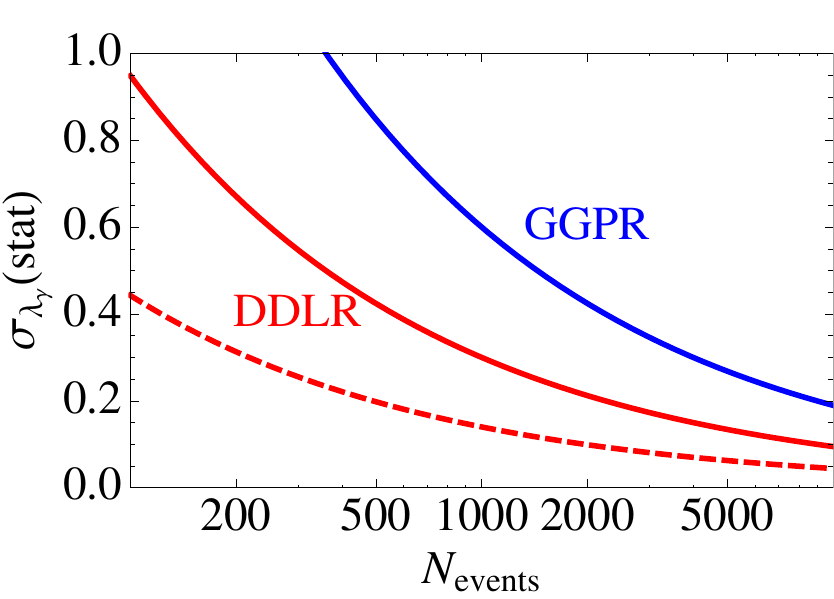}
	\caption{\footnotesize Dependence of the statistical error $\sigma_{\lambda_\gamma}$ on the total number of signal events of the decays $B^+\to(K^+\pi^-\pi^+)_{K_1(1270)}\gamma$ and $B^0\to(K^0\pi^+\pi^-)_{K_1(1270)}\gamma$ depending on the $\lambda_\gamma$ determination method: the error of $\lambda_\gamma$ which is determined by using the DDLR method (red solid) and the error determined from the up-down asymmetry (blue). Red dashed curve corresponds to the error of $\lambda_\gamma$ determined by the DDLR method for $B^+\to(K^0\pi^+\pi^0)_{K_1(1270)}\gamma$ and $B^0\to(K^+\pi^-\pi^0)_{K_1(1270)}\gamma$ decays.}
	\label{fig:Nevents_sigmalambda}
\end{figure}

In practice, we use numerical Monte Carlo method to simulate $W^{\,\prime}(\omega)$ and compare the statistical errors of two methods, DDLR and GGPR. Note that the expected number of $B\to K_1(1270)\gamma$ events is of the order of several thousands at 2~fb$^{-1}$. One can see from Fig.~\ref{fig:Nevents_sigmalambda} that {\it the inclusion of full Dalitz information improves the sensitivity of $\lambda_\gamma$ determination by typically a factor of two compared to the pure angular fit (or equivalently the up-down asymmetry measurement)}.

The polarization measurement through the $\omega$-moments in the DDLR method is sensitive to several uncertainties in the modelling of $K_1\to K\pi\pi$ decays. We estimated the theoretical errors of the hadronic model to be $\sigma_{\lambda_\gamma}\lesssim0.2$ (for more details see Ref.~\cite{Tayduganov:2011phd}). It must be emphasized that in our study \cite{Kou:2010kn} we used the hadronic model only for the illustration and demonstration of the DDLR method. These systematic uncertainties can be significantly reduced by an accurate, model-independent, partial wave analysis of the $K_1$-decays, in particular using the $B\to J/\psi K_1$'s decays observed by the Belle collaboration~\cite{Guler:2010if}.

\subsection{The angular analysis of $B\to K^*(\to K\pi)\ell^+\ell^-$ \label{sec:semileptonic}}

In Refs.~\cite{Melikhov:1998cd,Kruger:2005ep}, it was proposed to test the NP effects by studying the angular distributions of the four-body final state in the $B^0\to K^{*0}(\to K^-\pi^+)\ell^+\ell^-$ decay.

Written in terms of four kinematic variables, the differential decay rate can be written as
\begin{equation}
	\begin{split}
		&\frac{d^4\Gamma(\Bbar^0\to\Kbar^{*0}\ell^+\ell^-)}{dq^2\,d\cos\theta_\ell\,d\cos\theta_K\,d\phi} = \frac{9}{32\pi}\biggl\{I_1^s(q^2)\sin^2\theta_K+I_1^c(q^2)\cos^2\theta_K\biggr. \\
		&\quad\quad + [I_2^s(q^2)\sin^2\theta_K+I_2^c(q^2)\cos^2\theta_K]\cos2\theta_\ell + I_3(q^2)\sin^2\theta_K\sin^2\theta_\ell\cos2\phi \\
		&\quad\quad +I_4(q^2)\sin2\theta_K\sin2\theta_\ell\cos\phi + I_5(q^2)\sin2\theta_K\sin\theta_\ell\cos\phi \\
		&\quad\quad + [I_6^s(q^2)\sin^2\theta_K+I_6^c(q^2)\cos^2\theta_K]\cos\theta_\ell + I_7(q^2)\sin2\theta_K\sin\theta_\ell\sin\phi \\
		&\quad\quad + \biggl.I_8(q^2)\sin2\theta_K\sin2\theta_\ell\sin\phi+I_9(q^2)\sin^2\theta_K\sin^2\theta_\ell\sin2\phi \biggr\} \,,
	\end{split}
	\label{eq:B-Kstll_d4Gamma}
\end{equation}
where we use the notation adopted in Ref.~\cite{Altmannshofer:2008dz}. $I_i(q^2)$ can be expressed in terms of two transverse, $A_{\perp,\parallel}(q^2)$, one longitudinal, $A_0(q^2)$, amplitudes related to the spin state of the {\it on-shell} $K^*$, and one additional time-like amplitude, $A_t(q^2)$, related to the {\it off-shell} virtual gauge boson decaying into the lepton pair. All four amplitudes $A_{\perp,\parallel,0,t}(q^2)$ can be found in the Appendix of the present paper. In terms of these amplitudes~\cite{Altmannshofer:2008dz},
\begin{subequations}
	\begin{align}
		I_2^s(q^2) =& \frac{\beta_\ell^2}{4}\left[|A_\perp^{\ell_L}|^2+|A_\perp^{\ell_R}|^2+|A_\parallel^{\ell_L}|^2+|A_\parallel^{\ell_R}|^2\right] \,, \\
		I_3(q^2) =& \frac{\beta_\ell^2}{2}\left[|A_\perp^{\ell_L}|^2+|A_\perp^{\ell_R}|^2-|A_\parallel^{\ell_L}|^2-|A_\parallel^{\ell_R}|^2\right] \,, \\
		I_6^s(q^2) =& 2\beta_\ell\Re\left[A_\parallel^{\ell_L} A_\perp^{\ell_L*}-A_\parallel^{\ell_R}A_\perp^{\ell_R*}\right] \,, \\
		I_9(q^2) =& \beta_\ell^2\Im\left[A_\perp^{\ell_L}A_\parallel^{\ell_L*}+A_\perp^{\ell_R}A_\parallel^{\ell_R*}\right] \,.
	\end{align}
	\label{eq:Ii}
\end{subequations}

One of the most promising observables, that have a small impact from the theoretical uncertainties are the transverse asymmetries defined as~\cite{Kruger:2005ep,Becirevic:2011bp}\footnote{One has to pay attention that $\A_T^{\rm(im)}(q^2)$, we are using here, is different from $\A_{\rm im}(q^2)$, defined in Ref.~\cite{Egede:2008uy}: $\A_{\rm im}(q^2) = \frac{\Im[A_\perp^L(q^2)A_\parallel^{L*}(q^2)+A_\perp^R(q^2)A_\parallel^{R*}(q^2)]}{|A_\perp(q^2)|^2+|A_\parallel(q^2)|^2+|A_0(q^2)|^2}$.}
\begin{subequations}
	\begin{align}
		\A_T^{(2)}(q^2) =& \frac{I_3(q^2)}{2I_2^s(q^2)} \,, \\
		\A_T^{\rm(im)}(q^2) =& \frac{I_9(q^2)}{2I_2^s(q^2)} \,, \\
		\A_T^{\rm(re)}(q^2) =& \frac{\beta_\ell}{4}\frac{I_6^s(q^2)}{I_2^s(q^2)} \,.
	\end{align}
	\label{eq:AT2_ATim}
\end{subequations}
These asymmetries, as well as the other quantities introduced in the literature, can be extracted from the experimental angular decay distribution fitting $I_i(q^2)$. In particular, the measurement of $I_{3,9}(q^2)$ and $I_{2,6}^s(q^2)$ allows us to determine $\A_T^{\rm(2,\,im,\,re)}$ directly from the fit.

Note that $\A_T^{\rm(2,\,im,\,re)}(q^2)$ involve only $A_{\parallel,\perp}(q^2)$ and not the longitudinal and time-like amplitudes $A_{0,t}(q^2)$ (see Eq.~\eqref{eq:Kst_amplitudes}). As emphasized by one of the authors \cite{Becirevic:2011bp}, the advantage of using the quantities that include only $A_{\parallel,\perp}$ is that they do not require a detailed knowledge of hadronic form factors $T_{3}(q^2)$ and $A_{2,0}(q^2)$ which are quite hard to compute using the lattice QCD simulations. Moreover, as it was verified in Ref.~\cite{Becirevic:2011bp}, the ratios $A_1(q^2)/T_2(q^2)$ and $V(q^2)/T_1(q^2)$ are flat in the low $q^2$-region which makes the relevant hadronic uncertainties to be better controlled.

One can easily demonstrate that 
\begin{subequations}
	\begin{align}
		\lim_{q^2\to0}\A_T^{(2)}(q^2) =& \frac{2\Re[C_{7\gamma}\Cpc_{7\gamma}]}{|C_{7\gamma}|^2+|\Cp_{7\gamma}|^2} \,, \\
		\lim_{q^2\to0}\A_T^{\rm(im)}(q^2) =& \frac{2\Im[C_{7\gamma}\Cpc_{7\gamma}]}{|C_{7\gamma}|^2+|\Cp_{7\gamma}|^2} \,, \\
		\lim_{q^2\to0}\A_T^{\rm(re)}(q^2) =& 0 \,.
	\end{align}
	\label{eq:AT2_ATim_0}
\end{subequations}
This is the consequence of the fact that in the very low $\ell^+\ell^-$ invariant mass region the $\O_{7\gamma}$ operator is dominant with respect to the semileptonic $\O_{9,10}$ operators. Note that approximation of Eq.~\eqref{eq:AT2_ATim_0} is strictly valid only at $q^2=0$, and away from this point the expressions for $\A_T^{\rm(2,\,im,\,re)}(q^2)$ become more complicated due to the non-negligible contributions from the other terms proportional to $\CCp_{9,10}$ (see Eq.~\eqref{eq:Kst_amplitudes}). In practice, we work with binned experimental distributions within a range of $q^2$ and the full expression, involving $\CCp_{9,10}$, should be used. Note however that at low $q^2$ the impact of $C_{9,10}$ is very small.

Unlike $\A_T^{\rm(2,\,im)}(0)$ whose values can change considerably if NP affects the coefficients $\CCp_{7\gamma}$, the third asymmetry $\A_T^{\rm(re)}(0)$ remains insensitive to NP. The $q^2$-shapes of three asymmetries can give important hints of the presence of NP in some scenarios~\cite{Becirevic:2011bp}.

The new analysis of the $B\to K^*e^+e^-$ decay mode by the LHCb collaboration~\cite{Lefrancois:1179865} shows that one can expect an annual yield of 200 to 250 events for 2~fb$^{-1}$ in the region $30~{\rm MeV}<\sqrt{q^2}<1~{\rm GeV}$ which would amount to an error on $\A_T^{(2)}$ about
\begin{equation}
	\sigma^{\rm LHCb}(A_T^{(2)})\sim20\% \,.
	\label{eq:sigma_AT2_LHCb}
\end{equation}

\subsection{Comparison of the methods: advantages and disadvantages \label{sec:comparison}}

First, it should be noticed that the measurements of the time-dependent $CP$-asymmetry in $B\to K^*(\to K_S\pi^0)\gamma$ and of the two transverse asymmetries $\A_T^{\rm(2,\,im)}(q^2)$ in $B\to K^*\ell^+\ell^-$ are proportional to the absolute value of the ratio 
\begin{equation}
	r=\frac{\Cp_{7\gamma}}{C_{7\gamma}} \,,
	\label{eq:r-ratio}
\end{equation}
(up to order $O(|r|^2)$) {\it together with the complex phases}, $\phi_L+\phi_R$ and $\phi_L-\phi_R$ (Eqs.~\eqref{eq:Sfgamma} and \eqref{eq:AT2_ATim_0} respectively):
\begin{subequations}
	\begin{align}
		S_{K_S\pi^0\gamma} &\simeq -\frac{2|r|}{1+|r|^2}\sin(\phi_M-\phi_L-\phi_R) \,, \\
		A_T^{(2)}(0) &\simeq \frac{2|r|}{1+|r|^2}\cos(\phi_L-\phi_R) \,, \\
		A_T^{\rm(im)}(0) &\simeq \frac{2|r|}{1+|r|^2}\sin(\phi_L-\phi_R) \,,
	\end{align}
	\label{eq:combination}
\end{subequations}
where $\phi_L$ and $\phi_R$ are the relative $CP$-odd weak phases in the $b\to s\gamma$ process: $\phi_{L,R}=Arg[\CCp_{7\gamma}]$.

\begin{figure}[t!]\centering
	\includegraphics[width=0.45\textwidth]{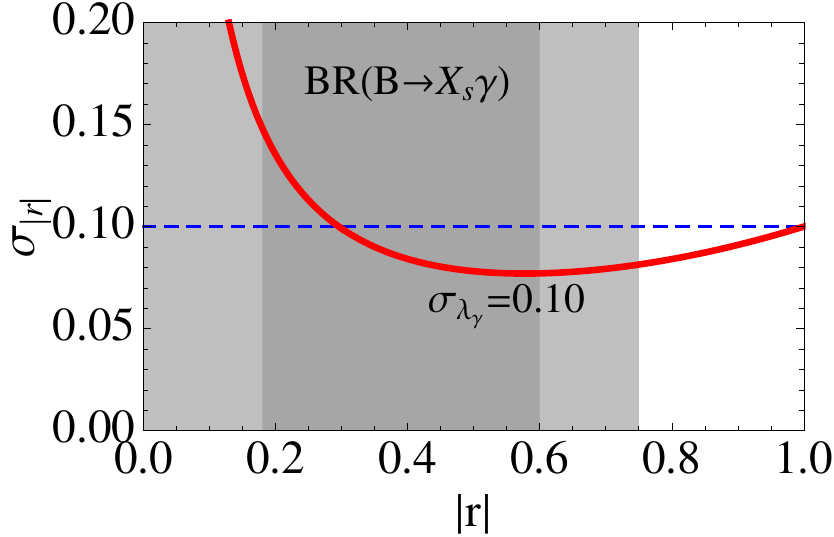}
	\caption{\footnotesize Comparison of the sensitivity of the two methods: the one directly determining $r=\Cp_{7\gamma}/C_{7\gamma}$ like in the time-dependent $CP$-asymmetry in $B\to K^*(\to K_S\pi^0)\gamma$ and transverse asymmetries $\A_T^{\rm(2,\,im)}$ in $B\to K^*\ell^+\ell^-$ and the other one determining $|r|^2$ such as $\lambda_\gamma$ in $B\to K_1(1270)\gamma$. Assuming $\sigma_{|r|}^{\rm LHCb}\approx0.1$ (blue dashed line) and $\sigma_{\lambda_\gamma}\approx0.1$ (red solid line), one can see that a better significance can be obtained with the later method for $|r|\gtrsim0.3$. Here, for illustration, we assumed both $\CCp_{7\gamma}$ to be real with $C_{7\gamma}$ being purely SM-like.}
	\label{fig:r-precision}
\end{figure}

On the other hand, from $B\to K_1\gamma$, a measurement of the polarization parameter $\lambda_\gamma$ would give for $\Bbar(B)$-decays
\begin{equation}
	\lambda_\gamma \simeq \pm\frac{|r|^2-1}{|r|^2+1} \,,
	\label{lamr}
\end{equation}
and is sensitive only to $|r|^2$. Supposing, for simplicity, that the $\CCp_{7\gamma}$ are real, the errors of these two type of methods can be compared using the following equation:
\begin{equation}
	\sigma_{|r|}=\frac{(1+|r|^2)^2}{4|r|}\sigma_{\lambda_\gamma} \,.
	\label{eq:sigmar}
\end{equation}
For $r\approx0$ the $B\to K^*$ decays are more advantageous since they are directly proportional to $r$, whereas our $\lambda_\gamma$ is in fact insensitive to $r\approx0$. Using $B\to K_1\gamma$ becomes more advantageous at LHCb for $|r|\gtrsim0.3$ as can be seen from Fig.~\ref{fig:r-precision} (this number corresponds to our estimated uncertainty on $\lambda_{\gamma}$, $\sigma_{\lambda_\gamma}^{\rm stat}\approx0.1$).

Therefore, for small $r$, it is better to rely on $S_{K_S\pi^0\gamma}$ and $\A_T^{\rm(2,\,im)}$. However, one must take into account the QCD corrections: what is really measured is not $r$, but the ratio of the helicity amplitudes,
\begin{equation}
	R_{K_{\rm res}}=\frac{\M(\Bbar\to\Kbar_{\rm res}\gamma_R)}{\M(\Bbar\to\Kbar_{\rm res}\gamma_L)} \,,
	\label{}
\end{equation}
and therefore discerning NP from QCD corrections will be very hard (if possible) if $r$ was small.

Note also that the $CP$-asymmetry depends on phases which are unknown, so that one must make additional assumptions on $\phi_{M,\,L,\,R}$. Instead, from $B\to K_1\gamma$ we can extract $|r|$ without requiring the knowledge of phases. This is where the asymmetries $\A_{T}^{\rm(2,\,im)}$ become more advantageous as they access to both $|r|$ and the relative phase $\phi_L-\phi_R$.

In summary, {\it the three methods considered in this paper should be viewed as complementary rather than competing, and should be combined}. This is what we do in Sec.~\ref{sec:NP_constraints}.

\section{Constraints on $\Cp_{7\gamma}$ combining various methods of the photon polarization determination \label{sec:NP_constraints}}

In this section we present an example of potential constraints for the right-handed current contribution to the photon polarization by combining three polarization measurement methods described in the previous section.

\subsection{Current constraint on $\Cp_{7\gamma}$ by $\BR(B\to X_s\gamma)$ and $S_{K_S\pi^0\gamma}$}

Since the SM and NP contributions are coherently added in the total left- and right-handed amplitudes, the branching ratio measurement of the inclusive process can provide only a partial information on the polarization or $r$. Indeed, if the right-handed amplitude $\Cp_{7\gamma}$ is of the same order as the SM prediction, or the enhancement of $\Cp_{7\gamma}$ goes along with the suppression of the left-handed amplitude $C_{7\gamma}$, the NP impact on the branching ratio
\begin{equation}
	\BR(B\to X_s\gamma)\propto|C_{7\gamma}|^2+|\Cp_{7\gamma}|^2 \approx |\CSM_{7\gamma}+\CNP_{7\gamma}|^2+|\CpNP_{7\gamma}|^2 \,,
	\label{}
\end{equation}
is small. In other words, there can be a substantial contribution of NP that is hardly discernable from the branching ratio alone. The same holds true for the branching ratio of the exclusive decays. Moreover, {\it the rates are not sensitive to the phases $\phi_{L,R}$, while the presence of non zero phases may be characteristic of certain NP models}. This is a reason why the multiplicity of methods for determination of $r$ can be useful to establish the presence of NP: one can see from Eq.~\eqref{eq:combination} that only a combination of the methods can yield both $|r|$ and the phases $\phi_{L,\,R}$.

\begin{figure}[p!]\centering
	\begin{subfigure}
		{\includegraphics[width=0.45\textwidth]{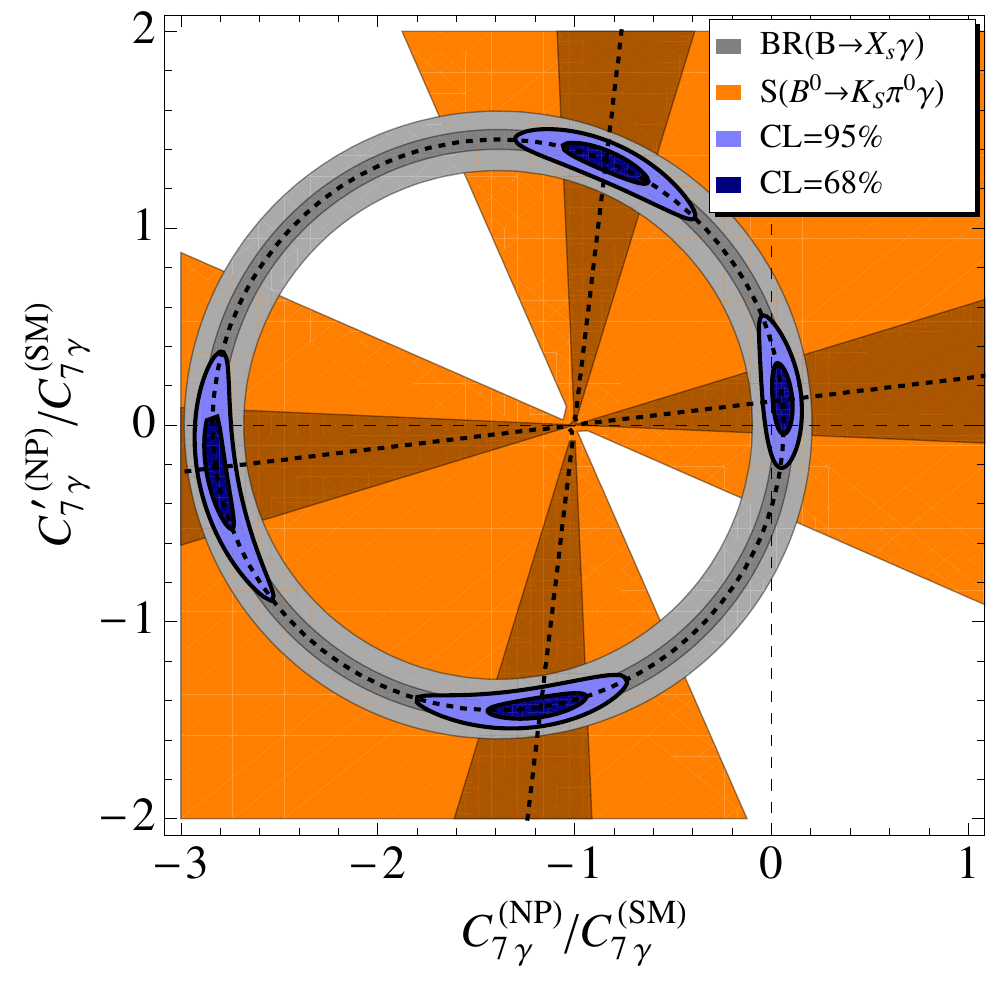}\label{fig:C7p_CL_NP_I}}
		\put(-95,205){(a)}
		\put(-160,185){\footnotesize\color{blue}$\bm{\CCpNP_{7\gamma}\in\mathbb{R}}$}
	\end{subfigure}
	\hspace{5mm}
	\begin{subfigure}
		{\includegraphics[width=0.45\textwidth]{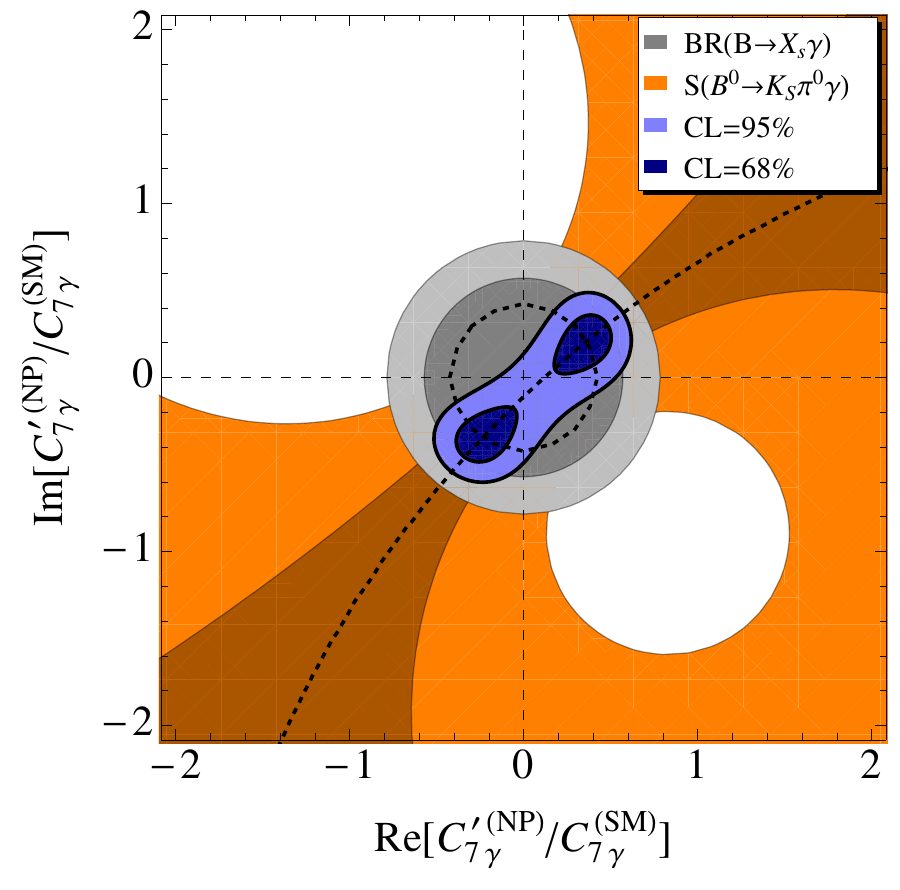}\label{fig:C7p_CL_NP_II}}
		\put(-95,205){(b)}
		\put(-160,185){\footnotesize\color{blue}$\bm{\CNP_{7\gamma}=0}$}
	\end{subfigure}
	\begin{subfigure}
		{\includegraphics[width=0.45\textwidth]{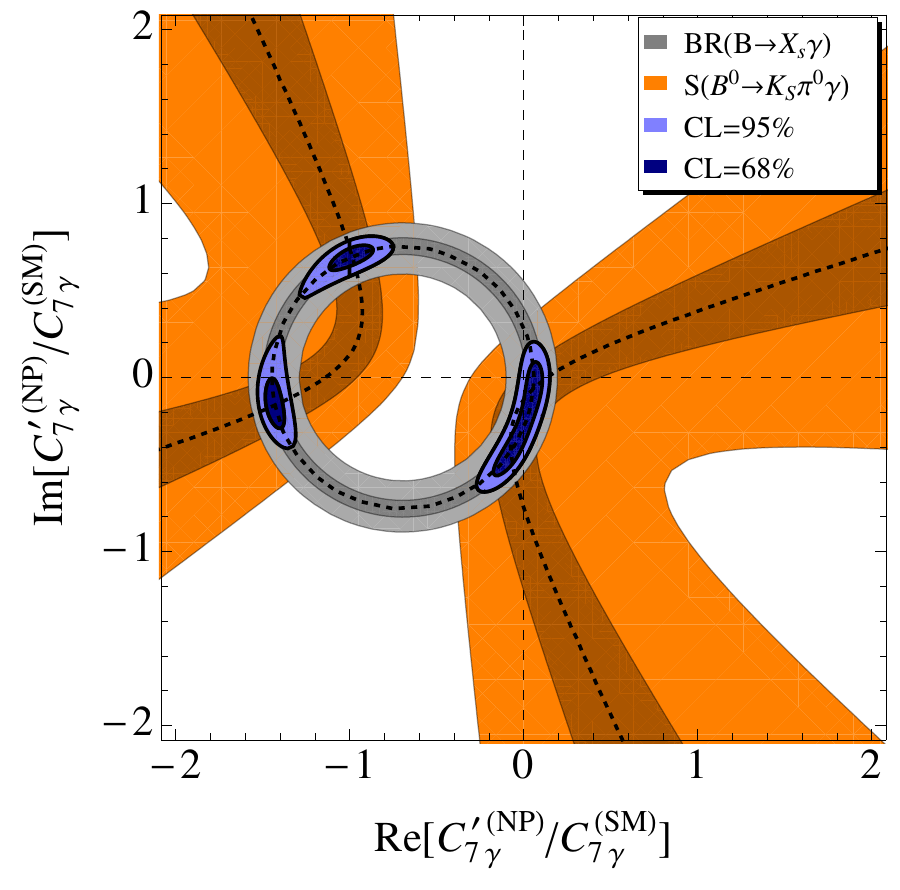}\label{fig:C7p_CL_NP_III}}
		\put(-95,205){(c)}
		\put(-160,185){\footnotesize\color{blue}$\bm{\CNP_{7\gamma}=\CpNP_{7\gamma}}$}
	\end{subfigure}
	\hspace{5mm}
	\begin{subfigure}
		{\includegraphics[width=0.45\textwidth]{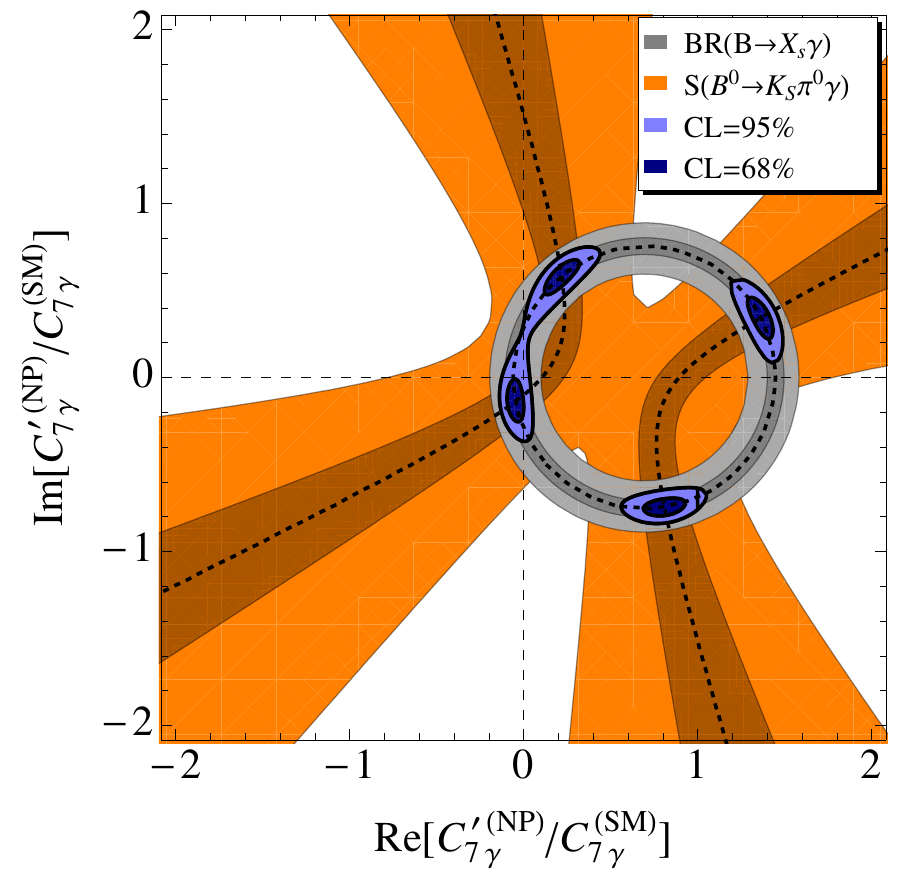}\label{fig:C7p_CL_NP_IV}}
		\put(-95,205){(d)}
		\put(-160,185){\footnotesize\color{blue}$\bm{\CNP_{7\gamma}=-\CpNP_{7\gamma}}$}
	\end{subfigure}
	\caption{\footnotesize Current constraints from the combination of the inclusive decay rate and the mixing-induced $CP$-asymmetry in $B\to K^*(\to K_S\pi^0)\gamma$. In Fig.~(a) we present the constraints in particular NP scenario where both $C_{7\gamma}$ and $\Cp_{7\gamma}$ are real. In Fig.~(b,\,c,\,d), for illustration, we consider several NP scenarios with the left-handed coefficient $\CNP_{7\gamma}=0,\,\CpNP_{7\gamma},\,-\CpNP_{7\gamma}$ respectively. Gray (dark gray) bound represents the $\pm3\sigma$ ($\pm1\sigma$) constraint from the $\BR(B\to X_s\gamma)$ measurement. Orange (dark orange) region represents the $\pm3\sigma$ ($\pm1\sigma$) constraint from the current $S_{K_S\pi^0\gamma}$ measurement. The light and dark blue regions correspond respectively to the 95\% and 68\%~CL bounds for $\CpNP_{7\gamma}$, obtained from the $\chi^2$-fit of the present measurements of $\BR(B\to X_s\gamma)$ and $S_{K_S\pi^0\gamma}$.}
	\label{fig:C7p_CL_NP}
\end{figure}
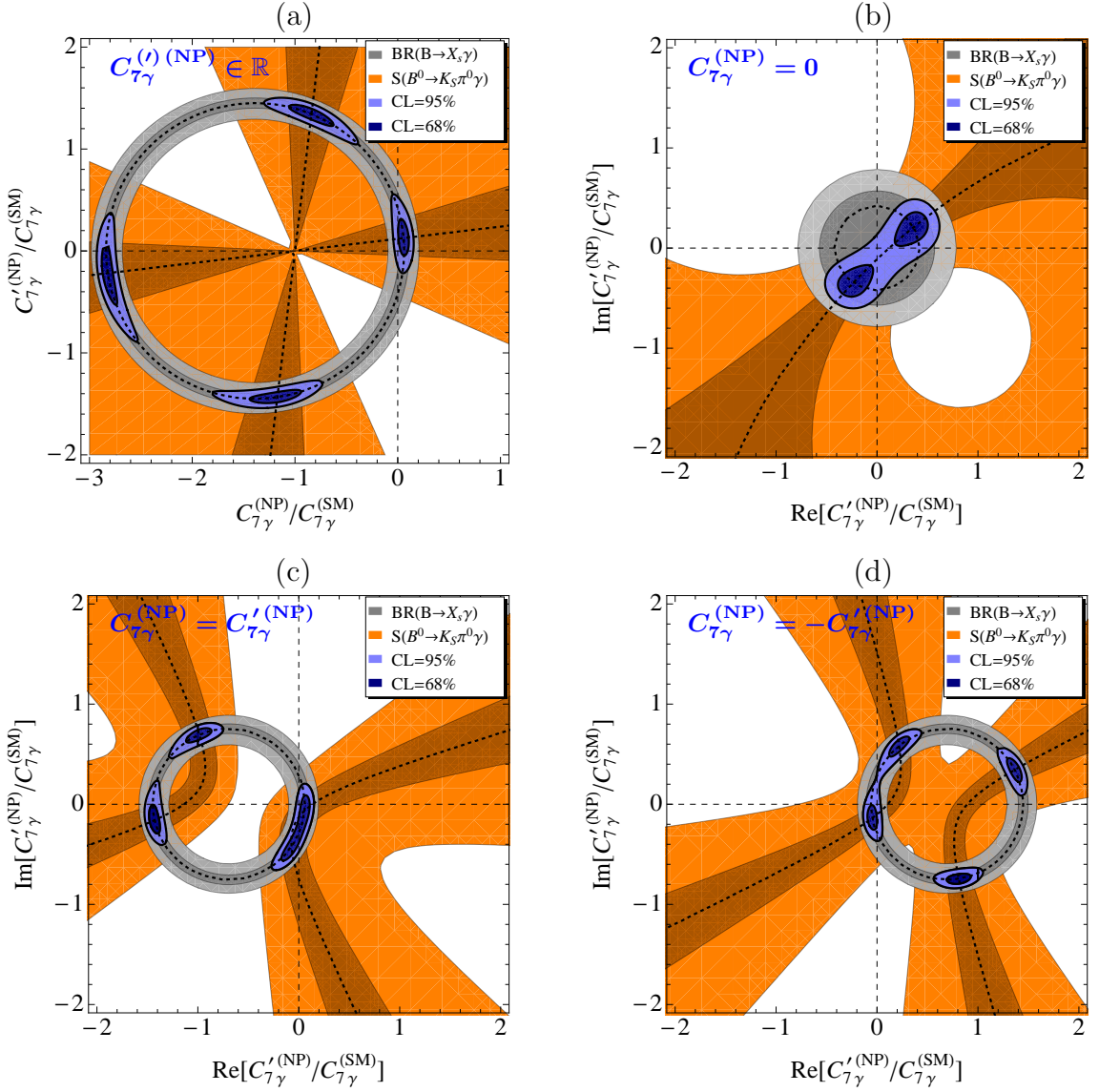

In Fig.~\ref{fig:C7p_CL_NP}--\ref{fig:C7p_AT2im_intq2} we show the constraints on $\Cp_{7\gamma}/C_{7\gamma}$ available at present and compare them with those that are planned to be obtained from the future measurements. For illustration, we consider four NP scenarios:

\begin{itemize}
	\item scenario~{\it I}: $\CNP_{7\gamma}\in\mathbb{R},~\CpNP_{7\gamma}\in\mathbb{R}$ ;
	\item scenario~{\it II}: $\CNP_{7\gamma}=0$, $\CpNP_{7\gamma}\in\mathbb{C}$ ;
	\item scenario~{\it III}: $\CNP_{7\gamma}=\CpNP_{7\gamma}\in\mathbb{C}$ ;
	\item scenario~{\it IV}: $\CNP_{7\gamma}=-\CpNP_{7\gamma}\in\mathbb{C}$ .
\end{itemize}

In all plots presented in Figs.~\ref{fig:C7p_CL_NP}--\ref{fig:C7p_AT2im_intq2}, we use the constraint from the inclusive rate. The region outside the gray (dark gray) circle is excluded at $3\sigma$ ($1\sigma$) level by the current measurement \cite{Asner:2010qj},
\begin{equation}
	\BR^{\rm exp}(B\to X_s\gamma)=(3.55\pm0.24)\times10^{-4} \,,
	\label{eq:BR_BXsgamma_exp}
\end{equation}
which we combined with the SM prediction given in Ref.~\cite{Kagan:1998ym}.

In Fig.~\ref{fig:C7p_CL_NP} we show the constraints from already measured $\BR(B\to X_s\gamma)$ and $S_{K_S\pi^0\gamma}$. Orange (dark orange) region represents the $\pm3\sigma$ ($\pm1\sigma$) region allowed by the current measurement of $S_{K_S\pi^0\gamma}$ \eqref{eq:SKSpi0gamma_exp}. Performing a $\chi^2$-fit of $\BR(B\to X_s\gamma)$ and $S_{K_S\pi^0\gamma}$, we obtain the 95\% and 68\%~CL regions for $\Cp_{7\gamma}$ for each considered NP scenario. One can see from the plots in Fig.~\ref{fig:C7p_CL_NP}, that there is still room for NP. Note, however, the apparent ambiguities in the $C_{7\gamma}-\Cp_{7\gamma}$ plane: in scenario~{\it I} it is fourfold in the $C_{7\gamma}-\Cp_{\gamma}$ plane and two-, three-, fourfold in the $\Re[\Cp_{7\gamma}]-\Im[\Cp_{7\gamma}]$ plane in scenarios~{\it II,\,III,\,IV}, respectively. Therefore, it is clear that additional observables are required to pin down the real and imaginary parts of $\CCp_{7\gamma}$.

\subsection{The expected sensitivity to $\Cp_{7\gamma}$ in the future measurements}

\begin{figure}[t!]\centering
	\begin{subfigure}
		{\includegraphics[width=0.45\textwidth]{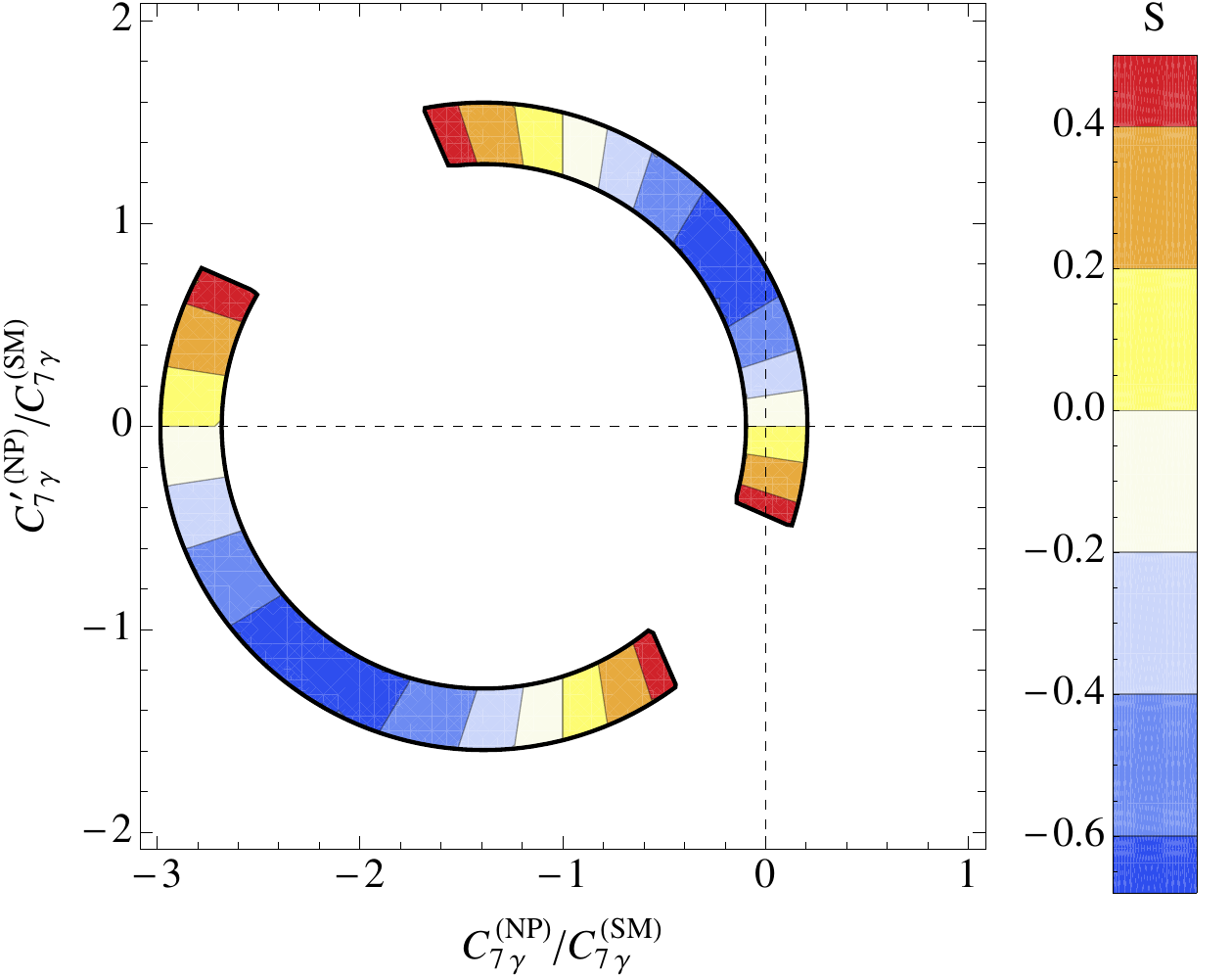}\label{fig:C7p_NP_Ia}}
		\put(-60,150){(a)}
		\put(-175,150){\footnotesize$\bm{B\to K_S\pi^0\gamma}$}
	\end{subfigure}
	\hspace{5mm}
	\begin{subfigure}
		{\includegraphics[width=0.45\textwidth]{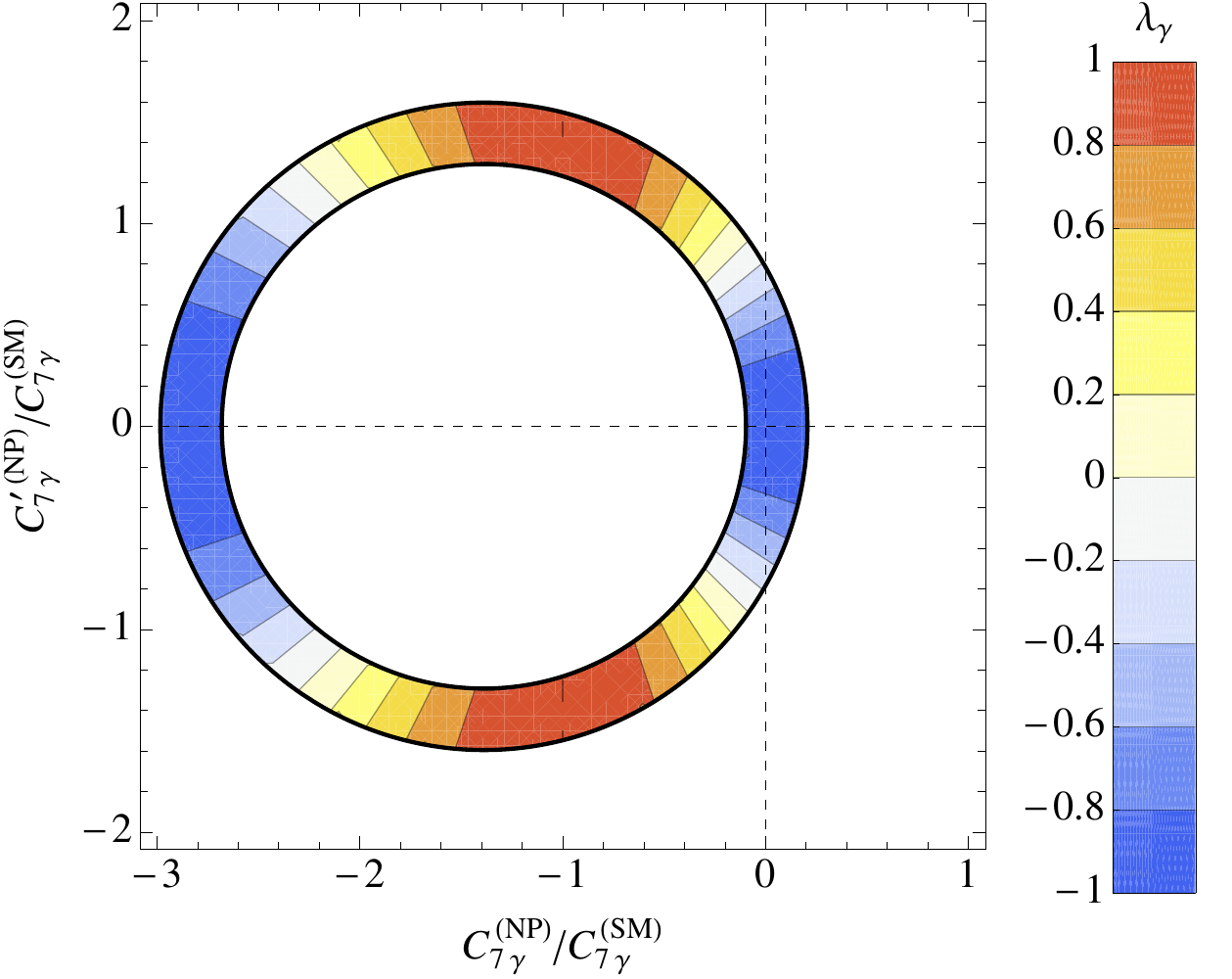}\label{fig:C7p_NP_Ib}}
		\put(-60,150){(b)}
		\put(-175,150){\footnotesize$\bm{B\to K_1\gamma}$}
	\end{subfigure}
	\begin{subfigure}
		{\includegraphics[width=0.45\textwidth]{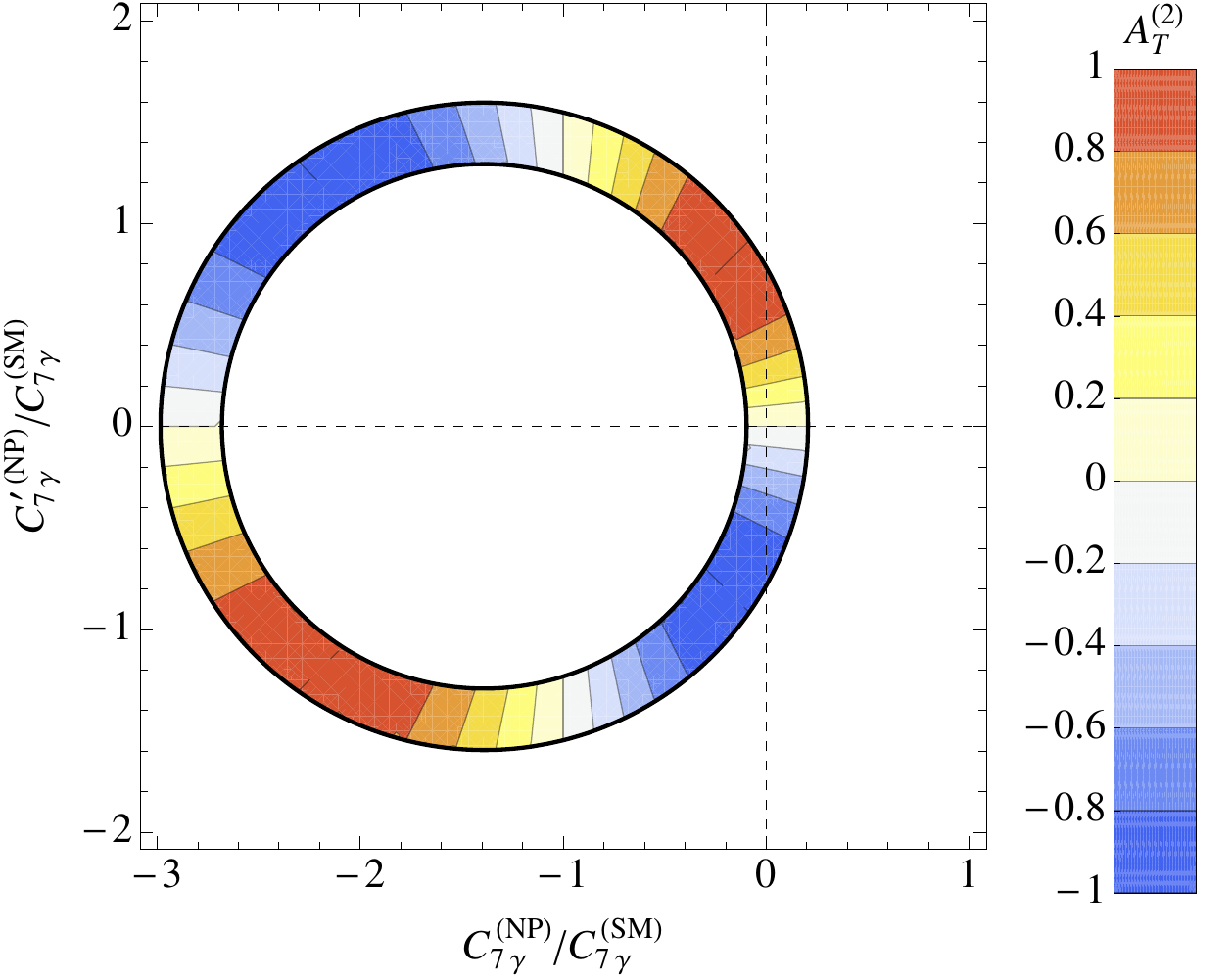}\label{fig:C7p_NP_Ic}}
		\put(-60,150){(c)}
		\put(-175,150){\footnotesize$\bm{B\to K^*\ell^+\ell^-}$}
	\end{subfigure}
	\caption{\footnotesize Prospect of the future constraints on $\CCp_{7\gamma}$ in the NP scenario~{\it I}: $\CNP_{7\gamma}$ and $\CpNP_{7\gamma}$ are both real. The contour colours in Fig.~(a,\,b,\,c) correspond respectively to $S_{K_S\pi^0\gamma}$, $\lambda_\gamma$ and $\A_T^{(2)}(0)$ allowed by a $\pm3\sigma$ error to the central value of $\BR^{\rm exp}(B\to X_s\gamma)$.}
	\label{fig:C7p_NP_I}
\end{figure}

\begin{figure}[t!]\centering
	\begin{subfigure}
		{\includegraphics[width=0.45\textwidth]{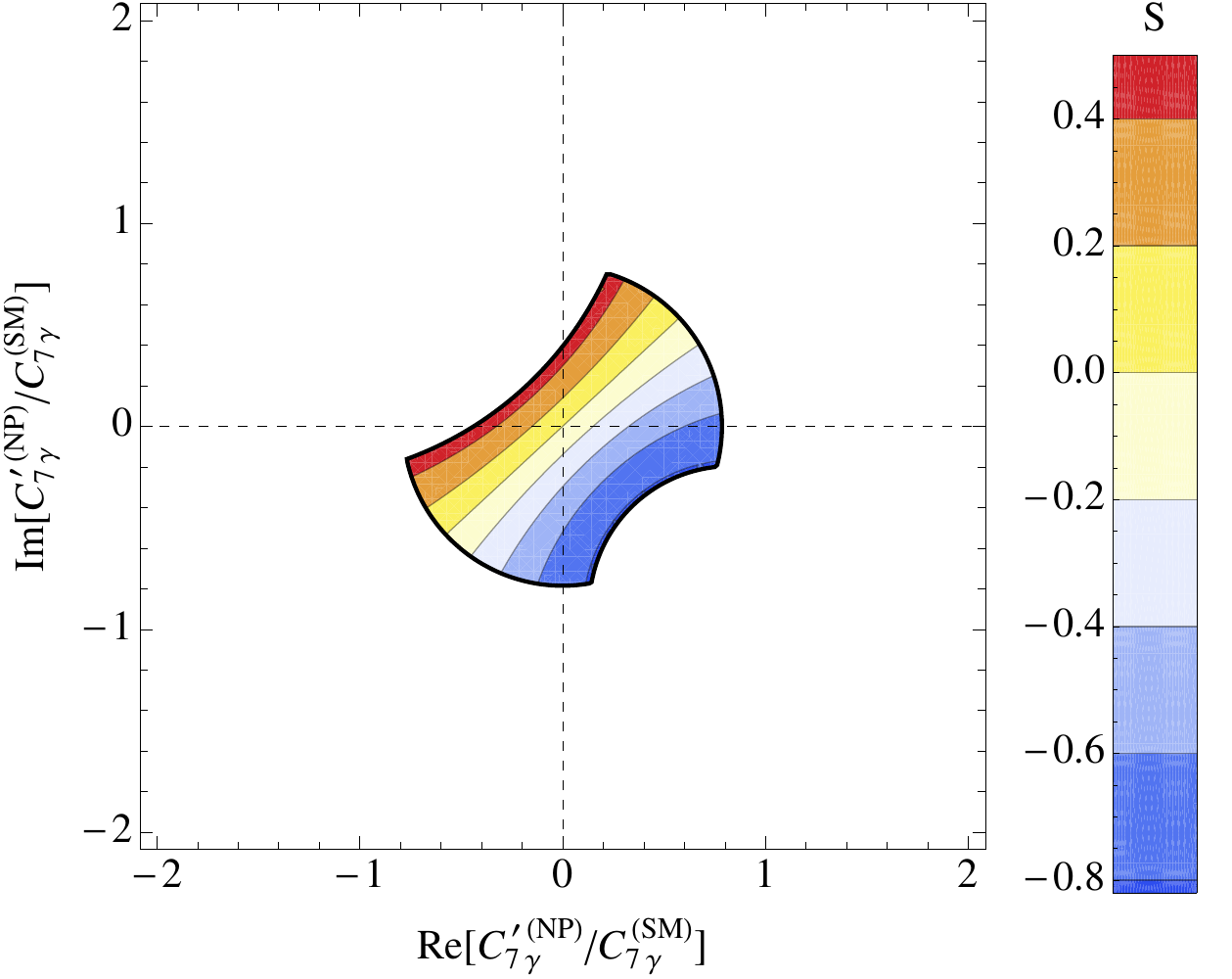}\label{fig:C7p_NP_IIa}}
		\put(-60,150){(a)}
		\put(-175,150){\footnotesize$\bm{B\to K_S\pi^0\gamma}$}
	\end{subfigure}
	\hspace{5mm}
	\begin{subfigure}
		{\includegraphics[width=0.45\textwidth]{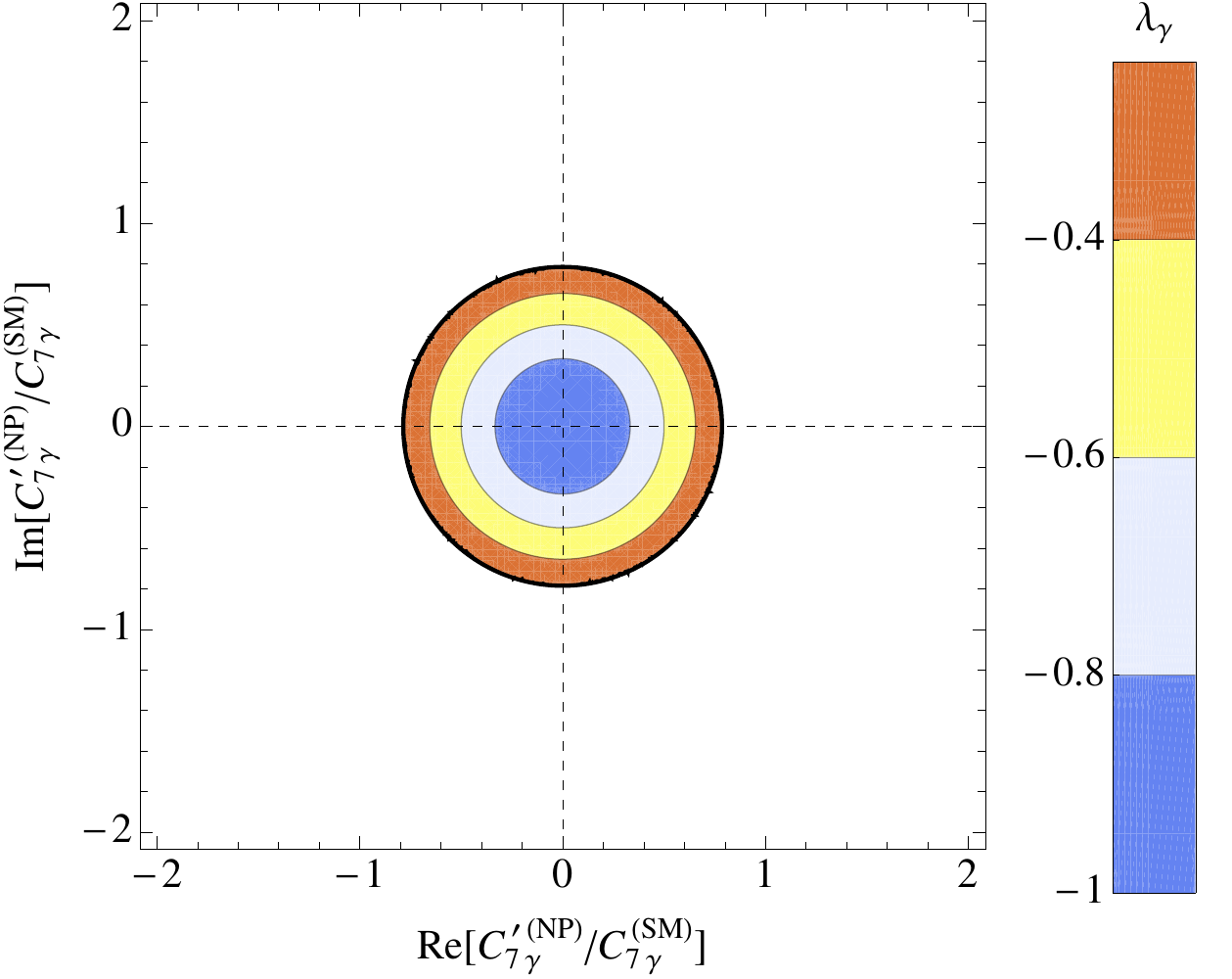}\label{fig:C7p_NP_IIb}}
		\put(-60,150){(b)}
		\put(-175,150){\footnotesize$\bm{B\to K_1\gamma}$}
	\end{subfigure}
	\begin{subfigure}
		{\includegraphics[width=0.45\textwidth]{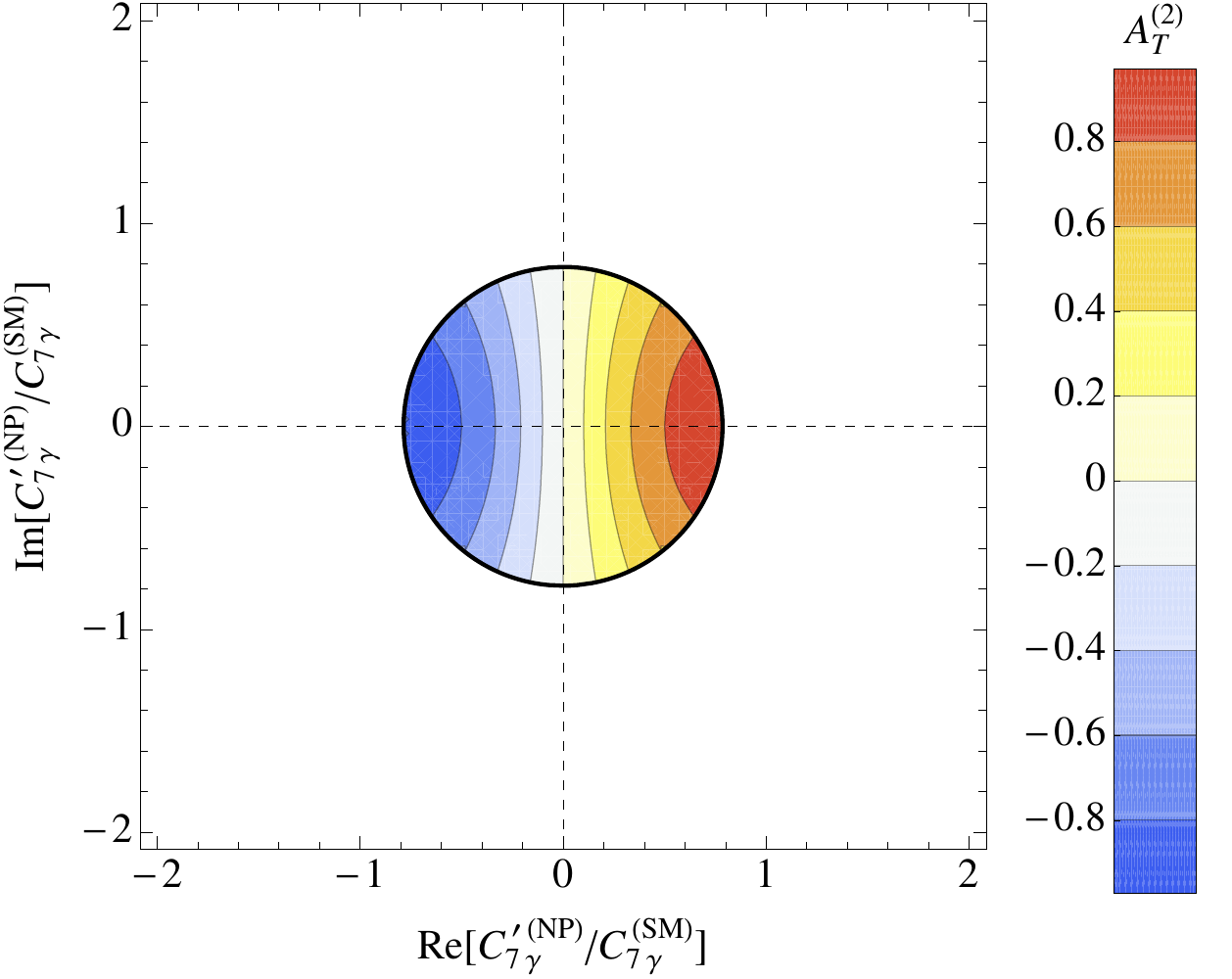}\label{fig:C7p_NP_IIc}}
		\put(-60,150){(c)}
		\put(-175,150){\footnotesize$\bm{B\to K^*\ell^+\ell^-}$}
	\end{subfigure}
	\hspace{5mm}
	\begin{subfigure}
		{\includegraphics[width=0.45\textwidth]{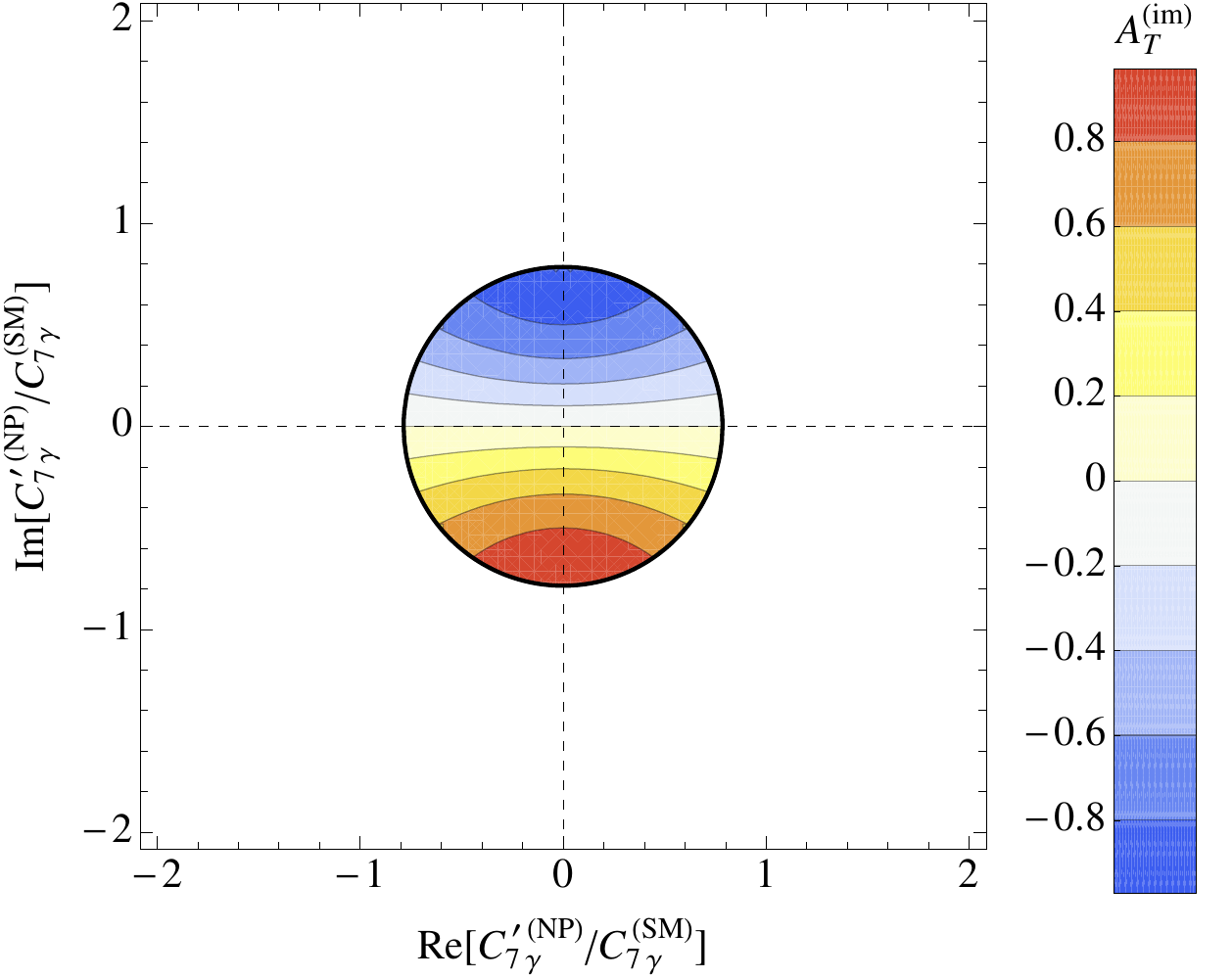}\label{fig:C7p_NP_IId}}
		\put(-60,150){(d)}
		\put(-175,150){\footnotesize$\bm{B\to K^*\ell^+\ell^-}$}
	\end{subfigure}
	\caption{\footnotesize Prospect of the future constraints on $\CCp_{7\gamma}$ in the NP scenario~{\it II}: $\CNP_{7\gamma}$ is purely SM-like, i.e. $\CNP_{7\gamma}=0$. The contour colours in Fig.~(a,\,b,\,c,\,d) correspond respectively to $S_{K_S\pi^0\gamma}$, $\lambda_\gamma$, $\A_T^{(2)}(0)$ and $\A_T^{\rm(im)}(0)$ allowed by a $\pm3\sigma$ error to the central value of $\BR^{\rm exp}(B\to X_s\gamma)$.}
	\label{fig:C7p_NP_II}
\end{figure}
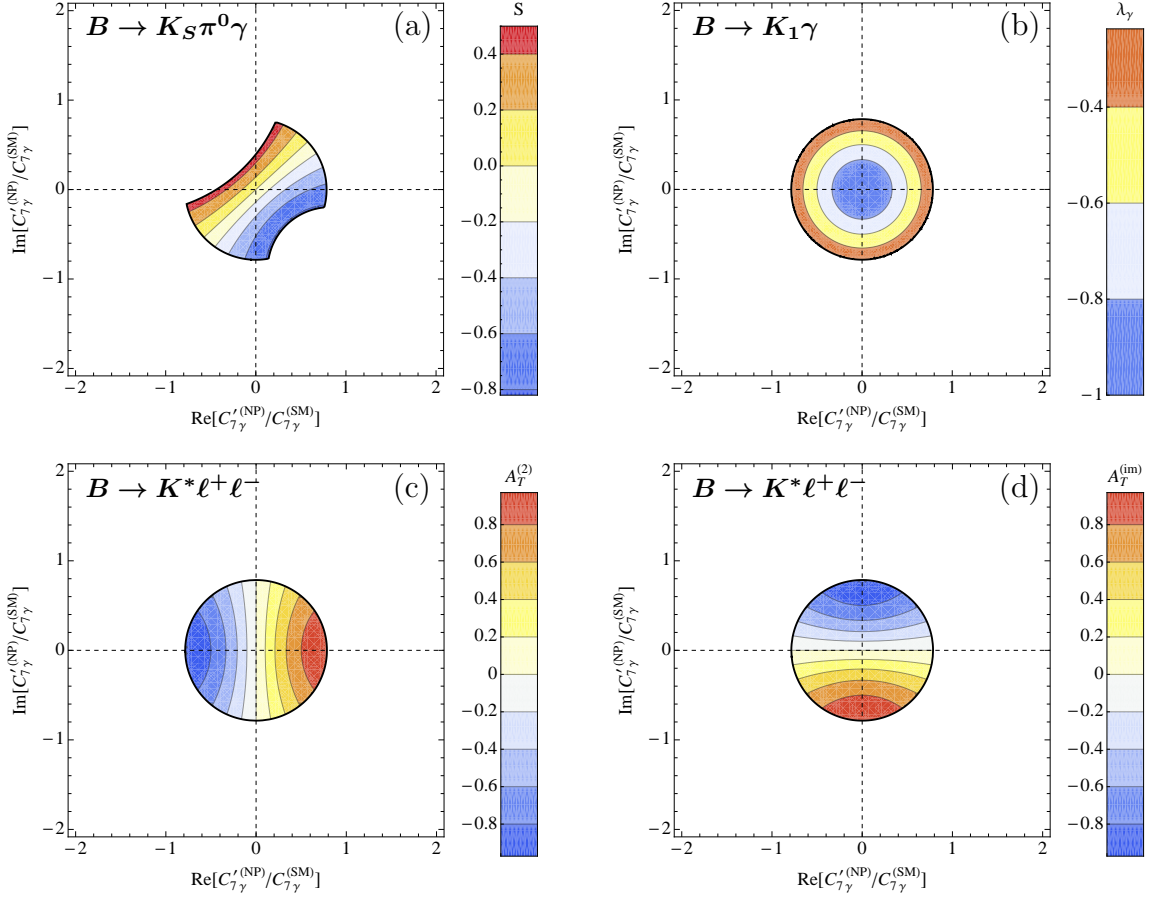

\begin{figure}[t!]\centering
	\begin{subfigure}
		{\includegraphics[width=0.45\textwidth]{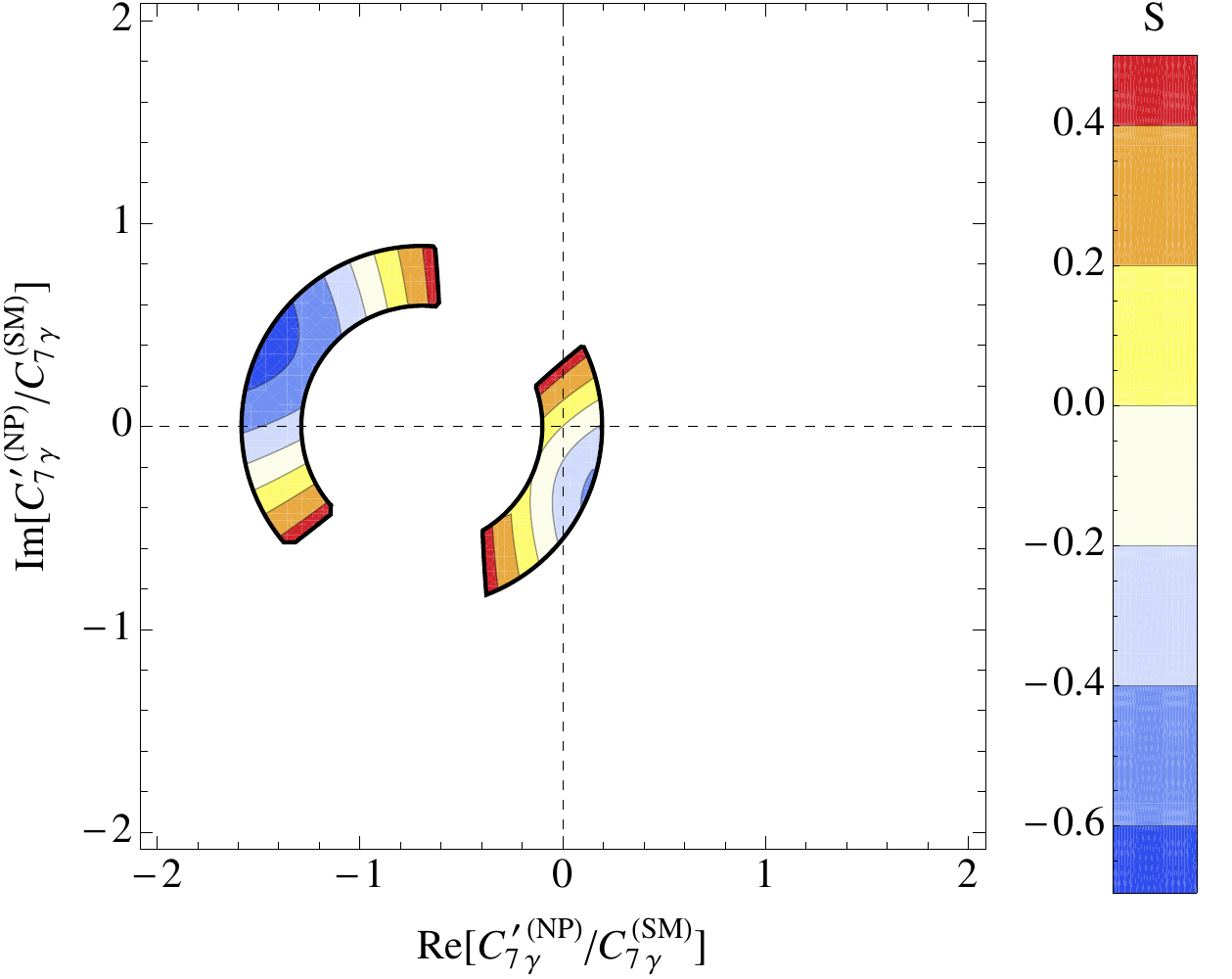}\label{fig:C7p_NP_IIIa}}
		\put(-60,150){(a)}
		\put(-175,150){\footnotesize$\bm{B\to K_S\pi^0\gamma}$}
	\end{subfigure}
	\hspace{5mm}
	\begin{subfigure}
		{\includegraphics[width=0.45\textwidth]{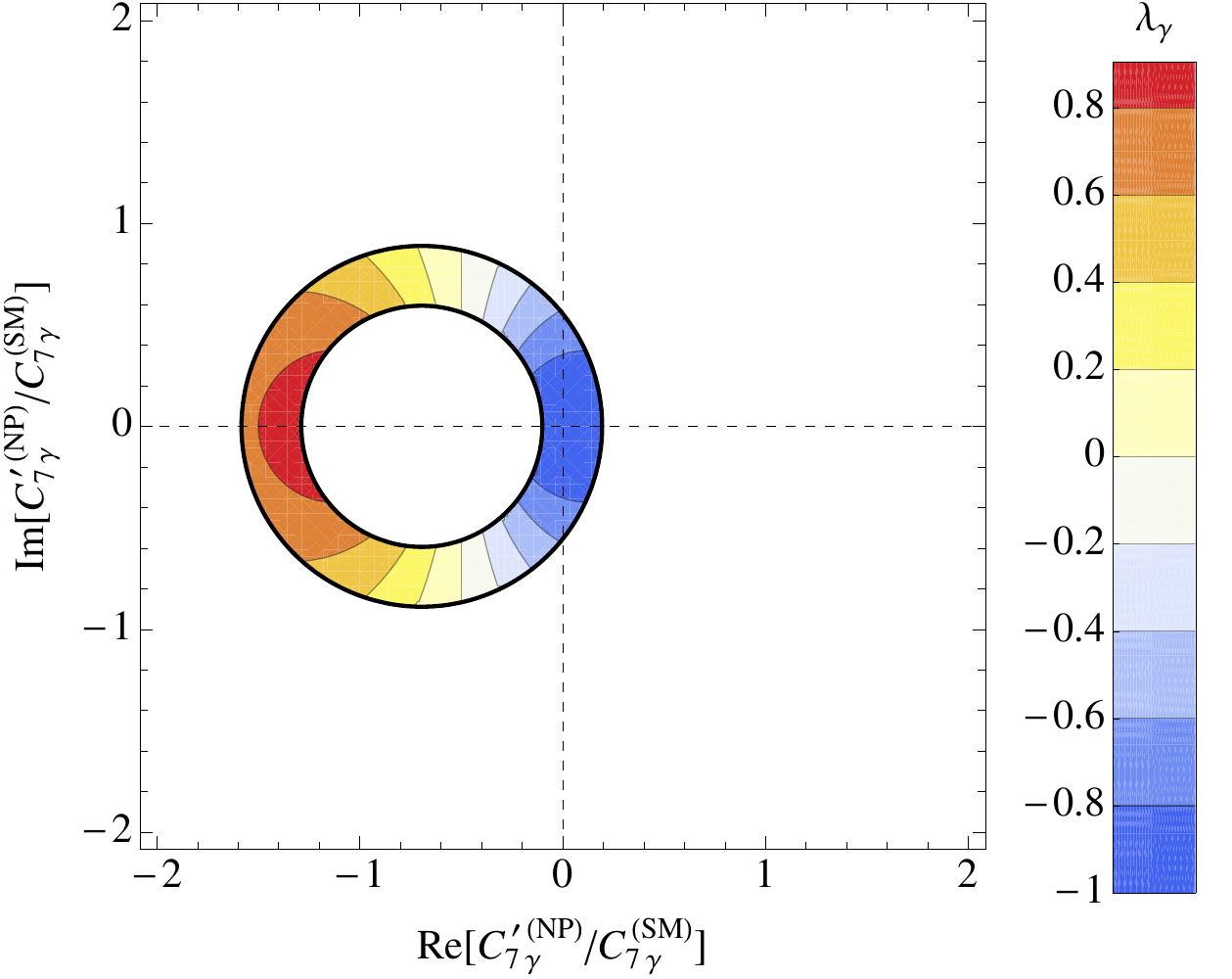}\label{fig:C7p_NP_IIIb}}
		\put(-60,150){(b)}
		\put(-175,150){\footnotesize$\bm{B\to K_1\gamma}$}
	\end{subfigure}
	\begin{subfigure}
		{\includegraphics[width=0.45\textwidth]{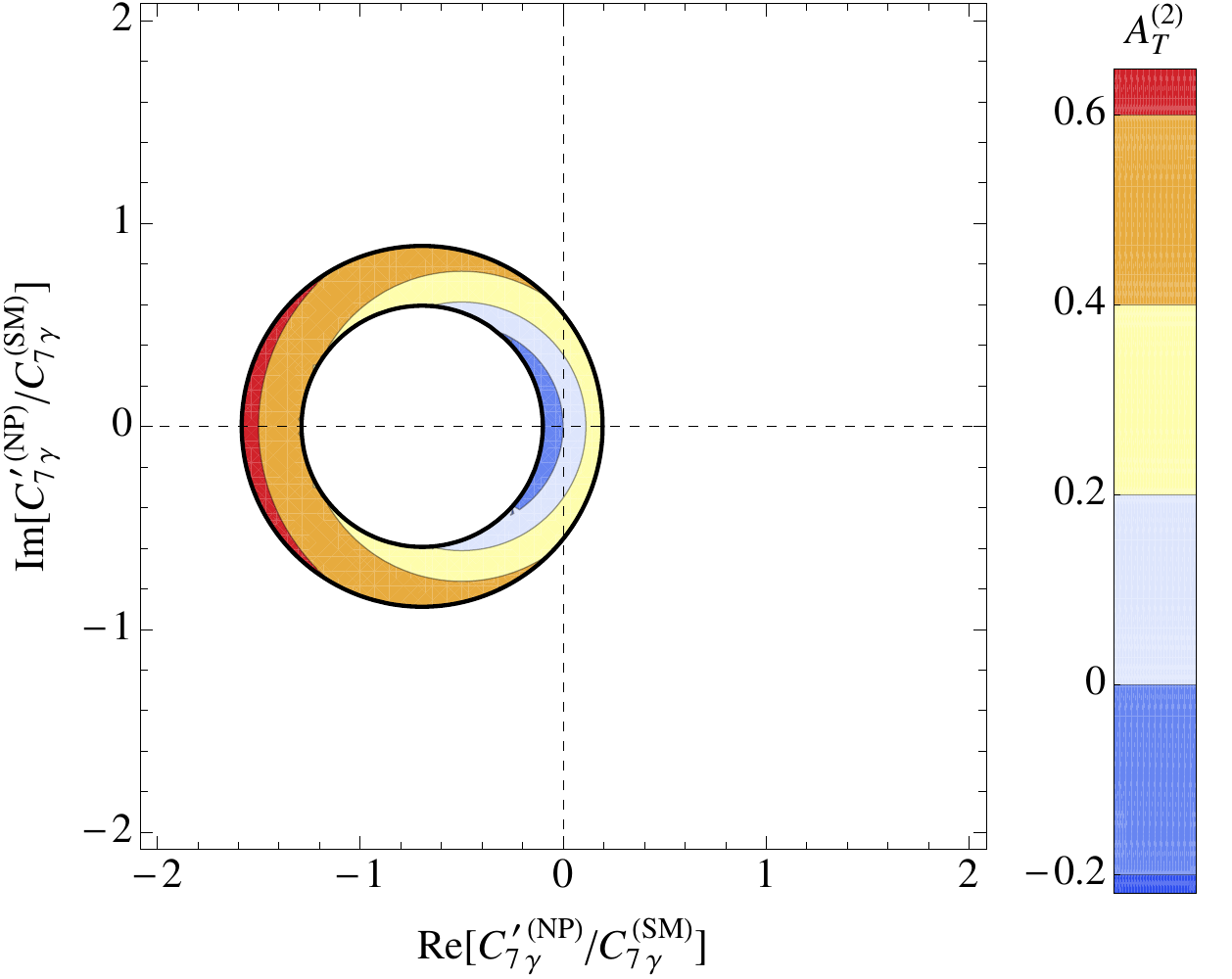}\label{fig:C7p_NP_IIIc}}
		\put(-60,150){(c)}
		\put(-175,150){\footnotesize$\bm{B\to K^*\ell^+\ell^-}$}
	\end{subfigure}
	\hspace{5mm}
	\begin{subfigure}
		{\includegraphics[width=0.45\textwidth]{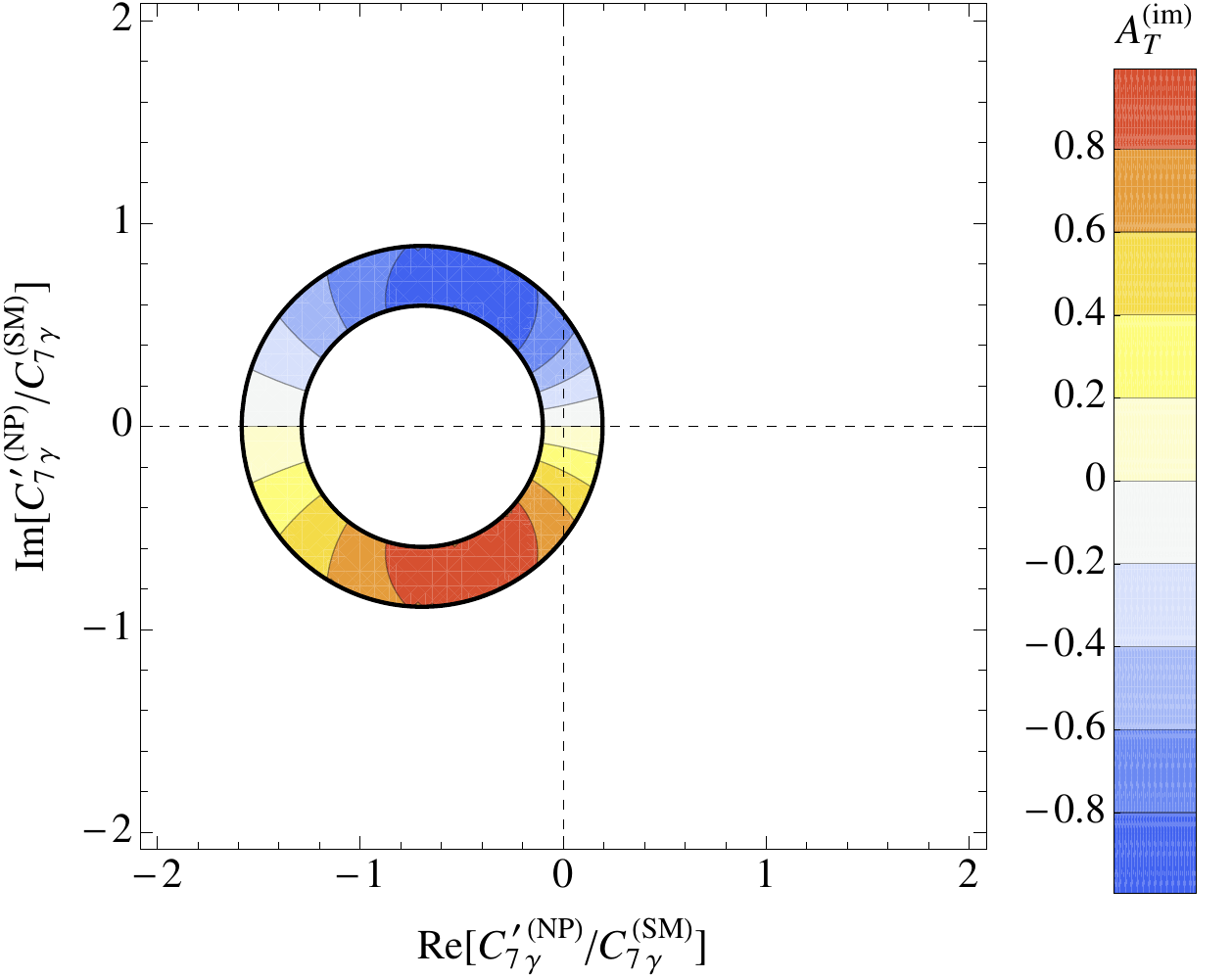}\label{fig:C7p_NP_IIId}}
		\put(-60,150){(d)}
		\put(-175,150){\footnotesize$\bm{B\to K^*\ell^+\ell^-}$}
	\end{subfigure}
	\caption{\footnotesize Prospect of the future constraints on $\CCp_{7\gamma}$ in the NP scenario~{\it III}: $\CNP_{7\gamma}=\CpNP_{7\gamma}$. The contour colours in Fig.~(a,\,b,\,c,\,d) correspond respectively to $S_{K_S\pi^0\gamma}$, $\lambda_\gamma$, $\A_T^{(2)}(0)$ and $\A_T^{\rm(im)}(0)$ allowed by a $\pm3\sigma$ error to the central value of $\BR^{\rm exp}(B\to X_s\gamma)$.}
	\label{fig:C7p_NP_III}
\end{figure}

\begin{figure}[h!]\centering
	\begin{subfigure}
		{\includegraphics[width=0.45\textwidth]{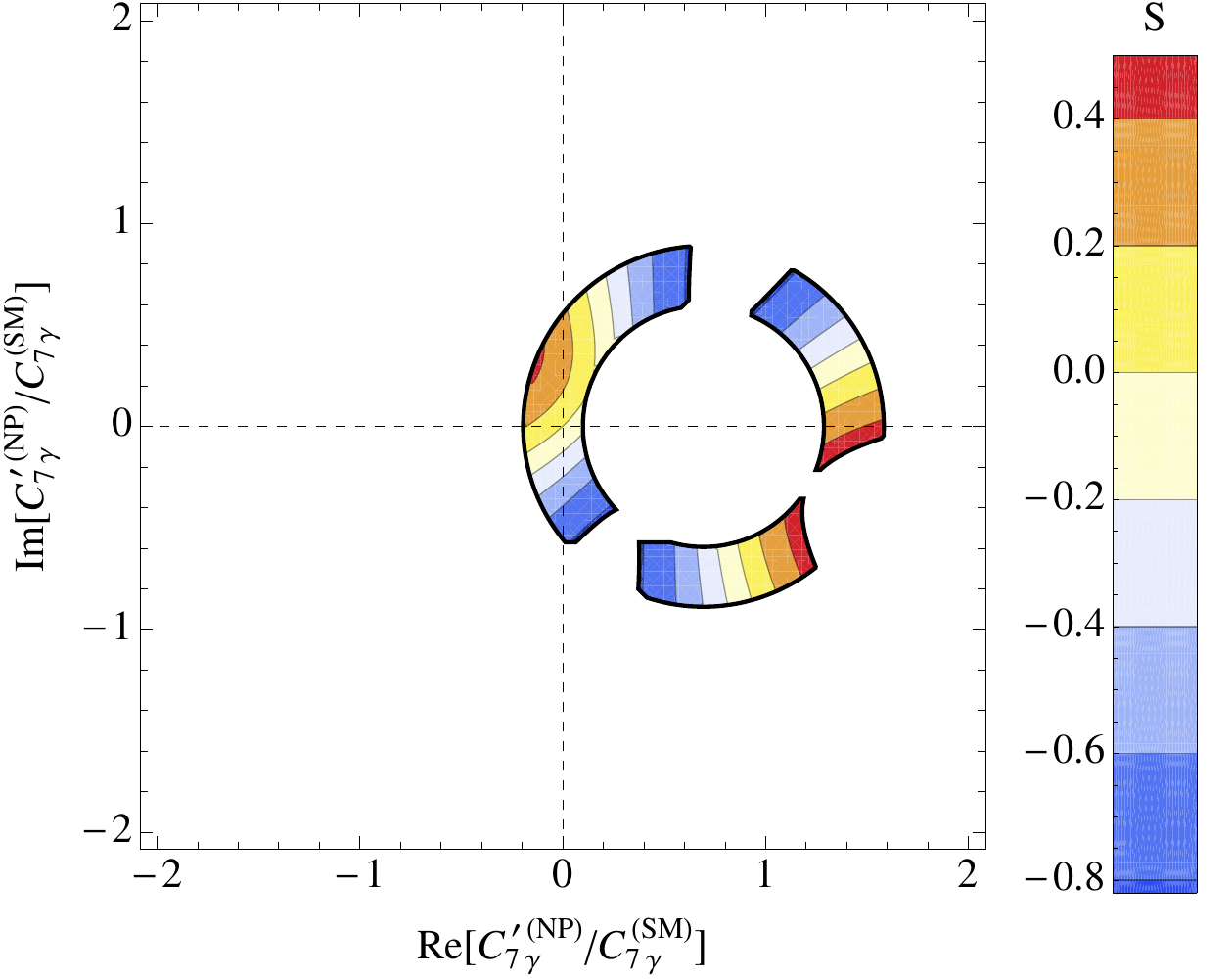}\label{fig:C7p_NP_IVa}}
		\put(-60,150){(a)}
		\put(-175,150){\footnotesize$\bm{B\to K_S\pi^0\gamma}$}
	\end{subfigure}
	\hspace{5mm}
	\begin{subfigure}
		{\includegraphics[width=0.45\textwidth]{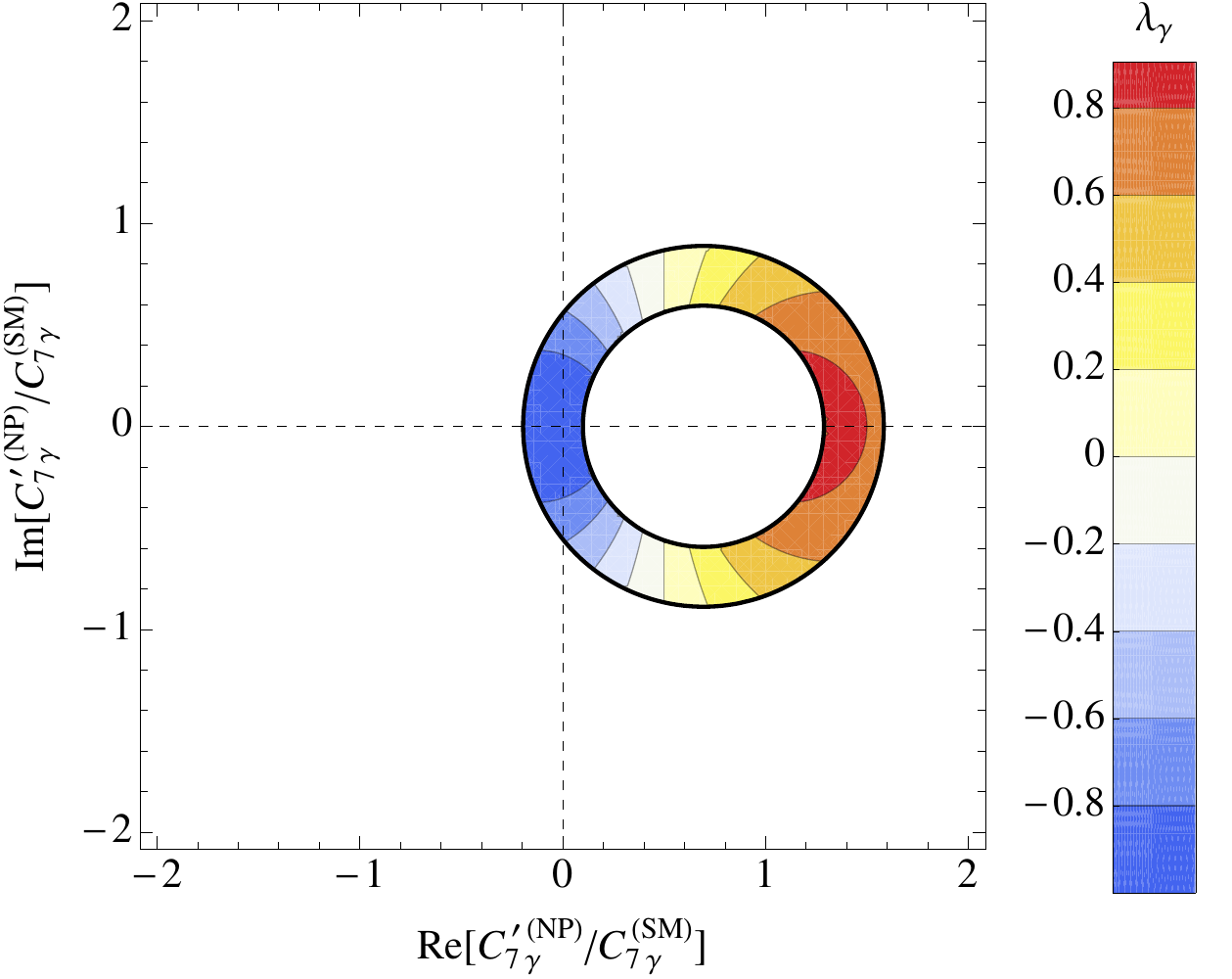}\label{fig:C7p_NP_IVb}}
		\put(-60,150){(b)}
		\put(-175,150){\footnotesize$\bm{B\to K_1\gamma}$}
	\end{subfigure}
	\begin{subfigure}
		{\includegraphics[width=0.45\textwidth]{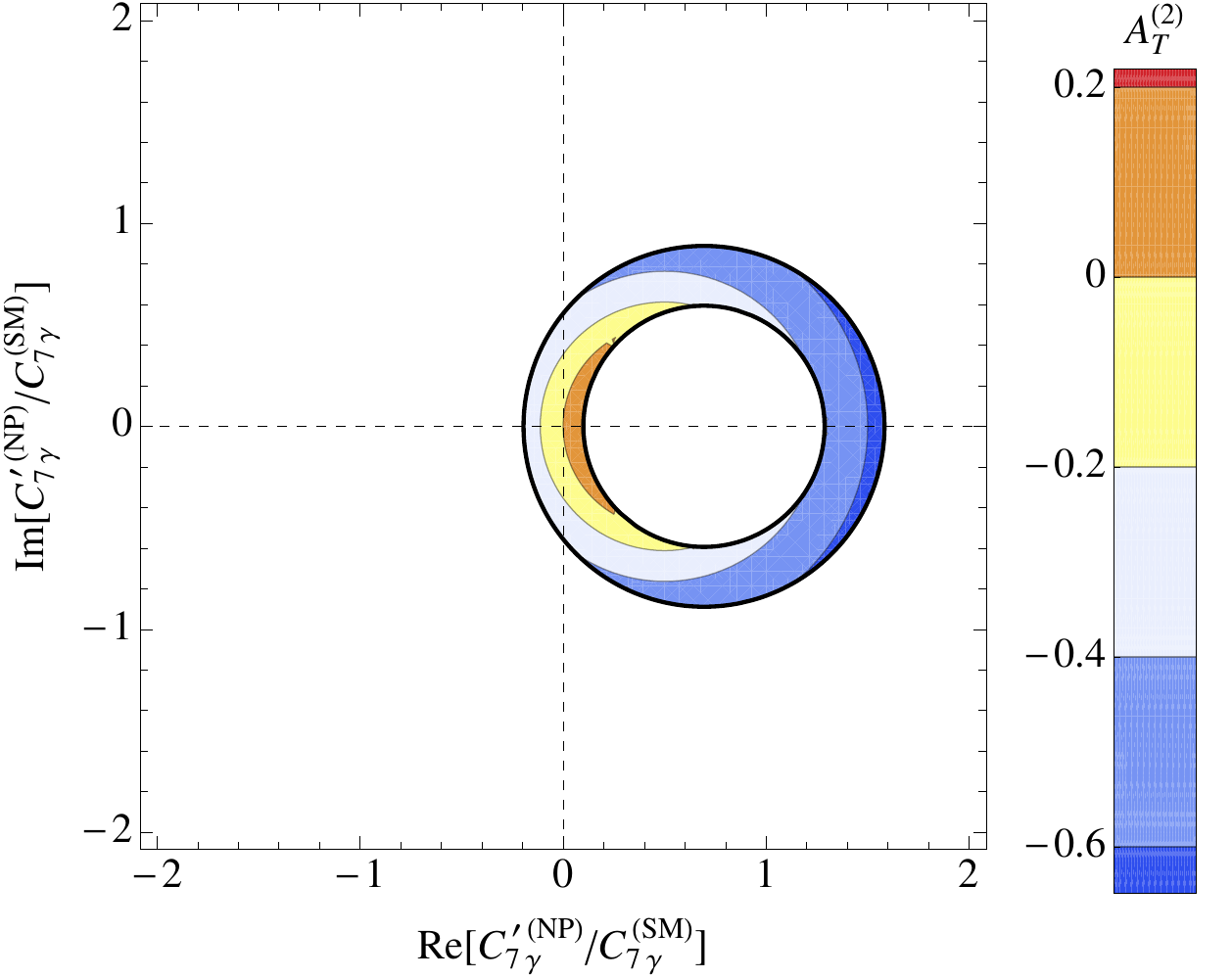}\label{fig:C7p_NP_IVc}}
		\put(-60,150){(c)}
		\put(-175,150){\footnotesize$\bm{B\to K^*\ell^+\ell^-}$}
	\end{subfigure}
	\hspace{5mm}
	\begin{subfigure}
		{\includegraphics[width=0.45\textwidth]{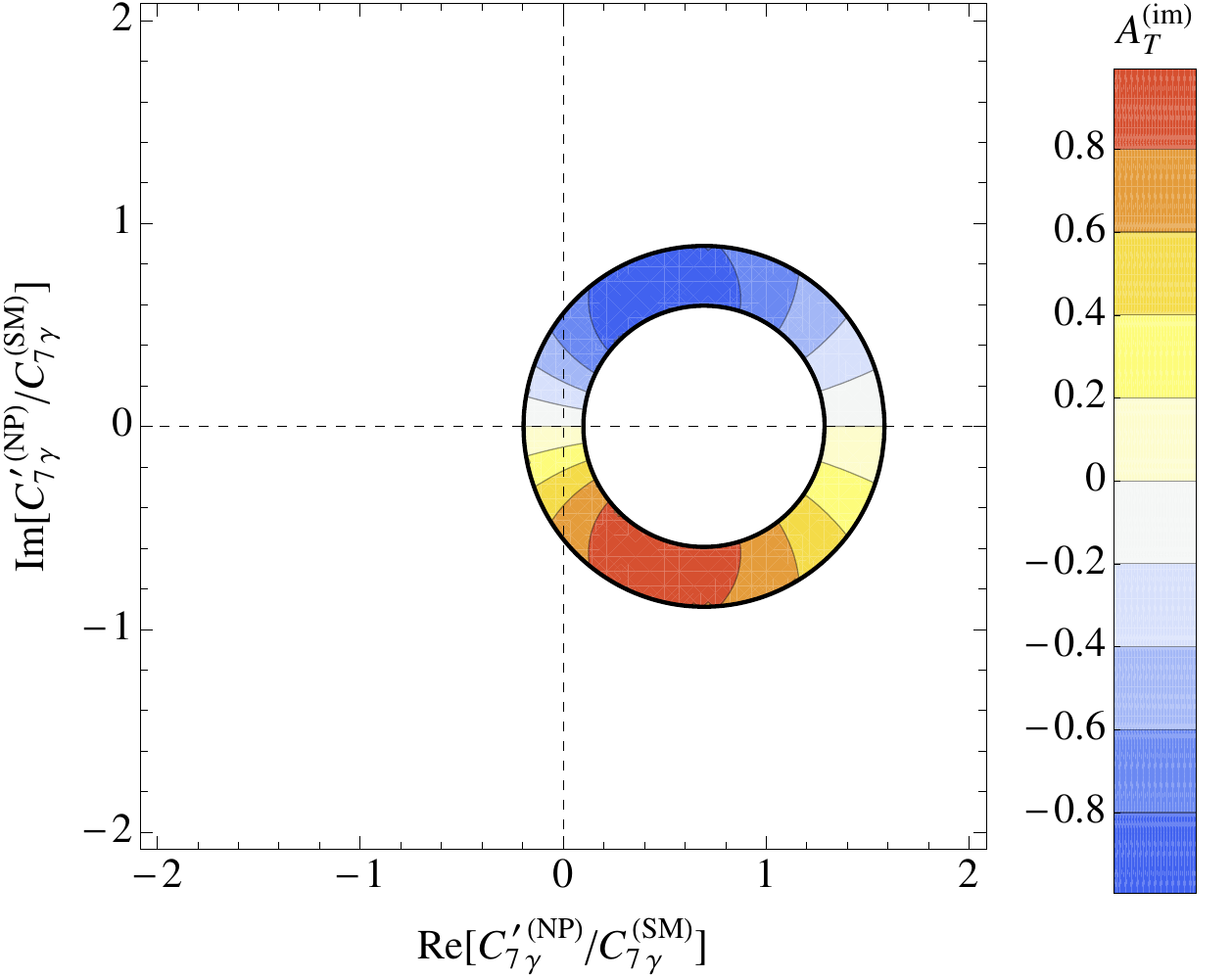}\label{fig:C7p_NP_IVd}}
		\put(-60,150){(d)}
		\put(-175,150){\footnotesize$\bm{B\to K^*\ell^+\ell^-}$}
	\end{subfigure}
	\caption{\footnotesize Prospect of the future constraints on $\CCp_{7\gamma}$ in the NP scenario~{\it IV}: $\CNP_{7\gamma}=-\CpNP_{7\gamma}$. The contour colours in Fig.~(a,\,b,\,c,\,d) correspond respectively to $S_{K_S\pi^0\gamma}$, $\lambda_\gamma$, $\A_T^{(2)}(0)$ and $\A_T^{\rm(im)}(0)$ allowed by a $\pm3\sigma$ error to the central value of $\BR^{\rm exp}(B\to X_s\gamma)$.}
	\label{fig:C7p_NP_IV}
\end{figure}

In Fig.~\ref{fig:C7p_NP_I}--\ref{fig:C7p_NP_IV}, we present a {\it future} prospect for constraining $\CCpNP_{7\gamma}$ in the four NP scenarios. The plots are obtained by assuming:

\begin{itemize}
	\item Improved measurement of the $CP$-asymmetry parameter $S_{K_S\pi^0\gamma}$ in $B\to K^*(\to K_S\pi^0)\gamma$. The contour colours correspond to $S_{K_S\pi^0\gamma}$ allowed by a $\pm3\sigma$ to the present world average \eqref{eq:SKSpi0gamma_exp}. Different colours are separated by the size of the current experimental error. That error will be significantly reduced at super $B$~factories.	
	\item Potential measurement of the polarization parameter $\lambda_\gamma$ in $\Bbar\to\Kbar_1(1270)\gamma$. The contour colours correspond to $\lambda_\gamma\in[-1,\,1]$. The spacing between contours is taken to be $\sigma_{\lambda_\gamma}=0.2$, which may be improved by the study of $K_1\to K\pi\pi$ decays. That can be made using a detailed experimental study of $B\to K_1\psi$ decay.
	\item Potential non-zero measurement of two transverse asymmetries, $\A_T^{\rm(2,\,im)}\in[-1,\,1]$, in $\Bbar\to\Kbar^*\ell^+\ell^-$. The contours correspond to $\A_T^{(2)}(0)$ and $\A_T^{\rm(im)}(0)$ respectively. The interval between the lines represents 20\% of uncertainty for each, which, in principle, can be achieved at LHCb. 
\end{itemize}
Note that in all these figures we applied the constraint from the measured $\BR(B\to X_s\gamma)$ as allowed by a $\pm3\sigma$ error to the central value \eqref{eq:BR_BXsgamma_exp}.

In Fig.~\ref{fig:C7p_NP_I}, we present our result for the \underline{scenario {\it I}}. The constraints from $S_{K_S\pi^0\gamma}$ and $\A_T^{(2)}$ look very similar in this scenario since both of them are proportional to $\frac{C_{7\gamma}\Cp_{7\gamma}}{C_{7\gamma}^2+C_{7\gamma}^{\,\prime\,2}}$ with $\CCpNP_{7\gamma}$ being real numbers. On the other hand, one can see that the shape of the constraint from $\lambda_\gamma$ is quite different. For example, the fourfold ambiguity in the constraints of $S_{K_S\pi^0\gamma}$ and $\A_T^{(2)}$ can be reduced to a twofold with the help of the $\lambda_\gamma$ measurement. In addition, one observes that the region around the line $\CNP_{7\gamma}=\CpNP_{7\gamma}$ is quite sensitive to the $\lambda_\gamma$ values, while it is not in the case of $S_{K_S\pi^0\gamma}$ and $\A_T^{(2)}$.

In Fig.~\ref{fig:C7p_NP_II}, we present our result for the \underline{scenario {\it II}}. The constraint from $S_{K_S\pi^0\gamma}$ is very strong (indeed, assuming that fact that the experimental error will be significantly reduced by super $B$~factories down to 2\%, the bound, i.e. the spacing between the adjacent contours will become about 10 times more narrow than those depicted in Fig.~\ref{fig:C7p_NP_IIa}) but it has an ambiguity along the diagonal. Note that this diagonal pattern of the constraint results from the fact that the observable is obtained from $r$ by a rotation in the complex plane,
\begin{equation}
	S_{K_S\pi^0\gamma}\simeq\frac{2|r|}{1+|r|^2}\sin(\phi_M-\phi_R)\simeq-\Re[r]\sin2\beta+\Im[r]\cos2\beta + O(|r|^2) \,,
	\label{}
\end{equation}
(see Eq.~\eqref{eq:combination} where $\phi_L$ is set to $\pi$ by assumption, since $C_{7\gamma}$ is real and negative in the SM; we also assumed that NP effect on $\phi_M$ is negligible, which is consistent with experiment). Thus, one finds that $S_{K_S\pi^0\gamma}$ is approximately a linear combination of $\Re[r]$ and $\Im[r]$ within the region allowed by $\BR(B\to X_s\gamma)$. One can make a general statement that if one experimentally finds $S_{K_S\pi^0\gamma}\simeq 0$, it will imply $|r|\simeq0$ or $\phi_L+\phi_R\simeq2\beta\simeq43^{\circ}$.

This problem can be partially solved by adding a constraint from $\lambda_\gamma$ which is a circle since $\lambda_\gamma$ is a function of $|r|^2$ and therefore is insensitive to the complex phases. The SM prediction corresponds to the central point $\CNP_{7\gamma}=\CpNP_{7\gamma}=(0,\,0)$. Near the center $\lambda_\gamma=\lambda_\gamma^{\rm SM}\simeq-1$, and the sensitivity to $\Cp_{7\gamma}$ is very low. For $\lambda_\gamma\simeq-0.8$ we have $|\CpNP_{7\gamma}/\CSM_{7\gamma}|\simeq0.3$ (i.e. one is clearly outside the SM prediction), but inside the circle one cannot distinguish the NP contribution from the SM one.

The combined measurement of $\A_T^{(2)}(q^2)$ and $\A_T^{\rm(im)}(q^2)$ can, in principle, constraint both $|r|$ and the relative phase $\phi_L-\phi_R$ (or equivalently, $\Re[r]$ and $\Im[r]$) independently on $S_{K_S\pi^0\gamma}$ and $\lambda_\gamma$. In contrast to $\lambda_\gamma$, it is also sensitive to the SM prediction.

In Fig.~\ref{fig:C7p_NP_III} and~\ref{fig:C7p_NP_IV}, we present our results for \underline{scenarios {\it III} and {\it IV}}. The combination of the $\A_T^{(2)}$ and $\A_T^{\rm(im)}$ measurements, contrary to the scenario~{\it II}, leaves a twofold ambiguity since the constraint from $\A_T^{(2)}$ becomes a circle. In these two scenarios, the three- and fourfold ambiguities of the $S_{K_S\pi^0\gamma}$ constraint can be removed by adding the $\lambda_\gamma$ and $\A_T^{\rm(2,\,im)}$ constraints.

A pleasant feature of Fig.~\ref{fig:C7p_NP_I}--\ref{fig:C7p_NP_IV} is that the shapes of the resulting plots are quite different in NP scenarios. The four constraints will overlap in scenarios compatible with measured $S_{K_S\pi^0\gamma}$, $\lambda_\lambda$ and $\A_T^{\rm(2,\,im)}$ and we will be able to extract $\CCp_{7\gamma}$ and their phases. In incompatible scenarios, the four constraints will not overlap.

\begin{figure}[t]\centering
	\begin{subfigure}
		{\includegraphics[width=0.4\textwidth]{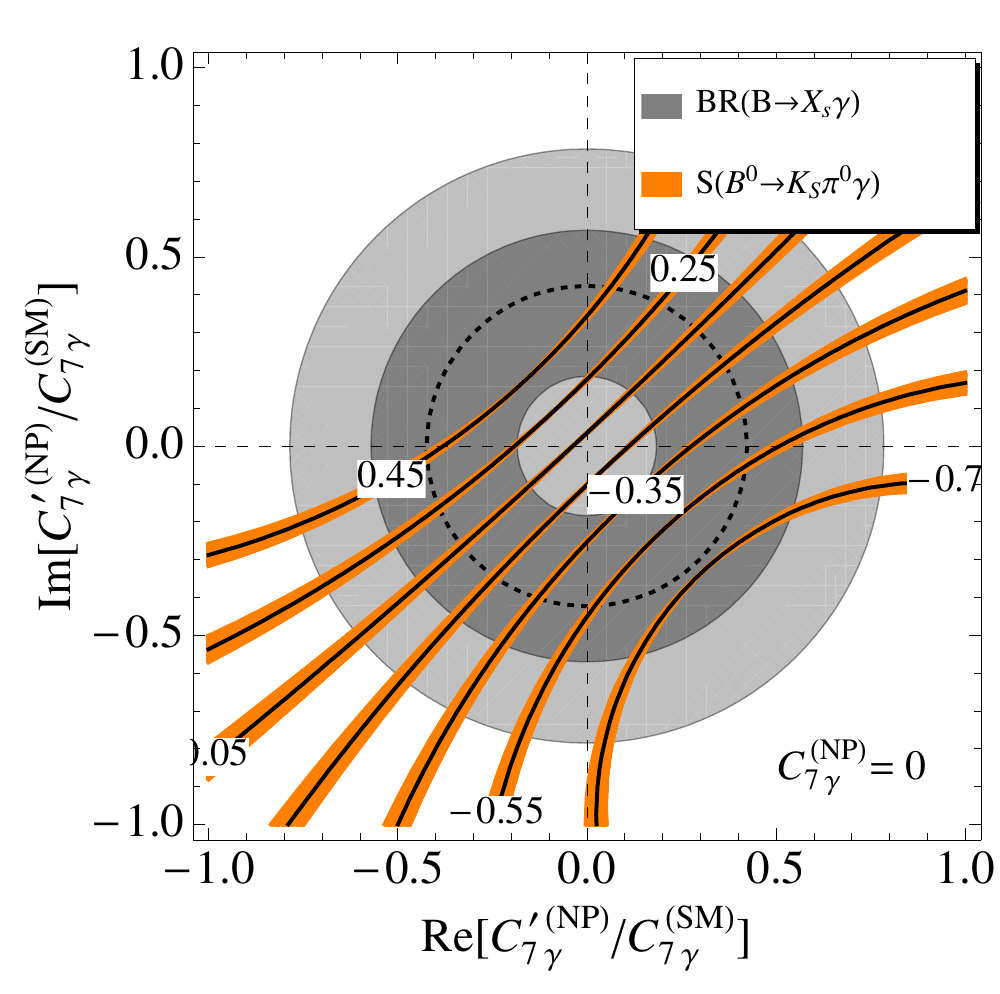}\label{fig:C7p_ACP_contours_a}}
		\put(-80,180){(a)}
	\end{subfigure}
	\hspace{5mm}
	\begin{subfigure}
		{\includegraphics[width=0.4\textwidth]{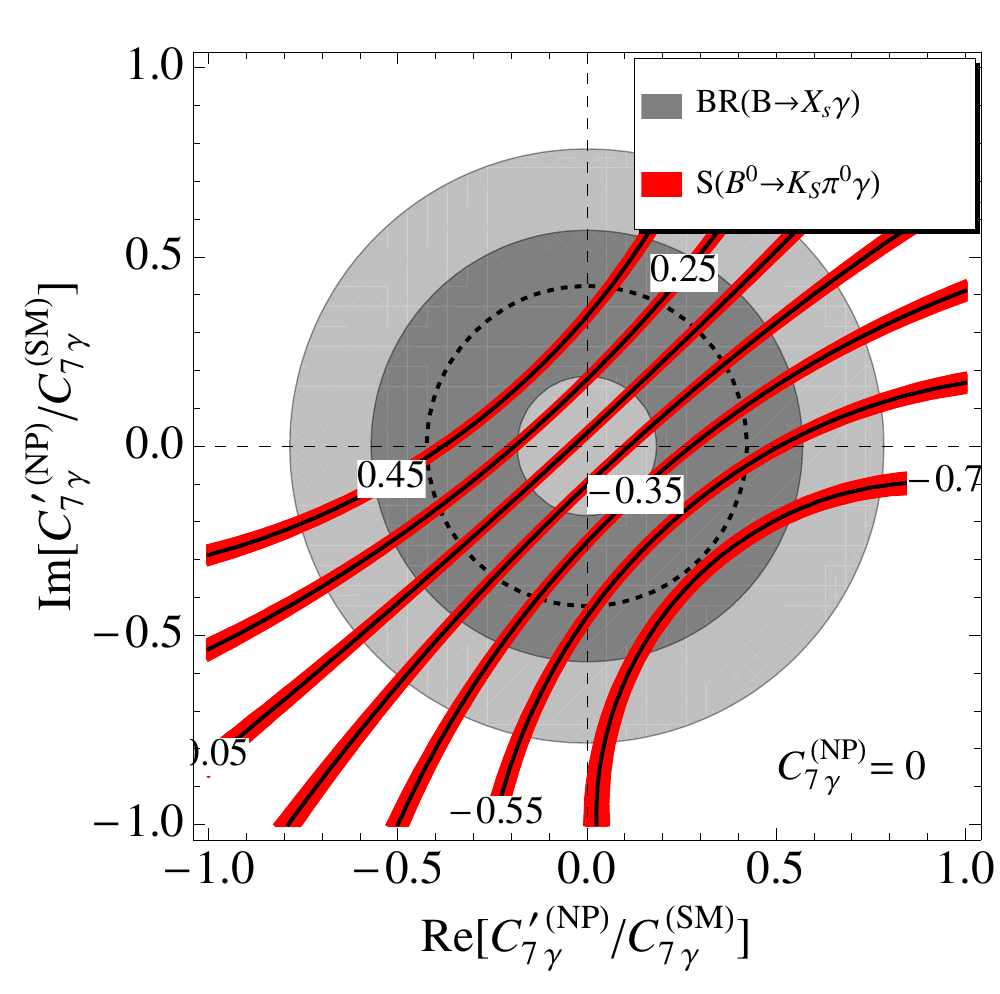}\label{fig:C7p_ACP_contours_b}}
		\put(-80,180){(b)}
	\end{subfigure}
	\caption{\footnotesize Prospect of the future constraints on $\Cp_{7\gamma}$ in the NP scenario~{\it II} with $\CNP_{7\gamma}=0$. The orange curves in Fig.~(a) represent the uncertainty of $S_{K_S\pi^0\gamma}$ related to the $B-\Bbar$ mixing phase $2\beta$. The red regions in Fig.~(b) represent the future bounds ($\pm1\sigma$) on $S_{K_S\pi^0\gamma}$ at super $B$~factories.}
	\label{fig:C7p_ACP_contours}
\end{figure}

\vspace{5mm}
Once again, we stress that we can determine $\Cp_{7\gamma}/C_{7\gamma}$ from $S_{K_S\pi^0\gamma}$ only in combination with the $B-\Bbar$ mixing phase, $\phi_M$. In this paper we assume that NP does not bring any significant contribution to the $B-\Bbar$ mixing box diagrams and use the currently measured value, $\sin2\beta=0.673\pm0.023$~\cite{PDG}. The impact of the uncertainty on $\sin2\beta$ is depicted in Fig.~\ref{fig:C7p_ACP_contours_a} with multiple orange bands, labeled with values of $S_{K_S\pi^0\gamma}$. In future, super $B$~factories will be able to measure the asymmetry within the 2\% error, which means that we will have a very thin constraint along one of the black lines in Fig.~\ref{fig:C7p_ACP_contours_b} within the red bands which represent $\pm1\sigma=\pm0.02$ regions. One can notice that theoretical uncertainty on $S_{K_S\pi^0\gamma}$, coming from the $B-\Bbar$ mixing phase determination, will be comparable to the experimental one at the super $B$~factories.

\vspace{5mm}
Keep in mind that $\A_T^{\rm(2,\,im)}(q^2)$ are going to be measured in a $q^2$-bin between 0 and~1~GeV$^2$, and to extract $\A_T^{\rm(2,\,im)}(0)$ one can use the following approximation
\begin{equation}
	\A_T^{\rm(2,\,im)}(q^2)=a_0^{\rm(2,\,im)}+a_1^{\rm(2,\,im)}q^2 +O(q^4) \,,
\end{equation}
with the intercepts and slopes simply being
\begin{equation}
	a_0^{\rm(2,\,im)} = \lim_{q^2\to0}\A_T^{\rm(2,\,im)}(q^2) \,, \quad a_1^{\rm(2,\,im)} = \left.\frac{\partial\A_T^{\rm(2,\,im)}(q^2)}{\partial q^2}\right|_{q^2=0} \,.
	\label{eq:AT2im_linear_approx}
\end{equation}
Their expressions in terms of Wilson coefficients are expressions in the Appendix.

\begin{figure}[h!]\centering
	\begin{subfigure}
		{\includegraphics[width=0.4\textwidth]{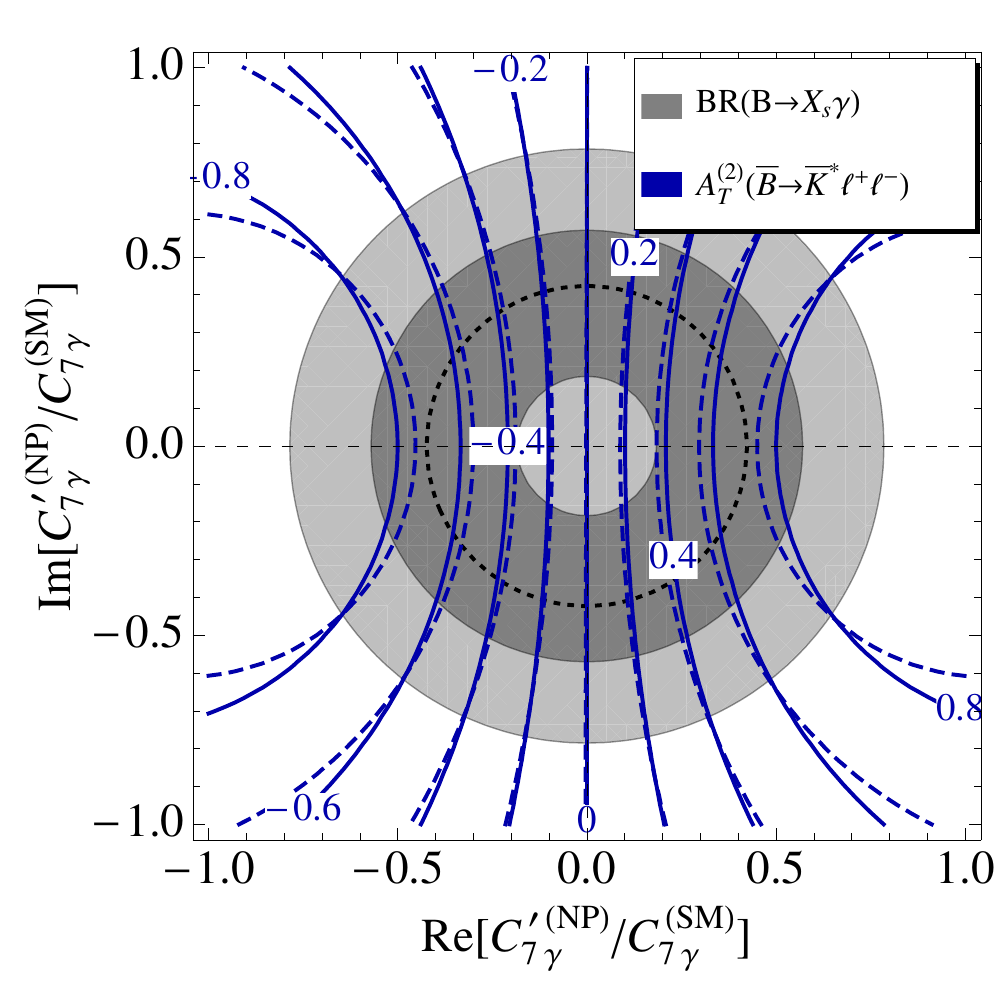}\label{fig:C7p_AT2im_intq2_a}}
		\put(-80,180){(a)}
	\end{subfigure}
	\hspace{5mm}
	\begin{subfigure}
		{\includegraphics[width=0.4\textwidth]{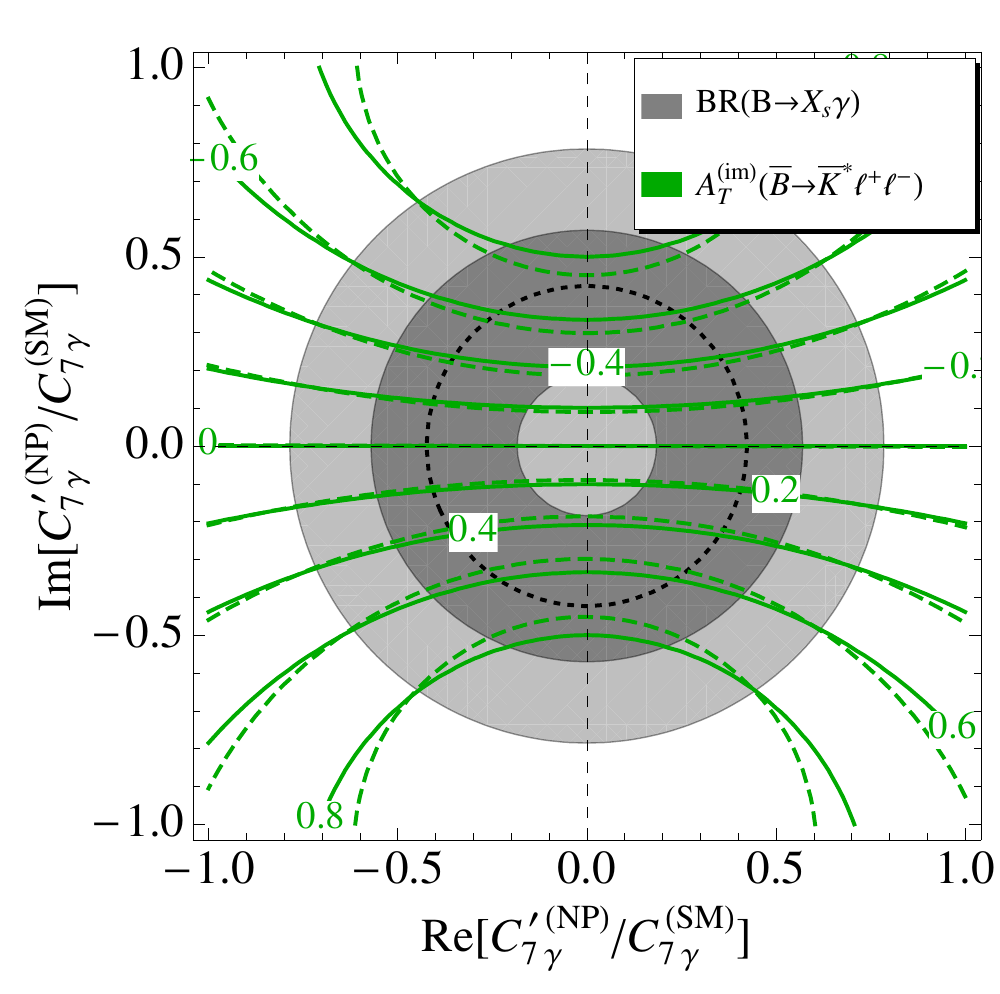}\label{fig:C7p_AT2im_intq2_b}}
		\put(-80,180){(b)}
	\end{subfigure}
	\begin{subfigure}
		{\includegraphics[width=0.4\textwidth]{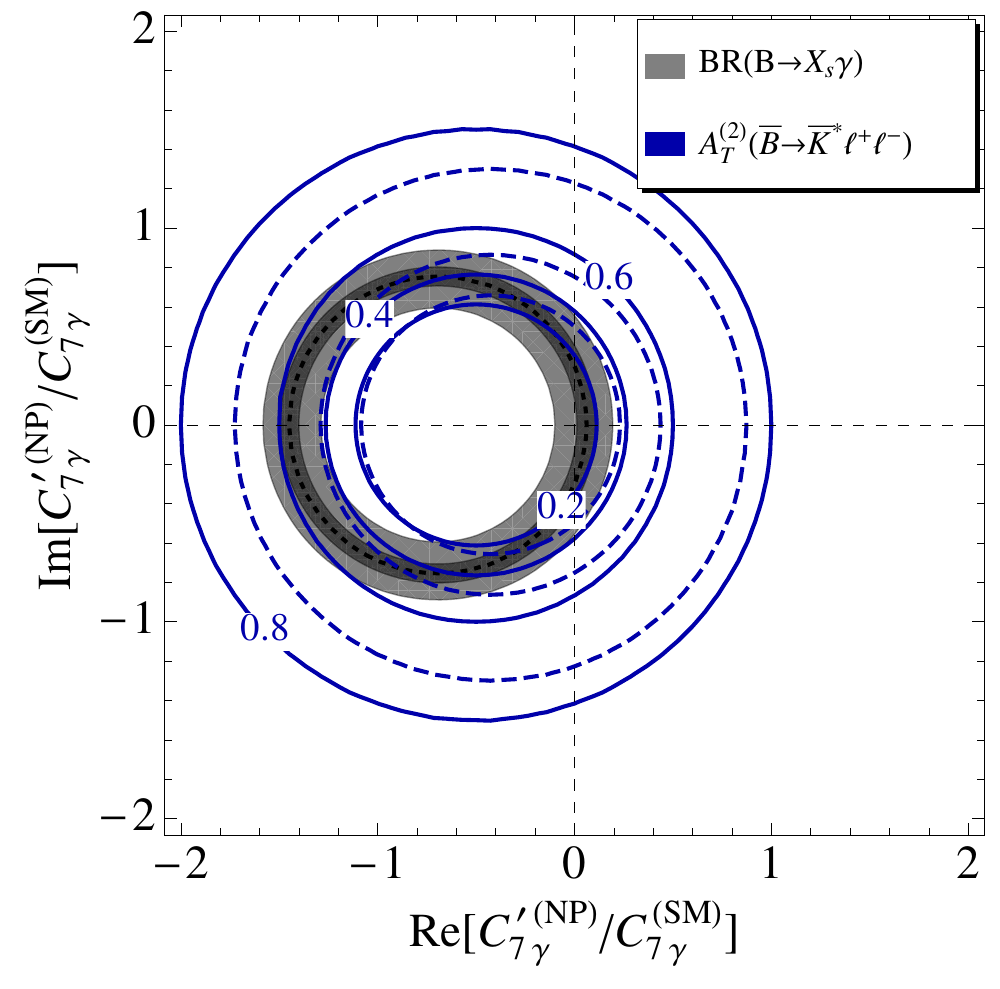}\label{fig:C7p_AT2im_intq2_c}}
		\put(-80,185){(c)}
	\end{subfigure}
	\hspace{5mm}
	\begin{subfigure}
		{\includegraphics[width=0.4\textwidth]{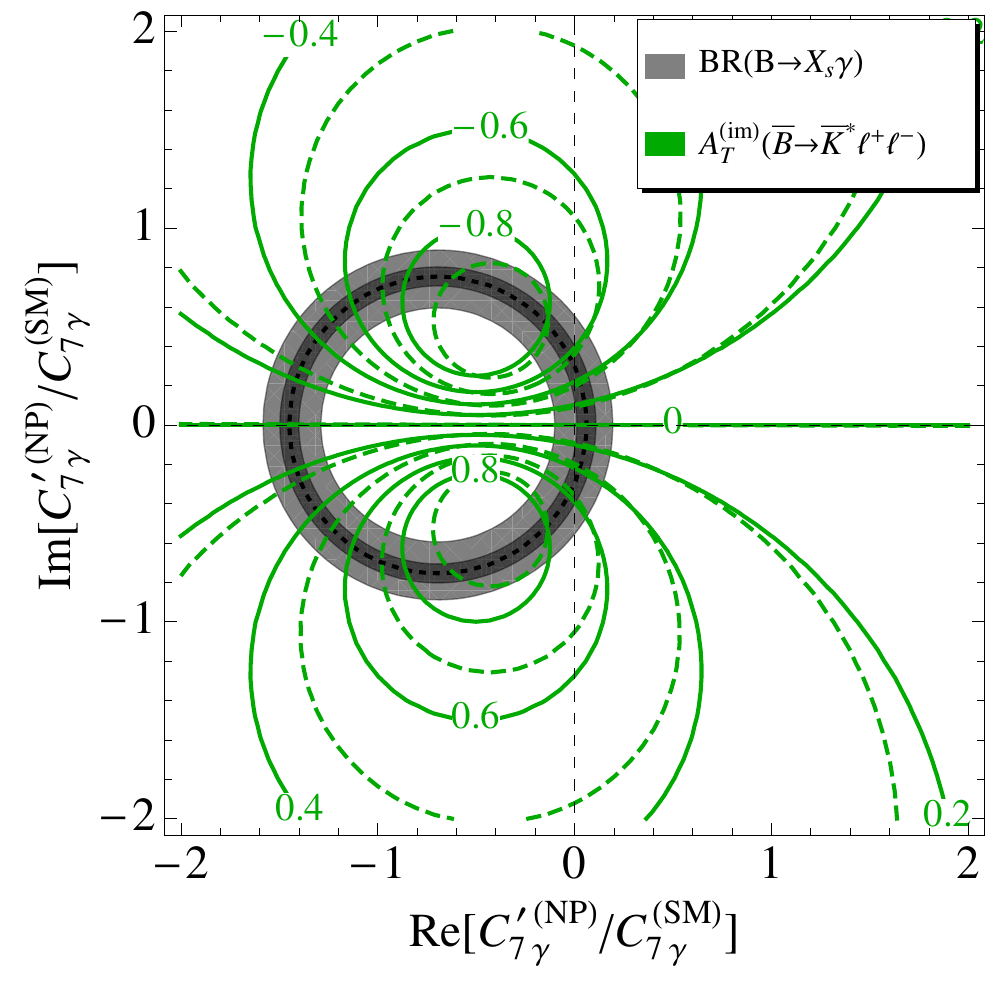}\label{fig:C7p_AT2im_intq2_d}}
		\put(-80,185){(d)}
	\end{subfigure}
	\caption{\footnotesize Prospect of the future constraints on $\Cp_{7\gamma}$ in the NP scenarios~{\it II}~(a,\,b) and {\it III}~(c,\,d). The solid and dashed curves correspond respectively to $\A_T^{\rm(2,\,im)}(0)$ and to $\A_T^{\rm(2,\,im)}$ integrated over $q^2$ in the $[4m_\ell^2,\,1~{\rm GeV}^2]$ range (for details see the text).}
	\label{fig:C7p_AT2im_intq2}
\end{figure}

In this work we do not use the slopes of $\A_T^{\rm(2,\,im)}(q^2)$ as constraints but one can envisage using them in the future. We test the approximation \eqref{eq:AT2_ATim_0} by a comparison to the asymmetries integrated within the lowest bin. The solid and dashed contours in Fig.~\ref{fig:C7p_AT2im_intq2}, labeled with values of $\A_T^{\rm(2,\,im)}$, correspond respectively to the asymmetries calculated at $q^2=0$ and integrated over $q^2$ up to 1~GeV$^2$. One can see from Fig.~\ref{fig:C7p_AT2im_intq2_a} and \ref{fig:C7p_AT2im_intq2_b} that the discrepancy between the solid and dashed lines within the allowed space is small compared to the spacing between the contours (i.e. the expected experimental error at LHCb). We can conclude that the slopes are reasonably small in the NP scenario~{\it II}. This is no longer valid for the case of scenarios~{\it III} or {\it IV} where this discrepancy is not negligible as can be seen from Fig.~\ref{fig:C7p_AT2im_intq2_c} and \ref{fig:C7p_AT2im_intq2_d}. Here we assumed $C_{9,10}$ to be SM-like, while these discrepancies can be larger or smaller depending on the NP effect on $\CCp_{9,10}$. Furthermore, in the future, when the refined measurement of $\A_T^{\rm(2,\,im,\,re)}(q^2)$ will be made, we will also be able to use the slopes of $\A_T^{\rm(2,\,im,\,re)}(q^2)$ to further constrain the NP models.

\subsection{SUSY models with large $(\delta_{RL}^d)_{23}$ mass insertion}

As a specific example of the above discussion we consider a SUSY motivated model. It is known that after the spontaneous symmetry breaking the squark mass can come from any combination of the left- and right-handed couplings in the soft SUSY breaking part of the Lagrangian:
\begin{equation}
	\begin{split}
		\mathcal{L}_{\rm soft}^{\rm squark~mass} =& (m_Q^2)_{ij}\tilde{q}_{Li}^\dagger \tilde{q}_{Lj}+(m_U^2)_{ij}\tilde{u}_{Ri}^\dagger \tilde{u}_{Rj}+(m_D^2)_{ij}\tilde{d}_{Ri}^\dagger \tilde{d}_{Rj} \\
		&+(\upsilon_2 A_U^{ij}\tilde{u}_{Ri}^\dagger \tilde{q}_{Lj}+\upsilon_1 A_D^{ij}\tilde{d}_{Ri}^\dagger \tilde{q}_{Lj}+h.c.) \,,
	\end{split}
	\label{}
\end{equation}
where $\upsilon_{1,2}$ are the vacuum expectation values of the Higgs fields and $i,j$ are the generation indices. {\it Since the squark mass matrices ($m_Q$, $m_U$, $m_D$) and the trilinear couplings ($A_U^{ij}$, $A_D^{ij}$) are not diagonal in the quark basis, the squark propagators can change flavour and chirality}. Once these new terms are introduced, the $b\to s\gamma$ process could receive a significant new contribution.

In organizing the soft SUSY breaking terms, the mass insertion approximation (MIA) \cite{Hall:1985dx} is often used. In the so-called super-CKM basis the couplings of fermions and sfermions to neutral gauginos are flavour diagonal, leaving the source of flavour violation in the off-diagonal terms of the sfermion mass matrix. These terms are described by $(\Delta_{AB}^q)_{ij}$, where $A$, $B$ denote the chirality ($L,R$) and $q$ indicates the ``up" or ``down" type. The sfermion propagator can then be expanded as \cite{Aushev:2010bq}
\begin{equation}
	\langle\tilde{q}_{Ai}\tilde{q}_{Bj}^*\rangle = i(k^2-m_{\tilde{q}}^2-\Delta_{AB}^q)_{ij}^{-1}\simeq\frac{i\delta_{ij}}{k^2-m_{\tilde{q}}^2}+\frac{i(\Delta_{AB}^q)_{ij}}{(k^2-m_{\tilde{q}}^2)^2}+\dots \,,
	\label{}
\end{equation}
where $m_{\tilde{q}}$ is the average squark mass. Assuming that $\Delta^2\ll m_{\tilde{q}}^2$, so that the first term in expansion is sufficient, the flavour violation can be parametrized in a {\it model independent} way by the dimensionless MIA parameters
\begin{equation}
	(\delta_{AB}^q)_{ij}=\frac{(\Delta_{AB}^q)_{ij}}{m_{\tilde{q}}^2} \,,
	\label{}
\end{equation}
the values of which can be constrained by various flavour experiments. 

Let us consider the dominant gluino contribution to the $\CCp_{7\gamma}$ Wilson coefficients\footnote{At leading order, both the charged Higgs and the chargino contributions to $\Cp_{7\gamma,\,8g}$ are suppressed by $m_s/m_b$. For simplicity, we do not present these last contributions here.}. The coefficients, evaluated at large scale $M_S$ can be written in terms of down-type MIA parameters, giving rise to the contribution from the insertion of the gluino mass and the one of a scalar mass term. They both violate chirality and flavour, and read\footnote{Here we do not consider the contribution from the gluino exchange with chirality violation coming from the $b$-quark mass which is suppressed by a factor $m_b/m_\sg$ compared to the dominant gluino MI contribution.}		
\begin{subequations}
	\begin{align}
		C_{7\gamma}^{\,(\sg)}(M_S) =& \frac{\sqrt2 \alpha_s\pi}{G_FV_{tb}V_{ts}^*m_\sq^2}\left[\frac{m_\sg}{m_b}(\delta_{LR}^d)_{23}g_7^{(1)}(x_\sg) + \frac{m_\sg\mu}{m_\sq^2}\frac{t_\beta}{1+\epsilon t_\beta}(\delta_{LL}^d)_{23}g_7^{(2)}(x_\sg)\right] \,, \\
		C_{7\gamma}^{\,\prime\,(\sg)}(M_S) =& \frac{\sqrt2 \alpha_s\pi}{G_FV_{tb}V_{ts}^*m_\sq^2}\left[\frac{m_\sg}{m_b}(\delta_{RL}^d)_{23}g_7^{(1)}(x_\sg) + \frac{m_\sg\mu^*}{m_\sq^2}\frac{t_\beta}{1+\epsilon t_\beta}(\delta_{RR}^d)_{23}g_7^{(2)}(x_\sg)\right] \,.
	\end{align}
	\label{eq:C7gluino_sgsg}
\end{subequations}
Here $x_\sg=m_{\tilde{g}}^2/m_{\tilde{q}}^2$, $t_\beta=\upsilon_1/\upsilon_2$, $\epsilon\sim10^{-2}$ for a degenerate SUSY spectrum, and the loop functions $g_7^{(1,2)}(x_\sg)$ can be found in Ref.~\cite{Altmannshofer:2009ne}\footnote{For comparison, see also Refs.~\cite{Gabbiani:1996hi} and \cite{Aushev:2010bq} where $M_{a,b}(x)$ correspond to $M_{1,2}(x)$ from Ref.~\cite{Hisano:2004tf}. In this case $g_7^{(1,2)}(x)=\mp\frac{4}{9}M_{1,a}(x)$, $g_8^{(1,2)}(x)=\mp\frac{1}{6}\left[M_{1,a}(x)+9M_{2,b}(x)\right]$ respectively.}.

We see that the SUSY models with large $(\delta_{RL}^d)_{23}$ can induce large $C_{7\gamma}^{\,\prime\,(\sg)}$. Since in this case the chirality flip occurs inside the loop, the factor $m_b$ of the SM is replaced by the internal gluino mass, i.e. from the first term in Eq.~\eqref{eq:C7gluino_sgsg}. This effect, often referred to as the chiral-enhancement, could lead to a dramatic increase of the right-handed photon emission in $b\to s\gamma$ processes. The last terms in Eq.~\eqref{eq:C7gluino_sgsg} come from the double MIA diagrams with $(\delta_{LR(RL)}^d)_{33}=-(m_b\mu^{(*)}t_\beta+\upsilon_1A_D^{33}/\sqrt2)/m_\sq^2$, and become important for large values of $t_\beta$ which we do not consider here.

{\it Note that the MIA parameters can, in general, be complex} (e.g. see the numbers quoted in Refs.~\cite{Khalil:2002fm,Ball:2003se}). 

\begin{figure}[t!p!]\centering
	\begin{subfigure}
		{\includegraphics[width=0.4\textwidth]{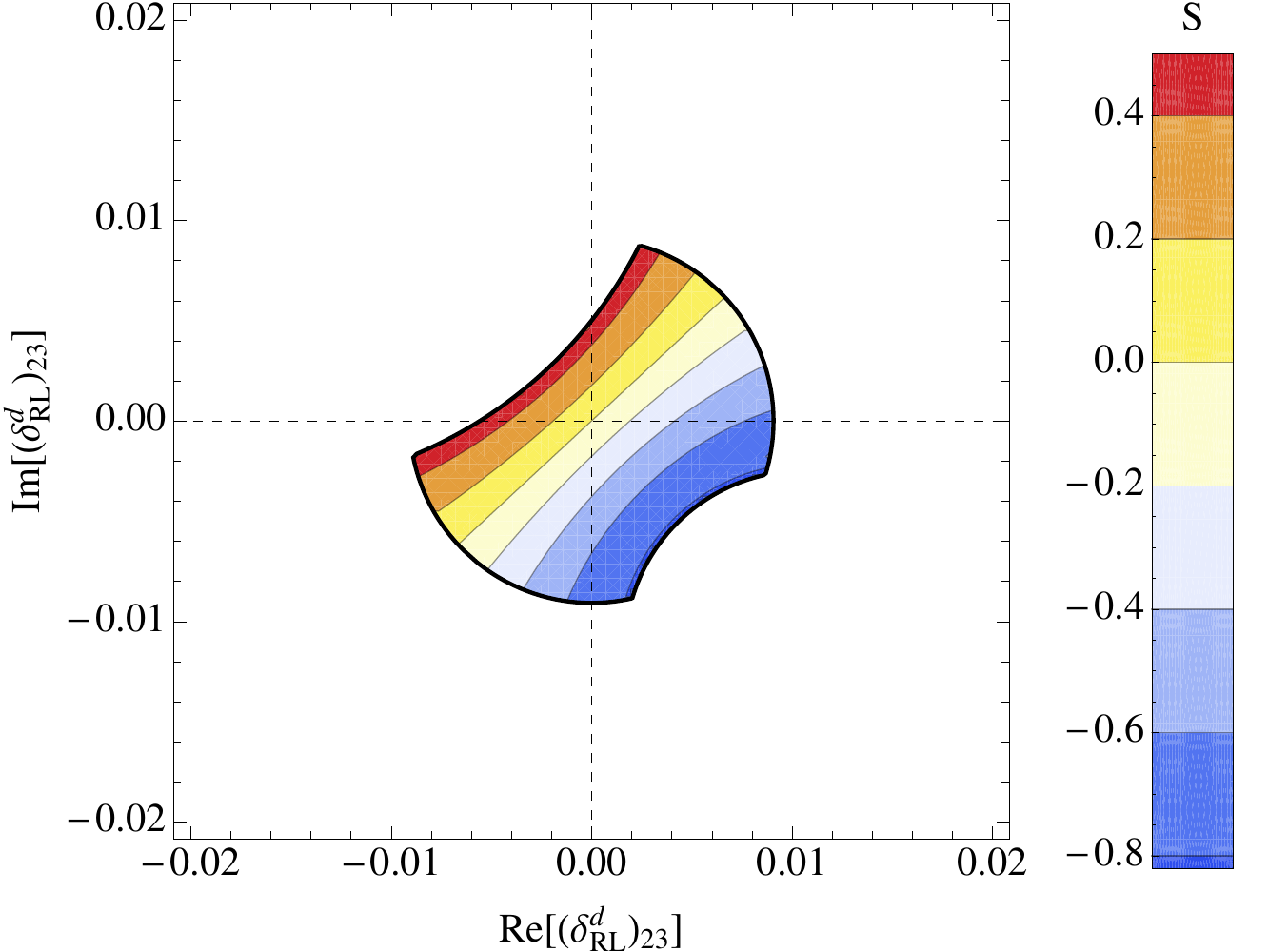}}
		\put(-140,145){\footnotesize$m_\sq\simeq m_\sg=500$~GeV}
		\put(-150,130){\tiny$\bm{B\to K_S\pi^0\gamma}$}
	\end{subfigure}
	\hspace{5mm}
	\begin{subfigure}
		{\includegraphics[width=0.4\textwidth]{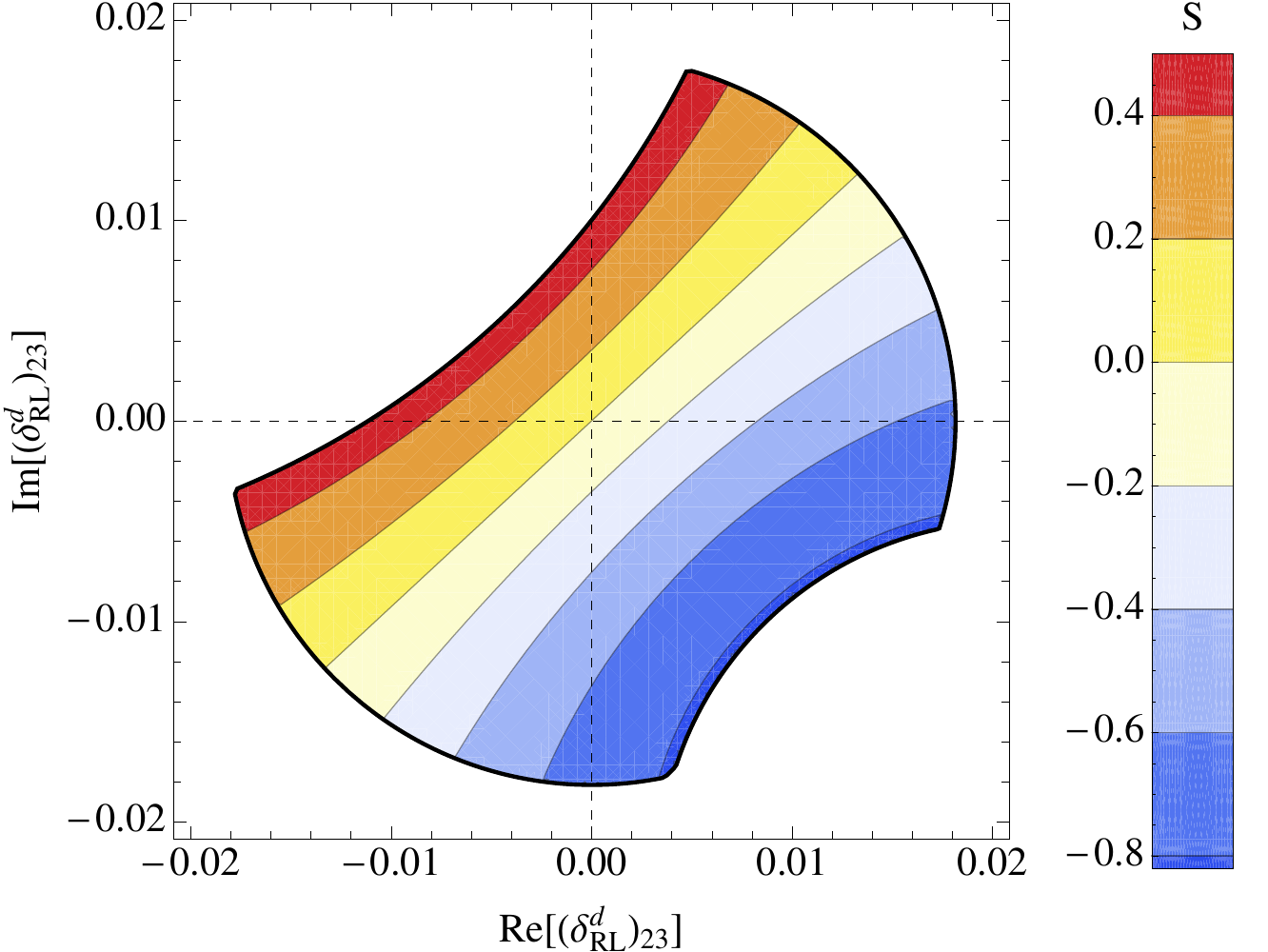}}
		\put(-140,145){\footnotesize$m_\sq\simeq m_\sg=1000$~GeV}
		\put(-150,130){\tiny$\bm{B\to K_S\pi^0\gamma}$}
	\end{subfigure}
	\begin{subfigure}
		{\includegraphics[width=0.4\textwidth]{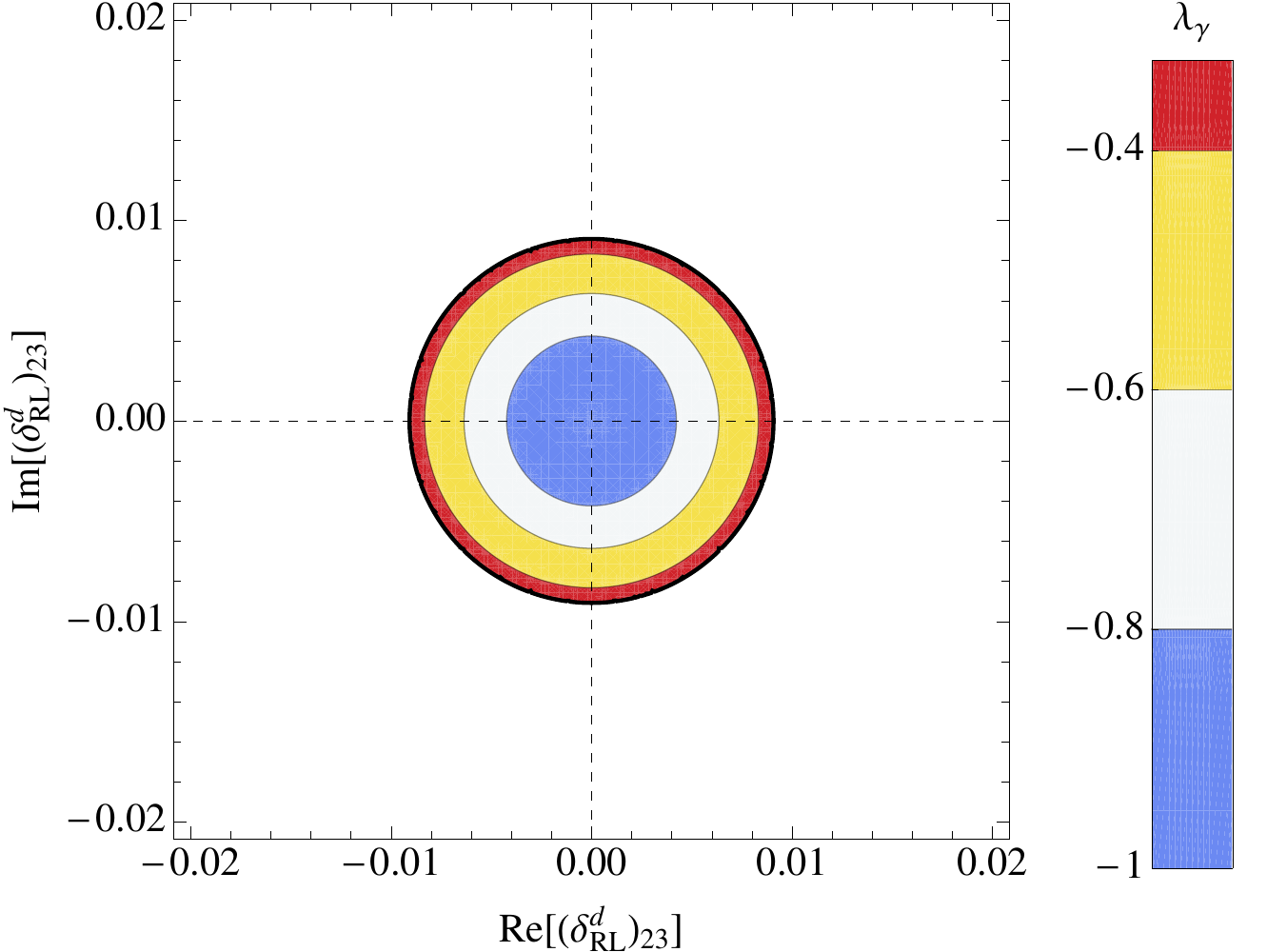}}
		\put(-150,130){\tiny$\bm{B\to K_1\gamma}$}
	\end{subfigure}
	\hspace{5mm}
	\begin{subfigure}
		{\includegraphics[width=0.4\textwidth]{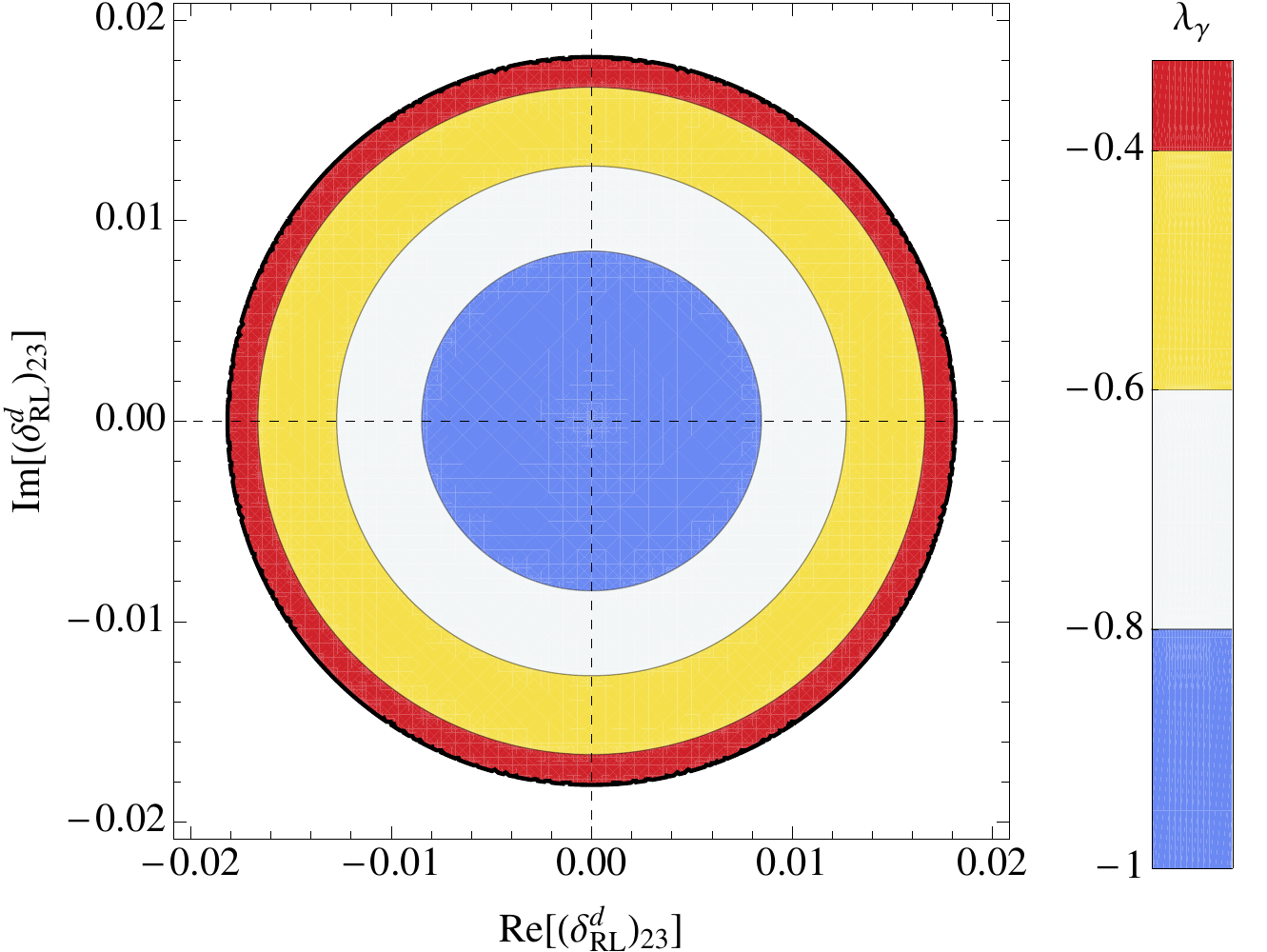}}
		\put(-150,130){\tiny$\bm{B\to K_1\gamma}$}
	\end{subfigure}
	\begin{subfigure}
		{\includegraphics[width=0.4\textwidth]{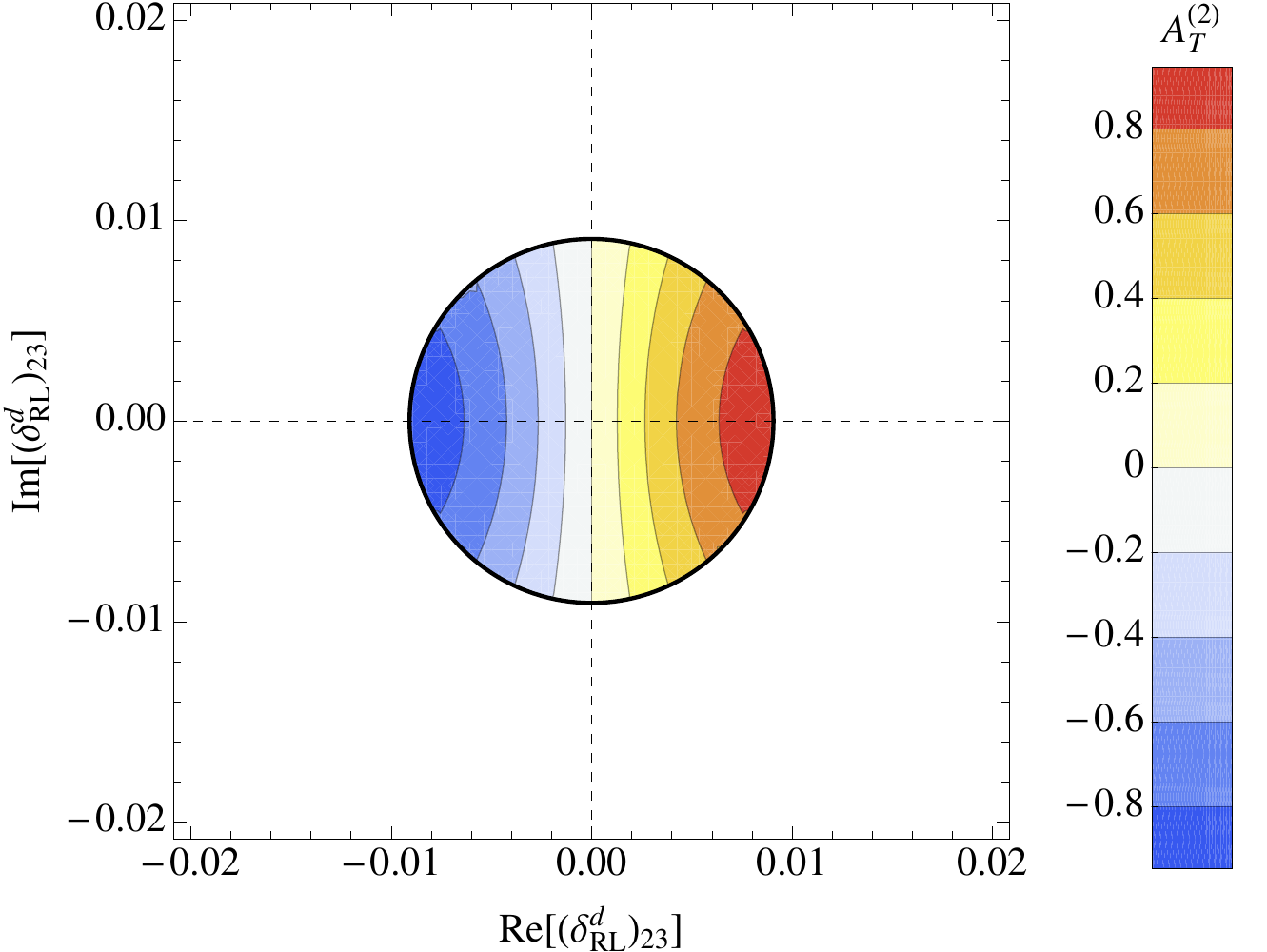}}
		\put(-150,130){\tiny$\bm{B\to K^*\ell^+\ell^-}$}
	\end{subfigure}
	\hspace{5mm}
	\begin{subfigure}
		{\includegraphics[width=0.4\textwidth]{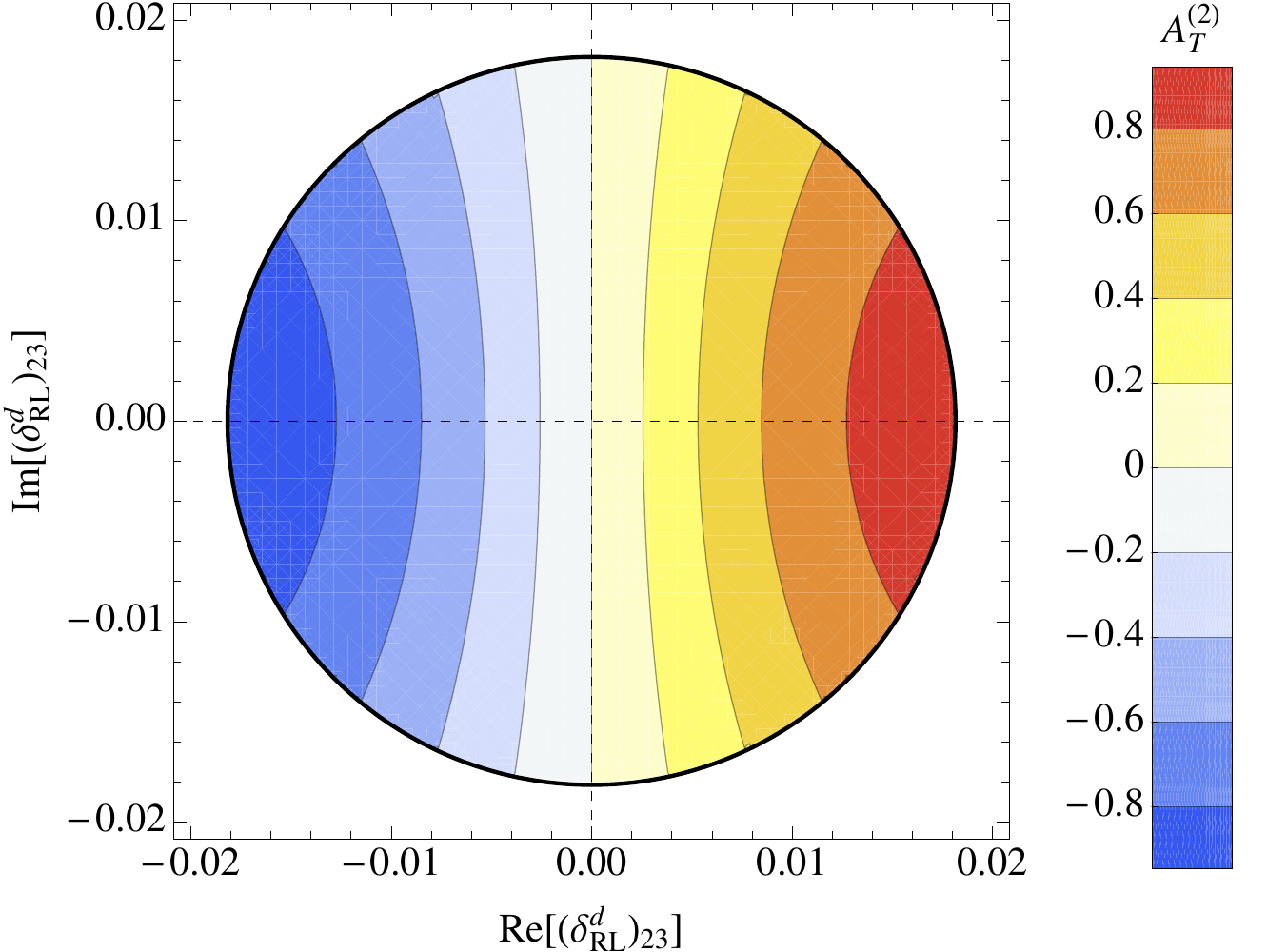}}
		\put(-150,130){\tiny$\bm{B\to K^*\ell^+\ell^-}$}
	\end{subfigure}
	\begin{subfigure}
		{\includegraphics[width=0.4\textwidth]{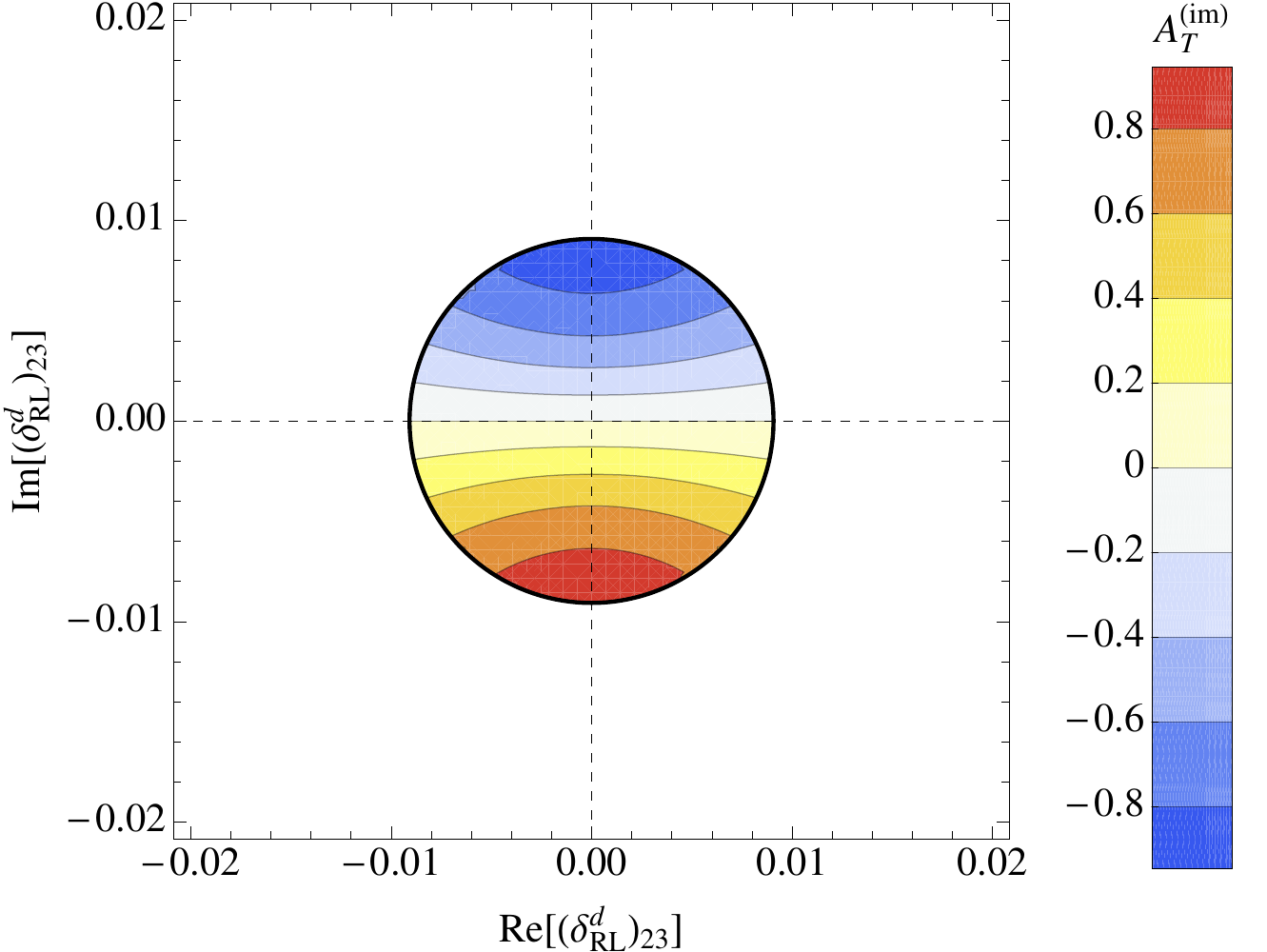}}
		\put(-150,130){\tiny$\bm{B\to K^*\ell^+\ell^-}$}
	\end{subfigure}
	\hspace{5mm}
	\begin{subfigure}
		{\includegraphics[width=0.4\textwidth]{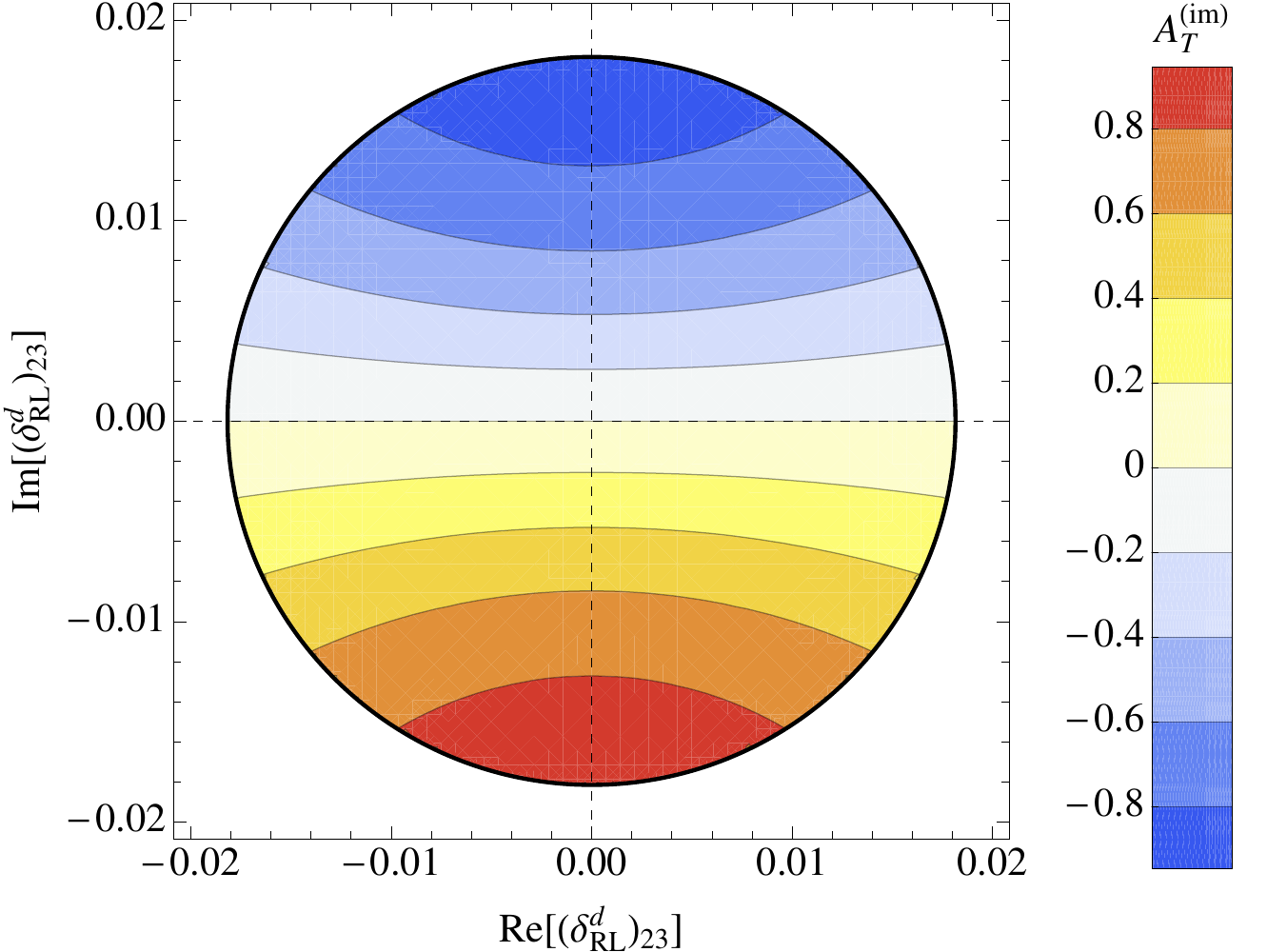}}
		\put(-150,130){\tiny$\bm{B\to K^*\ell^+\ell^-}$}
	\end{subfigure}
	\caption{\footnotesize Prospect of the future constraints on the real and imaginary parts of $(\delta_{RL}^d)_{23}$ for $m_\sq\simeq m_\sg=500$~GeV (on the left) and 1000~GeV (on the right), assuming the other mass insertion parameters in Eq.~\eqref{eq:C7gluino_sgsg} to be negligibly small. The contour colours correspond to $S_{K_S\pi^0\gamma}$, $\lambda_\gamma$, $\A_T^{(2)}(0)$ and $\A_T^{\rm(im)}(0)$ allowed by a $\pm3\sigma$ error to the central value of $\BR^{\rm exp}(B\to X_s\gamma)$.}
	\label{fig:dRL23}
\end{figure}

Using the anomalous dimension matrix from Ref.~\cite{Borzumati:1999qt} and running the coefficients $C_{7\gamma}^{\,(\prime)\,(\sg)}$ from the SUSY scale $M_S$ (See Eq.~\eqref{eq:C7gluino_sgsg}) to the low scale $\mu_b=m_{b,{\rm pole}}$, in Fig.~\ref{fig:dRL23} we show the potential constraints on the dominant $(\delta_{RL}^d)_{23}$ MIA parameter for $m_\sq\simeq m_\sg=500$~GeV (plots on the left) and 1000~GeV (plots on the right) respectively. One can notice that the bounds decrease as $1/m_\sq$. One can see that the future precise measurement of $S_{K_S\pi^0\gamma}$, $\lambda_\gamma$, $\A_T^{(2)}$ and $\A_T^{\rm(im)}$ will allow us to pin down both the real and imaginary parts of $(\delta_{RL}^d)_{23}$.

\section{Discussion of the $\O_2$ contribution to the ``wrong" helicity amplitude \label{sec:QCDcorrections}}

It must be emphasized that due to the QCD effects the right-handed helicity amplitude can receive a non negligible contribution from the operators other than standard electromagnetic penguin operators. Up to now, we have neglected them for simplicity but since it is crucial to know the error on $\Cp_{7\gamma}$, it is now useful to discuss this contribution as well. There are very different estimates, and the discrepancy among them has not been explained in the literature. 

In Refs.~\cite{Grinstein:2004uu,Grinstein:2005nu}, the authors give a general discussion in the framework of the Soft Collinear Effective Theory and end up with two important conclusions: 1) the ``wrong" helicity amplitude is suppressed by a factor $O(\Lambda_{\rm QCD}/m_b)$, 2) it comes mainly from the $\O_2$ operator. The conclusion is a parametric estimate of the ratio
\begin{equation}
	\frac{\M(\Bbar\to\Kbar^*\gamma_R)}{\M(\Bbar\to\Kbar^*\gamma_L)}\sim \frac{(C_2/3)}{C_{7\gamma}}\frac{\Lambda_{\rm QCD}}{m_b}\sim10\%\,.
	\label{eq:CRoverCL_Grinstein}
\end{equation}
{\it This number is not a quantitative estimate since the matrix elements are not known}. Only a rough order of magnitude estimate of the matrix element of the local operator is made. The actual result could be larger or smaller. Furthermore, the above result corresponds to the approximation of zero charm quark mass, $m_c=0$. 

Another quantitative estimate of this ``wrong" chirality contamination is offered with the method of QCD sum rules in the work of Khodjamirian {\it et al.}~\cite{Khodjamirian:1997tg}, and Ball~{\it et~al.}~\cite{Ball:2006cva}. They roughly agree in that the non-perturbative contribution of the $\O_2$ operator\footnote{A complementary estimate using LCSR with $B$ meson wave functions has been given in \cite{Khodjamirian:2010vf}.} is very small, which hardly modifies the tree-level estimate, $m_s/m_b$. The result of Ref.~\cite{Ball:2006cva} is
\begin{equation}
	\frac{\M(\Bbar\to\Kbar^*\gamma_R)}{\M(\Bbar\to\Kbar^*\gamma_L)}\simeq \frac{m_s}{m_b}\times(0.8\pm0.2)\simeq2\% \,.
	\label{eq:CRoverCL_Ball}
\end{equation}
The large numerical discrepancy between Eq.~\eqref{eq:CRoverCL_Grinstein} and \eqref{eq:CRoverCL_Ball} is surprising, since they seem to come from the same basic effect. 

In terms of the effective Hamiltonian the decay amplitude for the $\Bbar\to\Kbar^*\gamma$ decay, can be written as
\begin{equation}
	\begin{split}
		\M(\Bbar\to\Kbar^*\gamma) =& -\frac{4G_F}{\sqrt2}V_{tb}V_{ts}^*\langle\Kbar^*\gamma|C_{7\gamma}\O_{7\gamma}+\Cp_{7\gamma}\Op_{7\gamma} \\
		&+i\varepsilon_\gamma^{\mu*}\sum_{i\neq7\gamma}C_i\int d^4xe^{iqx}T\{j_\mu^{\rm e.m.}(x)\O_i(0)\}|\Bbar\rangle \,,
	\end{split}
	\label{eq:Amp_B-Kstgamma_1}
\end{equation}
and the numbers in Eqs.~\eqref{eq:CRoverCL_Grinstein} and \eqref{eq:CRoverCL_Ball} are estimates of this quantity, with the same basic idea of attaching the electromagnetic current and a soft gluon to a $c$-quark loop starting from the four-fermion operator $\O_2$. A possibility to relieve the helicity suppression of right-handed photons is indeed to consider an additional gluon emission resulting in the three-particle final state $b\to s\gamma g$~\footnote{In the case of the three-particle final state the argument of the helicity conservation in the footnote~1 is no longer valid.}.

Of course, one could explain the discrepancy simply by invoking the fact that the estimate \eqref{eq:CRoverCL_Grinstein} is very approximate, while the other \eqref{eq:CRoverCL_Ball} is based on QCD sum rules. A more careful analysis allows to be more specific. The result in Eq.~\eqref{eq:CRoverCL_Grinstein} comes with the assumption $m_c=0$ in
the loop function 
\begin{equation}
	\kappa(z) = \frac{1}{2}-\frac{2}{z}\arctan^2\left[\sqrt{\frac{z}{4-z}}\right] \,,
	\label{eq:kappaz}
\end{equation}
which takes the value $\kappa(\infty)=1/2$. Indeed, in this expression, $z$ is an {\it operator} acting on the fields, namely
\begin{equation}
	z=\frac{m_b}{m_c^2}(iD_+) \,,
	\label{}
\end{equation} 
where $D$ is the covariant derivative.

For an arbitrary $m_c$, instead, $\kappa(z)$ is a non local operator or a series of local operators with increasing number of additional covariant derivatives, corresponding to the powers of $z$, and with coefficients of order $(\Lambda m_b/m_c^2)^n$, where $\Lambda$ is a hadronic scale. More specifically, the expansion of $\kappa(z)$ is
\begin{equation}
	\kappa(z) = -\frac{z}{24}-\frac{z^2}{180}-\frac{z^3}{1120}+\dots \,.
	\label{eq:kappaz_developpement}
\end{equation} 

The other estimate in Eq.~\eqref{eq:CRoverCL_Ball}, initiated by the work of Khodjamirian {\it et al.}, uses a short distance expansion of the $T-$product appearing in Eq.~\eqref{eq:Amp_B-Kstgamma_1} and retains the lowest order in the expansion, proportional to $1/m_c^2$.

Let us then consider the first term in the expansion \eqref{eq:kappaz_developpement}. We see that the first local operator in the series will have one additional derivative with respect to the local operator at $m_c=0$, and look like the operator $\O_F$ (defined in Ref.~\cite{Khodjamirian:1997tg}), with a coefficient $-1/24(\Lambda m_b/m_c^2)$ instead of $1/2$ for the original non local operator at $m_c=0$. Therefore, we retrieve the power $1/m_c^2$, and a small coefficient for the operator $\O_F$. As a tentative estimate of the derivative operator one can use the standard recipe and replace each derivative by a factor $\Lambda$. Therefore, by setting $z\to\Lambda m_b/m_c^2$, the original estimate in Eq.~\eqref{eq:CRoverCL_Grinstein} becomes\footnote{We accounted for the factor of $2$ used in Refs.~\cite{Grinstein:2004uu,Grinstein:2005nu}, which gives $2\times 1/24$ in Eq.~\eqref{eq:gri}.}
\begin{equation}
	\frac{\M(\Bbar\to\Kbar^*\gamma_R)}{\M(\Bbar\to\Kbar^*\gamma_L)}\sim\frac{(C_2/3)}{C_{7\gamma}}\frac{\Lambda}{m_b}\times\frac{1}{12}\frac{\Lambda m_b}{m_c^2} \,.
	\label{eq:gri}
\end{equation} 
 This new estimate is obviously much smaller than the one given in Eq.~\eqref{eq:CRoverCL_Grinstein}, and explains the discrepancy between Eqs.~\eqref{eq:CRoverCL_Grinstein} and \eqref{eq:CRoverCL_Ball} seems to reside in a rather strong dependence on the charm quark mass when varied from $m_c=0$ to the physical value.

The crucial question is the validity of the limited expansion to the first order in $1/m_c^2$. We can notice that $z\to\Lambda m_b/m_c^2$ is not very small; it is close to 1, so that retaining the first term in the expansion, as done in the sum rules approach, is probably not safe. For $\Lambda=\Lambda_{\rm QCD}$, $z$ is close to $1/m_c^2$. For $\Lambda= \bar \Lambda \simeq 0.5~$GeV of HQET, instead, $z$ is close to 1. We can notice that even when $z=1$, $\kappa(z)\simeq0.05$, which is still $1/10$ of the value $1/2$ on which the numerical estimate of Eq.~\eqref{eq:CRoverCL_Grinstein} seems to be based. {\it We therefore tend to believe that this $\O_2$ contribution to the ``wrong" helicity remains really small}. Nevertheless, one must be aware that this conclusion relies on a highly qualitative feeling of how to estimate $z^n$, which means how to estimate the matrix elements of the local operators. For example, $\Lambda$ could be well replaced by equally reasonable and most naive $1~$GeV. For the latter, $z\simeq3$ and $\kappa(z)$ would be much larger, which would invalidate the short distance expansion. {\it In summary, it seems that once one takes into account the charm quark mass effect, the non perturbative contribution has a rather strong dependence on the scale of the momentum of gluons}.

In addition, the contributions calculated in Ref.~\cite{Matsumori:2005ax} can bring an effect of the order of $(30\div40)\%$ with respect to the leading ($O(\alpha_s^0)$) $m_s/m_b$ contribution. This is larger than the estimate made in Ref.~\cite{Ball:2006cva} (see Eq.~\eqref{eq:CRoverCL_Ball}).

{\it Note also that no calculation for the ratio $\M(\Bbar\to\Kbar_1\gamma_R)/\M(\Bbar\to\Kbar_1\gamma_L)$ has been provided so far.} In general, this ratio should be different from that of $B\to K^*\gamma$ due to the difference in the $B\to K_1$ and $B\to K^*$ hadronic form factors and due to the unknown contribution of the long-distance effects of the $\O_2$ operator. 

On the other hand, {\it for the sake of clarity and simplicity, we have decided not to take into account the long-distance $\O_2$ effects}. However, one must keep in mind that this could entail a theoretical uncertainty~$\sim(2\div10)\%$ on the ratio of the right-handed polarization amplitude over the left-handed one. In other words we are not dealing with high precision tests of the SM and the NP effects can be established only if the deviation from the SM is sufficiently large.

\section{Conclusions}

We have studied the prospects for determining the Wilson coefficients $C_{7\gamma}$ and $\Cp_{7\gamma}$ from the future measurements at LHCb and super $B$~factories. $\Cp_{7\gamma}$ probes the right-handed structure of the New Physics models which enter the $b\to s\gamma$ processes. In order to determine $\Cp_{7\gamma}$, we have used four observables: $S_{K_S\pi^0\gamma}$, $\lambda_\gamma$, $\A_T^{(2)}$ and $\A_T^{\rm(im)}$. 

\begin{itemize}
	\item The current experimental error on the mixing-induced $CP$-asymmetry parameter $S_{K_S\pi^0\gamma}=-0.16\pm0.22$~\cite{Asner:2010qj} will be reduced to $\pm 0.02$, at the super $B$~factories, at 75~ab$^{-1}$~\cite{Bona:2007qt}. 
	\item A direct method to measure the photon polarization in $B\to K_1(\to K\pi\pi)\gamma$ decay was proposed in Refs.~\cite{Gronau:2001ng,*Gronau:2002rz} and improved in~\cite{Kou:2010kn}. Our study shows that the photon polarization parameter $\lambda_\gamma$ can be measured at super $B$~factories with an accuracy $\sim 20\%$, with integrated luminosity of 75~ab$^{-1}$. Instead, at the LHCb one can reach the $\sim 10\%$ precision with only 2~fb$^{-1}$~\cite{Kou:2010kn}. 
	\item From the angular analysis of $B\to K^*(\to K\pi)\ell^+\ell^-$ decay at low dilepton invariant mass one can extract information on the photon polarization as well. The transverse asymmetries $\A_T^{(2)}$ and $\A_T^{\rm(im)}$, are particularly interesting since they will soon be measured to a good accuracy at LHCb. The estimated accuracy of $\A_T^{(2)}$ and $\A_T^{\rm(im)}$ is expected to be $\sim 0.2$ at integrated luminosity of 2~fb$^{-1}$~\cite{Lefrancois:1179865}.
\end{itemize}
In principle, these four observables can unambiguously constrain the New Physics contribution to $\Cp_{7\gamma}$ and $C_{7\gamma}$, even when these Wilson coefficients are complex numbers.

We studied four different NP scenarios of $\CCp_{7\gamma}\neq 0$ and presented the current constraints provided by $\BR(B\to X_s\gamma)$ and $S_{K_S\pi^0\gamma}$. Those constraints are still either loose and/or ambiguous. We then showed that the future measurements of $S_{K_S\pi^0\gamma}$, $\lambda_\gamma$, $\A_T^{(2)}$ and $\A_T^{\rm(im)}$ will not only restrain the allowed range of values for $\CCp_{7\gamma}$, but also solve or partially solve the ambiguities in the complex $(C_{7\gamma},\Cp_{7\gamma})$ plane. 

We should emphasize that each of the above quantities has its own advantages and disadvantages depending on the NP scenario. In the scenario~{\it I}, we found that the bounds coming from $S_{K_S\pi^0\gamma}$ and from $\A_T^{(2)}$ are similar. To disentangle the discrete fourfold ambiguity arising from these two constraints, the measurement of $\lambda_\gamma$ could help and reduce this ambiguity to twofold. 

In the scenario~{\it II}, $\lambda_\gamma$ plays an important role: although $S_{K_S\pi^0\gamma}$ bound will be extremely constraining at super $B$~factories, the resulting diagonal ambiguity could be at least partly solved by a constraint provided by the measured $\lambda_\gamma$. $\A_T^{(2)}$ and $\A_T^{\rm(im)}$ are very important since their combination can, in principle, constrain both $\Re[\Cp_{7\gamma}]$ and $\Im[\Cp_{7\gamma}]$ independently on $S_{K_S\pi^0\gamma}$ and $\lambda_\gamma$. 

In contrast to the scenario~{\it II}, in the scenarios~{\it III} and {\it IV}, constraints provided by $\A_T^{(2)}$ and $\A_T^{\rm(im)}$ leave a twofold ambiguity if $\Cp_{7\gamma}$ is large. This can be removed by adding the constraint coming from the measured $\lambda_\gamma$.

We also discussed the impact of the potential long distance contributions of the $\O_2$ operator that might plague right-handed polarization amplitude. Its contribution, which is estimated to be between $(2\div10)\%$, should be taken into account. Its current estimate is not safe yet, and more effort is needed to assess its value. For that reason the New Physics can be established from the decays studied in this work only if the deviations from the SM are significantly large.

\section*{Acknowledgements}

We would like to thank M.-H.~Schune, J.~Lefran\c{c}ois, F.~Le~Diberder, Y.~Sakai, K.~Trabelsi, M.~Nakao and S.~Hashimoto for very useful discussions and provided information. This work was supported in part by the ANR contract ``LFV-CPV-LHC" ANR-NT09-508531 and France-Japan corporation of IN2P3/CNRS TYL-LIA.

\appendix

\section{Spin amplitudes in the $\Bbar\to\Kbar^*(\to\Kbar\pi)\ell^+\ell^-$ decay}

Using the naive factorization, the matrix element of the effective Hamiltonian for the decay $\Bbar\to\Kbar^*\ell^+\ell^-$ can be written as
\begin{equation}
	\begin{split}
		\M(\Bbar\to&\Kbar^*\ell^+\ell^-) = \frac{G_F\alpha_{\rm em}}{\sqrt2\pi}V_{tb}V_{ts}^* \\
		\times & \biggl\{\biggr.\biggl[(C_9-C_{10})\langle\Kbar^*|\sbar_L\gamma^\mu b_L|\Bbar\rangle+(\Cp_9-\Cp_{10})\langle\Kbar^*|\sbar_R\gamma^\mu b_R|\Bbar\rangle \biggr. \\
		& \left.-\frac{2m_b}{q^2}\left(C_{7\gamma}\langle\Kbar|\sbar_Li\sigma^{\mu\nu}q_\nu b_R|\Bbar\rangle+\Cp_{7\gamma}\langle\Kbar|\sbar_Ri\sigma^{\mu\nu}q_\nu b_L|\Bbar\rangle\right)\right](\lbar_L\gamma_\mu\ell_L) \\
		& +\biggl[(C_9+C_{10})\langle\Kbar^*|\sbar_L\gamma^\mu b_L|\Bbar\rangle+(\Cp_9+\Cp_{10})\langle\Kbar^*|\sbar_R\gamma^\mu b_R|\Bbar\rangle \biggr. \\
		& \left.\left.-\frac{2m_b}{q^2}\left(C_{7\gamma}\langle\Kbar|\sbar_Li\sigma^{\mu\nu}q_\nu b_R|\Bbar\rangle+\Cp_{7\gamma}\langle\Kbar|\sbar_Ri\sigma^{\mu\nu}q_\nu b_L|\Bbar\rangle\right)\right](\lbar_R\gamma_\mu\ell_R)\right\} \,.
	\end{split}
	\label{eq:semileptonic_amp}
\end{equation}

Working in the transversity basis of amplitudes, one can obtain from Eq.~\eqref{eq:semileptonic_amp} the well-established in the literature expressions for the four possible amplitudes \cite{Altmannshofer:2008dz}:
\begin{subequations}
	\begin{align}
		\begin{split}
			A_\perp^{\ell_{L,R}}(q^2) =& N(q^2)\sqrt{2\lambda(q^2)}\left\{\frac{2m_b}{q^2}(C_{7\gamma}+\Cp_{7\gamma})T_1(q^2) \right. \\
			&+ \left.\left[(C_9+C_9)\mp(C_{10}+\Cp_{10})\right]\frac{V(q^2)}{m_B+m_{K^*}} \right\} \,,
		\end{split} \\
		\begin{split}
			A_\parallel^{\ell_{L,R}}(q^2) =& -N(q^2)\sqrt{2}(m_B^2-m_{K^*}^2)\left\{\frac{2m_b}{q^2}(C_{7\gamma}-\Cp_{7\gamma})T_2(q^2) \right. \\
			&+ \left.\left[(C_9-\Cp_9)\mp(C_{10}-\Cp_{10})\right]\frac{A_1(q^2)}{m_B-m_{K^*}} \right\} \,,
		\end{split} \\
		\begin{split}
			A_0^{\ell_{L,R}}(q^2) =& -\frac{N(q^2)}{2m_{K^*}\sqrt{q^2}}\biggl\{\left[(C_9-\Cp_9)\mp(C_{10}-\Cp_{10})\right]\times\biggr. \\
			& \left[(m_B^2-m_{K^*}^2-q^2)(m_B+m_{K^*})A_1(q^2)-\lambda(q^2)\frac{A_2(q^2)}{m_B+m_{K^*}}\right] \\
			+& \left.2m_b(C_{7\gamma}-\Cp_{7\gamma})\left[(m_B^2+3m_{K^*}^2-q^2)T_2(q^2)-\frac{\lambda(q^2)}{m_B^2-m_{K^*}^2}T_3(q^2)\right]\right\} \label{eq:A0LR} \,,
		\end{split} \\
		\begin{split}
			A_t(q^2) =& \frac{2N(q^2)\sqrt{\lambda(q^2)}}{\sqrt{q^2}}(C_{10}-\Cp_{10})A_0(q^2) \label{eq:At} \,,
		\end{split}
	\end{align}
	\label{eq:Kst_amplitudes}
\end{subequations}
where
\begin{subequations}
	\begin{align}
		N(q^2) =& V_{tb}V_{ts}^*\left[\frac{G_F^2\alpha_{\rm em}^2}{2^{10}\pi^5m_B^3}\frac{\beta_\ell(q^2)}{3}q^2\sqrt{\lambda(q^2)}\right]^{1/2} \,, \\
		\beta_\ell(q^2) =& \sqrt{1-\frac{4m_\ell^2}{q^2}} \,, \\
		\lambda(q^2) =& [q^2-(m_B+m_{K^*})^2][q^2-(m_B-m_{K^*})^2] \,,
	\end{align}
	\label{}
\end{subequations}
and $V(q^2)$, $A_{0,1,2}(q^2)$, $T_{1,2,3}(q^2)$ are the form factors which parametrize the hadronic matrix elements in Eq.~\eqref{eq:semileptonic_amp}.

In order to avoid possible confusion of the reader, it is worth to mention that the superscripts $L,R$ in the notation $A_{\parallel,\perp,0}^{L,R}$, which are commonly used in the literature (see Ref.~\cite{Kruger:2005ep} and all subsequent works), are {\it not} related to the $K^*$ or the virtual photon helicity/chirality amplitudes; instead they must be identified with the lepton chirality as in Eq.~\eqref{eq:semileptonic_amp} (e.g. see Ref.~\cite{Kim:2000dq}). Therefore we modified this notation by adding the lepton index: $A_{\parallel,\perp,0}^{L,R}\to A_{\parallel,\perp,0}^{\ell_{L,R}}$. One has to point out that there exist two different independent $L,R$ amplitudes since vector and axial vector couplings of the $Z$-boson to the leptonic current, which are contained in the $C_{9,10}$ coefficients, are different.

\section{Intercepts and slopes of $\A_T^{\rm(2,\,im)}(q^2)$ at low $q^2$}

In the limit of vanishing $q^2$ we get the intercepts and the slopes in $q^2$ of the transverse asymmetries defined in Eq.~\eqref{eq:AT2im_linear_approx}
\begin{subequations}
	\begin{align}
		a_0^{(2)} =& \frac{2\Re[C_{7\gamma}\Cpc_{7\gamma}]}{|C_{7\gamma}|^2+|\Cp_{7\gamma}|^2} \,, \\
		a_0^{\rm(im)} =& \frac{2\Im[C_{7\gamma}\Cpc_{7\gamma}]}{|C_{7\gamma}|^2+|\Cp_{7\gamma}|^2} \,,
	\end{align}
	\label{eq:intersepts}
\end{subequations}

\begin{subequations}
	\begin{align}
		\begin{split}
			a_1^{(2)} =& -\left(\frac{|C_{7\gamma}-\Cp_{7\gamma}||C_{7\gamma}+\Cp_{7\gamma}|}{|C_{7\gamma}|^2+|\Cp_{7\gamma}|^2}\right)^2 \left[\frac{m_B^2+m_{K^*}^2}{(m_B+m_{K^*})^2}+z\right] \\
			& -\frac{1}{2m_b}\left(\frac{|C_{7\gamma}+\Cp_{7\gamma}|^2}{|C_{7\gamma}|^2+|\Cp_{7\gamma}|^2}\right)^2\Re[(C_{7\gamma}-\Cp_{7\gamma})(C_9-\Cp_9)]\frac{A_1/T_2}{m_B-m_{K^*}} \\
			& +\frac{1}{2m_b}\left(\frac{|C_{7\gamma}-\Cp_{7\gamma}|^2}{|C_{7\gamma}|^2+|\Cp_{7\gamma}|^2}\right)^2\Re[(C_{7\gamma}+\Cp_{7\gamma})(C_9+\Cp_9)]\frac{V/T_1}{m_B+m_{K^*}} \,,
		\end{split} \\
		& \nonumber \\
		\begin{split}
			a_1^{\rm(im)} =& \frac{4\Re[C_{7\gamma}\Cpc_{7\gamma}]\Im[C_{7\gamma}\Cpc_{7\gamma}]}{(|C_{7\gamma}|^2+|\Cp_{7\gamma}|^2)^2} \left[\frac{m_B^2+m_{K^*}^2}{(m_B+m_{K^*})^2}+z\right] \\
			& -\frac{1}{m_b}\frac{\Im[C_{7\gamma}\Cpc_{7\gamma}]}{(|C_{7\gamma}|^2+|\Cp_{7\gamma}|^2)^2}\Re\left[(C_{7\gamma}-\Cp_{7\gamma})(C_9-\Cp_9)^*\frac{A_1/T_2}{m_B-m_{K^*}} \right. \\
			& \left.+(C_{7\gamma}+\Cp_{7\gamma})(C_9+\Cp_9)^*\frac{V/T_1}{m_B+m_{K^*}} \right] \\
			& -\frac{1}{2m_b}\frac{\Im[(C_{7\gamma}+\Cp_{7\gamma})(C_9-\Cp_9)^*]}{|C_{7\gamma}|^2+|\Cp_{7\gamma}|^2}\frac{A_1/T_2}{m_B-m_{K^*}} \\
			& +\frac{1}{2m_b}\frac{\Im[(C_{7\gamma}-\Cp_{7\gamma})(C_9+\Cp_9)^*]}{|C_{7\gamma}|^2+|\Cp_{7\gamma}|^2}\frac{V/T_1}{m_B+m_{K^*}} \,.
		\end{split}
	\end{align}
	\label{eq:slopes}
\end{subequations}

The ratios of the form factors that have similar $q^2$-behavior in the heavy quark limit and in the limit of large energy of $K^*$, are kept as constants~\cite{Charles:1998dr}, namely $A_1(q^2)/T_2(q^2)\equiv A_1/T_2=\rm{const}$ and $V(q^2)/T_1(q^2)\equiv A_1/T_2=\rm{const}$. These ratios satisfy the approximate relation~\cite{Becirevic:2011bp}
\begin{equation}
	\frac{A_1/T_2}{m_B-m_{K^*}}\approx \frac{V/T_1}{m_B+m_{K^*}}\approx0.2~\rm{GeV}^{-1} \,,
\end{equation}
which in practice we vary between $(0.17\div0.23)~{\rm GeV}^{-1}$~\cite{Ball:2004rg,Colangelo:1995jv,Becirevic:2006nm}. For the ratio of the tensor form factors we use the approximation
\begin{equation}
	\frac{T_2(q^2)}{T_1(q^2)}\approx 1+zq^2 \,,
\end{equation}
with $z=-0.030(3)$~\cite{Ball:2004rg,Becirevic:2006nm}.

\vspace{1cm}

\bibliographystyle{utphys}
\newpage
\bibliography{bibliography}

\end{document}